\documentclass[aps,superscriptaddress,twocolumn,dvips]{revtex4}
\usepackage{feynmp}
\usepackage{amsmath}
\usepackage{amssymb}
\usepackage{epsfig}
\usepackage{graphicx}
\usepackage{subfigure}
\usepackage{bbm}
\usepackage{color}
\usepackage{hyperref}
\usepackage{booktabs}

\newcommand{\Rmnum}[1]{\expandafter\@slowromancap\romannumeral #1@}
\allowdisplaybreaks[3]

\begin{document}

\title{Breakdown of Fermi liquid theory in topological multi-Weyl semimetals}

\author{Jing-Rong Wang}
\affiliation{Anhui Province Key Laboratory of Condensed Matter
Physics at Extreme Conditions, High Magnetic Field Laboratory of the
Chinese Academy of Science, Hefei, Anhui 230031, China}
\author{Guo-Zhu Liu}
\altaffiliation{Corresponding author: gzliu@ustc.edu.cn}
\affiliation{Department of Modern Physics, University of Science and
Technology of China, Hefei, Anhui 230026, China}
\author{Chang-Jin Zhang}
\altaffiliation{Corresponding author: zhangcj@hmfl.ac.cn}
\affiliation{Anhui Province Key Laboratory of Condensed Matter
Physics at Extreme Conditions, High Magnetic Field Laboratory of the
Chinese Academy of Science, Hefei, Anhui 230031, China}
\affiliation{Institute of Physical Science and Information
Technology, Anhui University, Hefei, Anhui 230601, China}

\begin{abstract}
Fermi liquid theory works very well in most normal metals, but is
found violated in many strongly correlated electron systems, such as
cuprate and heavy-fermion superconductors. A widely accepted
criterion is that, the Fermi liquid theory is valid when the
interaction-induced fermion damping rate approaches zero more
rapidly than the energy. Otherwise, it is invalid. Here, we
demonstrate that this criterion breaks down in topological double-
and triple-Weyl semimetals. Renormalization group analysis reveals
that, although the damping rate of double- and triple-Weyl fermions
induced by the Coulomb interaction approaches zero more rapidly than
the energy, the quasiparticle residue vanishes and the Fermi liquid
theory is invalid. This behavior indicates a weaker-than-marginal
violation of the Fermi liquid theory. Such an unconventional
non-Fermi liquid state originates from the special dispersion of
double- and triple-Weyl fermions, and is qualitatively different
from all the other Fermi-liquid and non-Fermi-liquid states. The
predicted properties of the fermion damping rate and the spectral
function can be probed by the angle-resolved photoemission
spectroscopy. The density of states, specific heat, and
conductivities are also calculated and analyzed after incorporating
the corrections induced by the Coulomb interaction.
\end{abstract}

\maketitle


\section{Introduction}

Fermi liquid (FL) theory \cite{GiulianiBook, ColemanBook},
originally proposed by Landau, provides a qualitatively correct and
even quantitatively precise description of a plethora of interacting
fermion systems. In a FL, the interaction-induced fermion damping
becomes progressively unimportant as the energy is lowered. The
fermionic excitations are long-lived near the Fermi surface, and can
be described by the model of free fermion gas. Usually, the only
quantum many-body effect is the regular renormalization of a few
number of parameters, such as the fermion mass, which greatly
simplifies theoretical treatment.

To judge whether the FL theory is applicable in an interacting
fermion system, one needs to first develop an effective criterion.
The inter-particle interaction leads to a fermion damping rate,
defined as $\Gamma(\omega) = |\mathrm{Im}\Sigma^R(\omega)|$, where
$\mathrm{Im}\Sigma^R(\omega)$ is the imaginary part of retarded
fermion self-energy \cite{GiulianiBook, ColemanBook, Schofield99,
Varma02}. Pauli's exclusion principle guarantees that
$\Gamma(\omega)$ must vanish at the Fermi surface, namely
$\Gamma(\omega\rightarrow 0) \rightarrow 0$. However,
$\Gamma(\omega)$ might go to zero quickly or slowly, depending on
the nature and the strength of the inter-particle interaction.
According to the traditional quantum many-body theory, the FL theory
is valid if the fermion damping rate approaches to zero more rapidly
than the energy $\omega$ in the limit $\omega \rightarrow 0$
\cite{Varma02}, namely
\begin{eqnarray}
\lim_{\omega\rightarrow 0} \frac{\Gamma(\omega)}{\omega} \rightarrow
0. \label{Eq:FLNJLcriterion}
\end{eqnarray}
If this criterion is not satisfied, the fermion damping is believed
to be strong enough to destroy the coherent quasiparticles, leading
to the breakdown of FL description \cite{Schofield99, Varma02}.

In ordinary metals, the Coulomb interaction is short-ranged due to
Debye screening, and only causes weak damping. It is
well-established \cite{GiulianiBook, ColemanBook, Schofield99,
Varma02} that the fermion damping rate behaves as $\Gamma(\omega)
\propto \omega^2$ in three-dimensional (3D) metals and
$\Gamma(\omega) \propto \omega^2\ln\left(\omega_0/\omega\right)$ in
two-dimensional (2D) metals, respectively. In both cases, the
criterion (\ref{Eq:FLNJLcriterion}) is satisfied, and the FL theory
works well. When fermions couple to some gapless bosonic mode, such
as U(1) gauge boson \cite{Halperin93, Lee06} or the quantum critical
fluctuation of a local order parameter \cite{Abanov03, Lohneysen07,
Fradkin10, Fradkin15, Shibauchi14}, the damping rate might take the
form $\Gamma(\omega) \propto \omega^a$ with $0<a\leq 1$. For
instance, the damping rate is $\Gamma(\omega) \propto \omega^{2/3}$
at the ferromagnetic (FM) quantum critical point (QCP)
\cite{Varma02, Lohneysen07} and nematic QCP \cite{Fradkin10}, and
$\Gamma(\omega) \propto \omega^{1/2}$ at the antiferromagnetic (AFM)
QCP \cite{Abanov03, Lohneysen07}. When $0<a\leq 1$, the criterion
(\ref{Eq:FLNJLcriterion}) is no longer satisfied and the FL theory
becomes invalid. The case of $a = 1$ is very special and defines a
marginal Fermi liquid (MFL) \cite{Varma02, Varma89, Keimer15}, which
offers a good phenomenological description of the unusual normal
state of cuprate superconductors. In a MFL, $\Gamma(\omega)$ and
$\omega$ approach to zero in the same way. MFL has long been
regarded as the weakest imaginable violation of FL theory
\cite{Varma89, Varma02, Sperstad11, Sutherland12}.

Although the criterion (\ref{Eq:FLNJLcriterion}) has been commonly
used to judge whether or not the FL theory is valid in all the
previously studied fermion systems, we will demonstrate in this
paper that this criterion is not always efficient. Remarkably, we
find that this criterion breaks down in both double- and triple-Weyl
semimetals (WSMs), which carry multiple monopole charges and are
thus topologically nontrivial \cite{Armitage18, YangNagaosa14, Xu11,
Fang12, Lai15, Jian15, Roy15, Chen16, Dai16, Bera16, Zhai16,
Lepori16, LiXiao16, Sbierski17, Mai17, Anh17, Park17, Roy17, Yan17,
Huang17, Hayata17, Sun17, Jian17, Gorbar17, WangLiuZhang17B,
Ezawa17, ZhangShiXin17, Mukherjee18, Lepori18}. In these two types
of semimetal (SM), the valence and conduction bands touch only at
isolated points. When the Fermi level is adjusted to the
band-touching points, the fermion density of states (DOS) vanishes
and the Coulomb interaction remains long-ranged. Owing to the
special dispersion of double- and triple-Weyl fermions, the Coulomb
interaction may induce highly unusual low-energy behaviors that do
not occur in other interacting fermion systems. To address this
issue, we perform a renormalization group (RG) analysis
\cite{Shankar94} beyond the instantaneous approximation to obtain
the scale dependence of all the model parameters. The fermion
damping rate $\Gamma(\omega)$ obtained in our RG analysis satisfies
the criterion (\ref{Eq:FLNJLcriterion}), thus the Landau damping of
double- and triple-Weyl fermions is weaker than that of MFL. Based
on the traditional criterion, one would naively conclude that the FL
theory is valid. This is actually incorrect. Indeed, we find that
the quasiparticle residue $Z_{f}$ decreases all the way upon
approaching to the Fermi level, albeit at a small speed. In the zero
energy limit, $Z_f$ vanishes. Obviously, there is no overlap between
the interacting and free fermions, and both double- and triple-WSMs
exhibit an unconventional non-FL ground state. This non-FL state is
qualitatively distinct from all the other FLs and non-FLs, and
provides an example of weaker-than-marginal violation of the FL
theory.

We thus see that, the widely adopted criterion for the validity of
FL theory is actually incomplete. It is necessary to develop a more
complete formalism to define and classify non-FL states. The
unconventional violation of FL theory originates from the special
energy dispersion of fermions in double- and triple-WSMs. Our work
provides important new insight into the intriguing quantum many-body
effects.

The unconventional non-FL state predicted in this work is
experimentally detectable. The damping rate and spectral function of
double- and triple-Weyl fermions exhibit unique and distinguishable
features, which could be explored by performing angle-resolved
photoemission spectroscopy (ARPES) experiments \cite{Damascelli03,
Valla99, Kaminski01, Richard09, Miao16, Siegel11, Pan12, Xu14,
Kondo15} in several candidate double- and triple-WSM materials,
provided that the system is tuned close to the band-touching point.
Moreover, the fermion DOS, specific heat, and conductivities can
also be used to characterize such a non-FL state. We will calculate
these observable quantities and analyze the interaction corrections.

The rest of the paper will be organized as follows. The model
Hamiltonian and the propagators for Weyl fermions are described in
Sec.~\ref{Sec:Hamiltonian}. In Sec.~\ref{Sec:RGAnalysis}, we present
the coupled RG flow equations for the model parameters, and analyze
the numerical results. On the basis of the solutions, we demonstrate
that double- and triple-Weyl fermions exhibit unconventional non-FL
behaviors, which is distinct from all the other types of fermion. We
then calculate a number of observable quantities in
Sec.~\ref{Sec:ObserQuant} and analyze their low-energy properties.
We discuss how to experimentally probe the observable effects of the
non-FL state and give a brief remark on the impact of finite
chemical potential in Sec.~\ref{Sec:ExpDetection}. We summarize the
main results of this paper and compare to previous relevant works in
Sec.~\ref{Sec:Summary}. All the computational details are presented
in
Appendices~\ref{App:Polarization}-\ref{App:ObservableQuantitiesInteraction}.

\section{Model Hamiltonian \label{Sec:Hamiltonian}}

The Hamiltonian for free double-Weyl fermions is
\begin{eqnarray}
H_{d} &=& \sum_{j=1}^{N}\int d^3\mathbf{x}
\psi_{d,j}^{\dag}(\mathbf{x}) \mathcal{H}_{d}(\mathbf{x})
\psi_{d,j}(\mathbf{x}),
\nonumber \\
\mathcal{H}_{d}(\mathbf{x}) &=& Ad_{1}(\mathbf{x})\sigma_{x} +
Ad_{2}(\mathbf{x})\sigma_{y} +vd_{3}(\mathbf{x})\sigma_{z},
\end{eqnarray}
where $d_{1}(\mathbf{x})=-\left(\partial_{x}^{2} -
\partial_{y}^{2}\right)$,
$d_{2}(\mathbf{x}) = -2\partial_{x}\partial_{y}$, and
$d_{3}(\mathbf{x}) = -i\partial_{z}$. The Hamiltonian for the free
triple-Weyl fermions is written as
\begin{eqnarray}
H_{t} &=& \sum_{j=1}^{N}\int d^3\mathbf{x}
\psi_{t,j}^{\dag}(\mathbf{x}) \mathcal{H}_{t}(\mathbf{x})
\psi_{t,j}(\mathbf{x}),
\nonumber \\
\mathcal{H}_{t}(\mathbf{x}) &=& Bg_{1}(\mathbf{x})\sigma_{x} +
Bg_{2}(\mathbf{x})\sigma_{y}+ vg_{3}(\mathbf{x})\sigma_{z},
\end{eqnarray}
where $g_{1}(\mathbf{x})=i\left(\partial_{x}^{3} - \partial_{x}
\partial_{y}^{2}\right)$, $g_{2}(\mathbf{x}) =
i\left(\partial_{y}^{3} - \partial_{y}\partial_{x}^{2}\right)$, and
$g_{3}(\mathbf{x})=-i\partial_{z}$. Here, we use $\psi_{d,j}$ and
$\psi_{t,j} $ to represent the two-component spinor fields for
double- and triple-Weyl fermions, respectively. The index $j$ is $j
= 1, 2, ..., N$ with $N$ being the number of fermion flavor.
$\sigma_{x, y, z}$ are the standard Pauli matrices. Two model
parameters $A$ and $v$ are introduced to characterize the spectrum
of double-Weyl fermions, satisfying
\begin{eqnarray}
E_{d}(\mathbf{k}) = \sqrt{A^{2}k_{\bot}^{4} + v^2k_{z}^{2}}.
\end{eqnarray}
The energy spectrum of triple-Weyl fermions is given by
\begin{eqnarray}
E_{t}(\mathbf{k}) = \sqrt{B^{2}k_{\bot}^{6} + v^2k_{z}^{2}},
\end{eqnarray}
where $B$ is another independent model parameter. For double- and
triple-Weyl fermions, the monopole charges are known to be $\pm 2$
and $\pm 3$ \cite{Armitage18, YangNagaosa14, Xu11, Fang12, Lai15,
Jian15, Roy15, Chen16, Dai16, Bera16, Zhai16, Lepori16, LiXiao16,
Sbierski17, Mai17, Anh17, Park17, Roy17, Yan17, Huang17, Hayata17,
Sun17, Jian17, Gorbar17, WangLiuZhang17B, Ezawa17, ZhangShiXin17,
Mukherjee18, Lepori18}, respectively. As a comparison, the monopole
changes are $\pm 1$ for usual Weyl fermions \cite{Armitage18}. The
interesting topological properties of double- and triple-WSMs are
directly related to the above fermion dispersions.

The fermions are subject to the long-range Coulomb interaction,
which is described by
\begin{eqnarray}
H_{\mathrm{C}} = \frac{1}{4\pi}\sum_{j=1}^{N}\int d^3\mathbf{x} d^3
\mathbf{x}'\rho_{j}(\mathbf{x}) \frac{e^2}{\epsilon\left|\mathbf{x} -
\mathbf{x}'\right|}\rho_{j}(\mathbf{x}'),
\end{eqnarray}
where the fermion density operator is given by $\rho_{j}(\mathbf{x})
= \psi_{j}^{\dag}(\mathbf{x}) \psi_{j}(\mathbf{x})$. We use $e$ to
denote electron charge and $\epsilon$ dielectric constant. The total
model will be treated by making perturbative expansion in powers of
$1/N$.

The free propagator of double-Weyl fermions reads
\begin{eqnarray}
G_{d0}(i\omega,\mathbf{k}) = \frac{1}{i\omega - A
d_{1}(\mathbf{k})\sigma_{x}-Ad_{2}(\mathbf{k})\sigma_{y}
-vk_{z}\sigma_{z}},
\label{Eq:FreeFermonPropagatorDoubleWeylDef}
\end{eqnarray}
where $d_{1}(\mathbf{k})=\left(k_{x}^{2}-k_{y}^{2}\right)$ and
$d_{2}(\mathbf{k})=2k_{x}k_{y}$, and the free propagator of
triple-Weyl fermions is
\begin{eqnarray}
G_{t0}(i\omega,\mathbf{k}) = \frac{1}{i\omega -
Bg_{1}(\mathbf{k})\sigma_{x} - Bg_{2}(\mathbf{k})\sigma_{y} -
vk_{z}\sigma_{z}},
\label{Eq:FreeFermonPropagatorTripleWeylDef}
\end{eqnarray}
where $g_{1}(\mathbf{k}) = \left(k_{x}^{3}-3k_{x}k_{y}^{2}\right)$
and $g_{2}(\mathbf{k}) = \left(k_{y}^{3}-3k_{y}k_{x}^{2}\right)$.
Independent of fermion dispersion, the bare Coulomb interaction
function has the form in momentum space:
\begin{eqnarray}
V_{0}(\mathbf{q}) = \frac{4\pi e^2}{\epsilon\left(q_{\bot}^{2}+\zeta
q_{z}^{2}\right)} = \frac{4\pi\alpha v}{q_{\bot}^{2}+\zeta
q_{z}^{2}},
\end{eqnarray}
where $\alpha = e^2/\epsilon v$ serves as an effective interaction
strength. Since the fermion dispersion is anisotropic, the momentum
components $q_{\bot}$ and $q_z$ should be re-scaled differently
under RG transformations. To facilitate RG calculation, we introduce
the parameter $\zeta$ and require that $q_{\bot}$ and $\sqrt{\zeta}
q_z$ scale in the same way. After including the dynamical screening
caused by the polarization function, one can write the dressed
Coulomb interaction function as
\begin{eqnarray}
V_{d,t}(i\Omega,\mathbf{q}) = \frac{1}{V_{0}^{-1}(\mathbf{q}) +
\Pi_{d,t}(i\Omega,\mathbf{q})},\label{Eq:DressedCoulomb}
\end{eqnarray}
in which the polarization function for double- and triple-Weyl
fermions are defined as
\begin{eqnarray}
\Pi_{d}(i\Omega,\mathbf{q}) &=& -N\int\frac{d\omega}{2\pi}
\frac{d^3\mathbf{k}}{(2\pi)^{3}} \mathrm{Tr}
\left[G_{d0}(i\omega,\mathbf{k})\right.\nonumber
\\
&&\times\left.G_{d0}(i\omega + i\Omega, \mathbf{k} +
\mathbf{q})\right]
\end{eqnarray}
and
\begin{eqnarray}
\Pi_{t}(i\Omega,\mathbf{q}) &=&
-N\int\frac{d\omega}{2\pi}\frac{d^3\mathbf{k}}{(2\pi)^{3}}
\mathrm{Tr}\left[G_{t0}(i\omega,\mathbf{k})\right.\nonumber
\\
&&\left.\times G_{t0}(i\omega + i\Omega,\mathbf{k} +
\mathbf{q})\right].
\end{eqnarray}
These two functions are computed in Appendix~\ref{App:Polarization}.
They can be well approximated by the following analytical
expressions:
\begin{eqnarray}
\Pi_{d}(i\Omega,q_{\bot},q_{z}) &=& N\left[\frac{q_{\bot}^{2}}{3
\pi^{2} v}\ln\left(\frac{\sqrt{A}\Lambda_{UV}} {\left(\Omega^2 +
A^{2}q_{\bot}^{4}\right)^{1/4}}+1\right)\right.\nonumber
\\
&&\left.+\frac{1}{64 A}\frac{vq_{z}^{2}}{\sqrt{\Omega^2 + v^{2}
q_{z}^{2}}}\right],\label{Eq:PolaDWMainText}
\end{eqnarray}
and
\begin{eqnarray}
\Pi_{t}(i\Omega,q_{\bot},q_{z}) &=& N\left[\frac{3q_{\bot}^{2}}{4
\pi^2v} \ln\left(\frac{B^{\frac{1}{3}}\Lambda_{UV} }{\left(\Omega^2
+ B^{2}q_{\bot}^{6}\right)^{1/6}}+1\right)\right.\nonumber
\\
&&\left.+c_{t}\frac{1}{B^{2/3}}\frac{vq_{z}^{2}}{\left(\Omega^2 +
v^2q_{z}^{2}\right)^{2/3}}\right],\label{Eq:PolaTWAnsatz3}
\end{eqnarray}
where $c_{t}$ is a constant satisfying
\begin{eqnarray}
c_{t} = \frac{2^{1/3}
\pi^{1/2}}{90\Gamma\left(5/6\right)
\Gamma(2/3)}.
\end{eqnarray}
The Coulomb interaction remains long-ranged because
$\Pi_{d,t}(0,\mathbf{q})$ vanish in the limit $\mathbf{q}\rightarrow
0$.

\section{Renormalization group study \label{Sec:RGAnalysis}}

To determine how model parameters flow with varying energy scale, we
will employ the perturbative RG method \cite{Shankar94, Lai15,
Jian15, ZhangShiXin17, Gonzalez99, Hofmann14, Goawami11, Hosur12,
Throckmorton15, Moon13, Herbut14, YangNatPhys14, Huh16, Isobe16,
WangLiuZhang17A}. Different from previous works on related topic
\cite{Lai15, Jian15, WangLiuZhang17B, ZhangShiXin17}, we will not
adopt the instantaneous approximation and incorporate the dynamical
screening of Coulomb interaction in our RG calculations.

\subsection{Flow equations}

The self-energy of double-Weyl fermions induced by the Coulomb
interaction is formally given by
\begin{eqnarray}
\Sigma_{d}(i\omega,\mathbf{k}) &=& \int'\frac{d\Omega}{2\pi}
\frac{d^{3}\mathbf{q}}{(2\pi)^{3}}G_{d0}(i\omega+i\Omega,\mathbf{k}
+ \mathbf{q}) \nonumber \\
&&\times V_{d}(i\Omega,\mathbf{q}),\label{Eq:DWFermionSelfEnergyMinText}
\end{eqnarray}
where the notation $\int'$ implies that a momentum shell will be
properly chosen in the calculation. Here it is convenient to choose
the momentum shell $b\Lambda< E_{d}(\mathbf{k})<\Lambda$, where
$b=e^{-\ell}$ with $\ell$ being a flow parameter. According to the
calculations detailed in Appendix~\ref{App:SelfEnergyDW},
$\Sigma_{d}$ can be approximated as
\begin{eqnarray}
\Sigma_{d}(i\omega,\mathbf{k}) &\approx& \left\{i\omega C_{d1} -
A\left[d_{1}(\mathbf{k})\sigma_{x} + d_{2}(\mathbf{k})
\sigma_{y}\right] C_{d2}\right.\nonumber
\\
&&\left.-vk_{z}\sigma_{z}C_{d3}\right\}\ell,
\label{Eq:DWSelfEnergyResultMainText}
\end{eqnarray}
to the leading order. The expressions of $C_{d1}$, $C_{d2}$, and
$C_{d3}$ are shown in
Eqs.~(\ref{Eq:ExpressionCd1})-(\ref{Eq:ExpressionGd}) in
Appendix~\ref{App:SelfEnergyDW}.

According to the calculations of
Appendix~\ref{App:DerivationRGEquationDW}, we find that the coupled
RG equations are
\begin{eqnarray}
\frac{dZ_{f}}{d\ell} &=& -C_{d1}Z_{f}, \label{Eq:RGDWZf}
\\
\frac{dA}{d\ell} &=& \left(C_{d2}-C_{d1}\right)A, \label{Eq:RGDWA}
\\
\frac{dv}{d\ell} &=& \left(C_{d3}-C_{d1}\right)v, \label{Eq:RGDWv}
\\
\frac{d\alpha}{d\ell} &=& \left(C_{d1} - C_{d3}\right)\alpha,
\label{Eq:RGDWAlpha}
\\
\frac{d\beta_{d}}{d\ell} &=& \left(C_{d1} + C_{d2} - 2C_{d3} -
1\right)\beta_{d}, \label{Eq:RGDWBeta}
\\
\frac{d\gamma_{d}}{d\ell} &=& \frac{1}{2}(C_{d2} -
C_{d1})\gamma_{d}. \label{Eq:RGDWGamma}
\end{eqnarray}
Here, the quasiparticle residue $Z_f$ measures the overlap between
free and interacting fermions, and $\alpha = e^{2}/v\epsilon$
characterizes the Coulomb interaction strength. The other two
parameters are defined as
\begin{eqnarray}
\beta_{d} = \frac{\zeta A\Lambda}{v^{2}}, \qquad  \gamma_{d} =
\frac{\sqrt{A}\Lambda_{UV}}{\sqrt{\Lambda}}.
\end{eqnarray}

As shown in Appendix~\ref{App:SelfEnergyTW}, the self-energy of
triple-Weyl fermions caused by Coulomb interaction can be written as
\begin{eqnarray}
\Sigma_{t}(i\omega,\mathbf{k}) &=& \int'\frac{d\Omega}{2\pi}
\frac{d^{3}\mathbf{q}}{(2\pi)^{3}} G_{t0}(i\omega+i\Omega,
\mathbf{k} + \mathbf{q})V_{t}(i\Omega,\mathbf{q})\nonumber
\\
&\approx&\left\{i\omega C_{t1} - B\left[g_{1}(\mathbf{k})\sigma_{x}
+ g_{2}(\mathbf{k})\sigma_{y}\right]C_{t2}\right.\nonumber
\\
&&\left.-vk_{z}\sigma_{z}C_{t3}\right\}\ell.
\label{Eq:TWSelfEnergyResultMainText}
\end{eqnarray}
The momentum shell is taken as $b\Lambda<E_{t}(\mathbf{k})<\Lambda$.
The expressions of $C_{t1}$, $C_{t2}$, and $C_{t3}$ are given by
Eqs.~(\ref{Eq:ExpressionCt1})-(\ref{Eq:ExpressionGt}) in
Appendix~\ref{App:SelfEnergyTW}. As shown in
Appendix~\ref{App:DerivationRGEquationTW}, the RG equations for
triple-Weyl fermions are
\begin{eqnarray}
\frac{dZ_{f}}{d\ell} &=& -C_{t1}Z_{f},\label{Eq:RGTWZf}
\\
\frac{dB}{d\ell}&=&(C_{t2}-C_{t1})B,\label{Eq:RGTWB}
\\
\frac{dv}{d\ell}&=&(C_{t3}-C_{t1})v, \label{Eq:RGTWVf}
\\
\frac{d\alpha}{d\ell}&=&\left(C_{t1}-C_{t3}\right)\alpha,
\label{Eq:RGTWAlpha}
\\
\frac{d\beta_{t}}{d\ell} &=& \left(\frac{4}{3}C_{t1} +
\frac{2}{3}C_{t2}-2C_{t3}-\frac{4}{3}\right)\beta_{t},
\label{Eq:RGTWBeta}
\\
\frac{d\gamma_{t}}{d\ell} &=& \frac{1}{3}(C_{t2} -
C_{t1})\gamma_{t}, \label{Eq:RGTWGamma}
\end{eqnarray}
where $\beta_{t}$ and $\gamma_{t}$ are defined as
\begin{eqnarray}
\beta_{t} = \frac{\zeta B^{2/3}\Lambda^{4/3}}{v^{2}}, \qquad
\gamma_{t} = \frac{B^{1/3}\Lambda_{UV}}{\Lambda^{1/3}},
\end{eqnarray}

\begin{figure}[htbp]
\center
\includegraphics[width=2.4in]{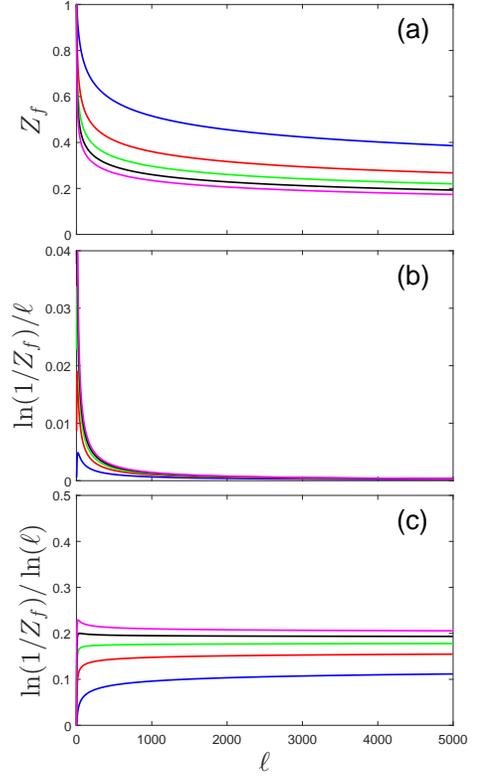}
\caption{Low-energy behavior of $Z_{f}$ for double-Weyl fermions.
The blue, red, green, black, and magenta lines correspond to
$\alpha_{0}=0.1$, $0.5$, $1$, $1.5$, and $2$ respectively.
$\beta_{d0}=1$ and $\gamma_{d0} = 0.2$. Here, $N=2$.
\label{Fig:VRGDW}}
\end{figure}

\subsection{Quasiparticle residue and damping rate}

The coupled flow equations can be solved numerically. The flow of
quasiparticle residue $Z_{f}(\ell)$ for double-Weyl fermions is
presented in Figs.~\ref{Fig:VRGDW}(a)-(c). According to
Fig.~\ref{Fig:VRGDW}(a), we find that $Z_{f}(\ell)$ eventually flows
to zero in the limit $\ell\rightarrow \infty$, albeit at a small
speed. Thus, the FL description is invalid, and the double-Weyl
fermions are not well-defined Landau-type quasiparticles.

\begin{figure}[htbp]
\center
\includegraphics[width=2.4in]{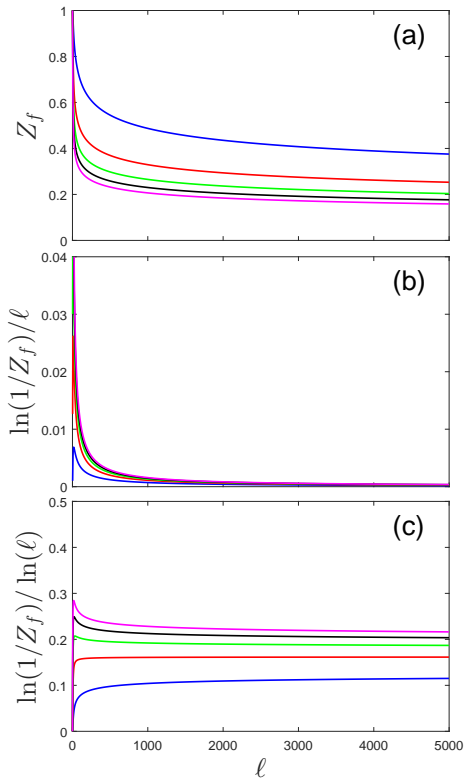}
\caption{Low-energy behavior of $Z_{f}$ for triple-Weyl fermions.
The blue, red, green, black, and magenta lines correspond to
$\alpha_{0}=0.1$, $0.5$, $1$, $1.5$, and $2$ respectively.
$\beta_{t0}=1$, and $\gamma_{t0} = 0.2$. Here, $N=2$.
\label{Fig:VRGTW}}
\end{figure}

It is necessary to determine how rapidly $Z_f(\ell)$ vanishes. We
observe from Fig.~\ref{Fig:VRGDW}(b) that
\begin{eqnarray}
\lim_{\ell\rightarrow \infty} \frac{\ln(1/Z_{f})}{\ell} \rightarrow
0.
\end{eqnarray}
Based on Fig.~\ref{Fig:VRGDW}(c), we find that $Z_{f}(\ell)$
exhibits the asymptotic behavior
\begin{eqnarray}
\lim_{\ell\rightarrow \infty} \frac{\ln(1/Z_{f})}{\ln(\ell)}
\rightarrow \eta,
\end{eqnarray}
where $\eta$ is a constant satisfying $0 < \eta < 1$. For physical
flavor $N = 2$, $\eta\approx 0.18$. For large values of $\ell$,
$Z_{f}(\ell)$ behaves as
\begin{eqnarray}
Z_{f}(\ell) \sim \ell^{-\eta}. \label{Eq:ZfLowEnergyRegime}
\end{eqnarray}
It is known that $Z_f$ is connected to the real part of the retarded
fermion self-energy $\mathrm{Re}\Sigma^{R}(\omega)$ via the relation
\begin{eqnarray}
Z_{f}(\omega) = \frac{1}{\left|1-\frac{\partial}{\partial\omega}
\mathrm{Re}\Sigma^{R}(\omega)\right|}. \label{Eq:ZfDefinition2}
\end{eqnarray}
The $\omega$-dependence of $Z_f$ can be obtained from $Z_f(\ell)$ by
making the transformation $\omega = \omega_{0} e^{-\ell}$, where
$\omega_{0}$ is some initial value of $\omega$. On the basis of
Eq.~(\ref{Eq:ZfDefinition2}), it is now easy to obtain
\begin{eqnarray}
\mathrm{Re}\Sigma^{R}(\omega)\sim \omega
\left[\ln\left(\frac{\omega_{0}}{\omega}\right)\right]^{\eta}.
\label{Eq:RealPartRTSelfEnergy}
\end{eqnarray}
Employing the Kramers-Kronig relation \cite{GiulianiBook}, we get
the imaginary part
\begin{eqnarray}
\mathrm{Im}\Sigma^{R}(\omega) \sim \frac{\omega}{\left[\ln
\left(\omega_{0}/\omega \right)\right]^{1-\eta}},
\label{Eq:ImPartRTSelfEnergy}
\end{eqnarray}
which directly gives the fermion damping rate. The two equations
(\ref{Eq:RealPartRTSelfEnergy}) and (\ref{Eq:ImPartRTSelfEnergy})
are the main results of this paper.

The asymptotic behavior of $Z_{f}(\ell)$ for triple-Weyl fermions
can be similarly obtained by solving the flow equations. Results are
shown in Figs.~\ref{Fig:VRGTW}(a)-(c). We find that $Z_{f}(\ell)$,
$\mathrm{Re}\Sigma^{R}(\omega)$, and $\mathrm{Im}\Sigma^{R}(\omega)$
exhibit almost the same qualitative low-energy behaviors as those of
double-Weyl fermions, namely they are also descried by
Eqs.~(\ref{Eq:ZfLowEnergyRegime}), (\ref{Eq:RealPartRTSelfEnergy}),
and (\ref{Eq:ImPartRTSelfEnergy}). For $N=2$, we find $\eta$ can be
also approximated by $\eta\approx 0.18$.

\begin{figure}[htbp]
\center
\includegraphics[width=2.3in]{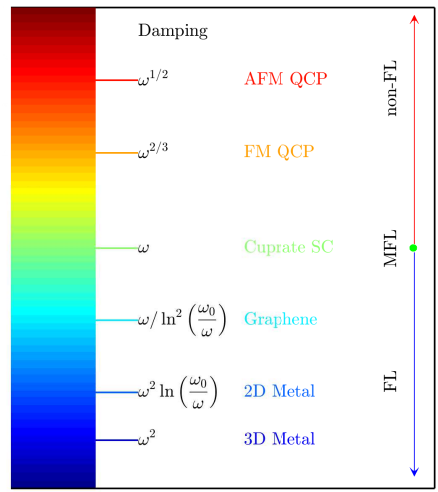}
\caption{Schematic illustration for traditional classification of
FL, non-FL, and MFL. Interacting fermion system with damping rate
weaker than MFL is believed to be a FL. \label{Fig:FLNFLDiagram}}
\end{figure}

\subsection{Fate of fermionic excitations in various interacting systems}

To gain a better understanding of the uniqueness of the non-FL
revealed in this work, we compare this state to a number of typical
FLs and non-FLs, as summarized in Fig.~\ref{Fig:FLNFLDiagram}. In
many cases, the fermion damping rate can be expressed by a
power-law, i.e., $\mathrm{Im}\Sigma^{R}(\omega) \sim \omega^{a}$.
One can verify that: (i) for $0 < a < 1$, $Z_{f} \sim
\omega^{1-a}\rightarrow 0$ as $\omega \rightarrow 0$; (ii) for
$a=1$, $Z_{f}\sim 1/\ln(\omega_{0}/\omega)\rightarrow 0$ as $\omega
\rightarrow 0$; (iii) if $a>1$, $Z_{f}$ flows to a nonzero constant
as $\omega \rightarrow 0$. An obvious conclusion is that, FL theory
is valid when $a > 1$, but breaks down when $0 < a \le 1$. In 2D
metal and graphene (2D DSM), the fermion damping rate is found to
respectively depend on $\omega$ as $\omega^2\ln(\omega_{0}/\omega)$
\cite{GiulianiBook} and $\omega/\ln^{2}(\omega_{0}/\omega)$
\cite{Gonzalez99}, which are actually not power functions. However,
the condition (\ref{Eq:FLNJLcriterion}) is still satisfied in these
two cases, and the corresponding residue $Z_f \neq 0$. Therefore,
the condition (\ref{Eq:FLNJLcriterion}) does provide an effective
criterion to judge the validity of FL theory in all previously
studied interacting fermion systems.

There is an interesting possibility that the damping rate vanishes
at exactly the same speed as the energy, namely $\Gamma(\omega) \sim
\omega$. The residue $Z_f$ decreases down to zero logarithmically as
$\omega \rightarrow 0$. This behavior is broadly identified as the
weakest violation of FL theory \cite{Varma89, Varma02, Sperstad11,
Sutherland12}, which is the reason why an interacting fermion system
displaying such a damping rate is called MFL. According to this
notion, any system in which the Landau damping effect is weaker than
MFL is usually regarded as a normal FL.

We emphasize here that, although the criterion
(\ref{Eq:FLNJLcriterion}) is widely utilized to judge the validity
of FL theory in various interacting fermion systems, its efficiency
has never been rigorously justified. There can be exceptions. It is
in principle possible for $Z_f$ to vanish more slowly than a
logarithmic decrease. In this work, we demonstrate that this
criterion breaks down in two concrete physical systems, namely
topological double- and triple-WSMs. Although the damping rate of
multi-Weyl fermions satisfies the criterion
(\ref{Eq:FLNJLcriterion}), the residue $Z_f$ vanishes in the limit
$\omega \rightarrow 0$ more slowly than the logarithmic decrease.
Thus, these two systems display a weaker-than-marginal violation of
FL theory. An important indication of this result is that the FL
theory could be spoiled even though the inter-particle interaction
does not induce a stronger-than-energy Landau damping. In this
regard, the ground state of double- and triple-WSMs is distinct from
all the other interacting fermion systems.

\begin{figure}[htbp]
\center
\includegraphics[width=2.8in]{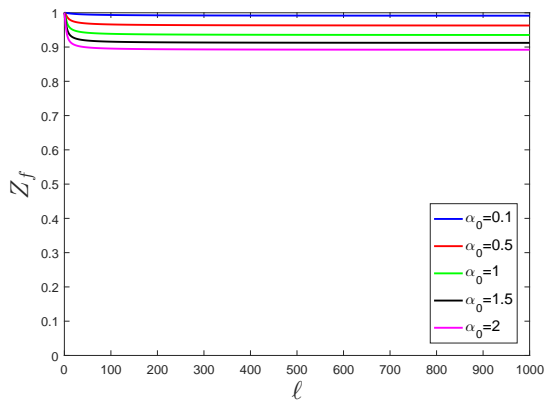}
\caption{The residue $Z_f$ of interacting usual Weyl fermions flows
to certain finite constant as $\ell \rightarrow +\infty$. Therefore,
the Coulomb interaction does not violate the FL theory.
\label{Fig:VRGZfUSM}}
\end{figure}

\subsection{Comparing with usual Weyl semimetal}

In this subsection, we study the residue and damping rate of
fermions in usual WSM, and compare the results with double- and
triple-WSMs.

The free propagator of usual Weyl fermions is
$G_{u0}^{-1}(i\omega,\mathbf{k}) = i\omega - v\mathbf{k}\cdot
\mathbf{\sigma}$. The dressed Coulomb interaction takes the form
\begin{eqnarray}
V_{u}(i\Omega,\mathbf{q}) = \frac{1}{V_{0}^{-1}(\mathbf{q}) +
\Pi_{u}(i\Omega,\mathbf{q})},
\end{eqnarray}
where $V_{0}(\mathbf{q})=\frac{4\pi\alpha v}{|\mathbf{q}|^{2}}$, and
the polarization is given by
\begin{eqnarray}
\Pi_{u}(i\Omega,\mathbf{q}) &=& -N\int\frac{d\omega}{2\pi}
\int\frac{d^3\mathbf{k}}{(2\pi)^{3}}
\mathrm{Tr}\left[G_{u0}(i\omega,\mathbf{k})\right.\nonumber
\\
&&\left.\times G_{u0}\left(i\omega+i\Omega,\mathbf{k} +
\mathbf{q}\right)\right]\nonumber
\\
&\approx& \frac{N\left|\mathbf{q}\right|^{2}}{12\pi^{2}v}
\ln\left(\frac{v\Lambda_{UV} + \sqrt{\Omega^{2} +
v^{2}\left|\mathbf{q}\right|^{2}}}{\sqrt{\Omega^{2} +
v^{2}\left|\mathbf{q}\right|^{2}}}\right).
\end{eqnarray}

Considering the self-energy induced by the Coulomb interaction
\begin{eqnarray}
\Sigma_{u}(i\omega,\mathbf{k}) &=& \int'\frac{d\Omega}{2\pi}
\frac{d^{3}\mathbf{q}}{(2\pi)^{3}}G_{u0}(i\omega+i\Omega,
\mathbf{k}+\mathbf{q})V_{u}(i\Omega,\mathbf{q})\nonumber
\\
&\approx& \left(i\omega C_{u1} - v\mathbf{k}\cdot
\mathbf{\sigma}C_{u2}\right)\ell,
\end{eqnarray}
and performing the RG analysis, we obtain the RG equations as
following
\begin{eqnarray}
\frac{dZ_{f}}{d\ell}&=&-C_{u1}Z_{f},
\\
\frac{dv}{d\ell}&=&\left(C_{u2}-C_{u1}\right)v,
\\
\frac{d\alpha}{d\ell}&=&\left(C_{u1}-C_{u2}\right)\alpha,
\\
\frac{d\gamma_{u}}{d\ell}&=&\left(C_{u2}-C_{u1}\right)\gamma_{u},
\end{eqnarray}
where
\begin{eqnarray}
C_{u1}&=&\frac{1}{4\pi^{3}}\int_{-\infty}^{+\infty} dx
\frac{x^{2}-1}{\left(x^{2}+1\right)^{2}}\mathcal{G}_{u}(x),
\\
C_{u2}&=&\frac{1}{4\pi^{3}}\int_{-\infty}^{+\infty} dx
\frac{x^{2}+\frac{1}{3}}{\left(x^{2}+1\right)^{2}}
\mathcal{G}_{u}(x),
\end{eqnarray}
with
\begin{eqnarray}
\mathcal{G}_{u}^{-1}(x) &=& \frac{1}{4\pi \alpha} +
\frac{N}{12\pi^{2}}\ln\left(\frac{\gamma_{u}e^{\ell} +
\sqrt{x^{2}+1}}{\sqrt{x^{2}+1}}\right).
\end{eqnarray}
The parameter $\gamma_{u}$  is given by $\gamma_{u} =
\frac{v\Lambda_{UV}}{\Lambda}$.

\begin{figure*}[htbp]
\center
\includegraphics[width=6.6in]{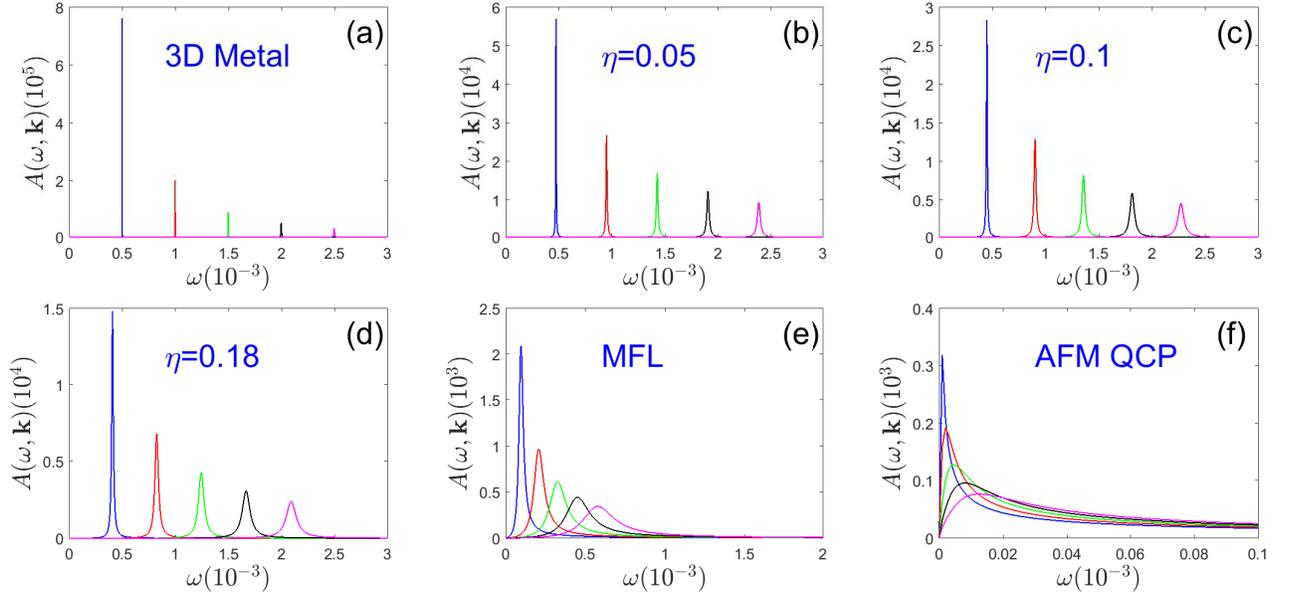}
\caption{Energy distribution curves in a number of FLs and non-FLs.
(a): 3D metal; (b)-(d): unconventional non-FL with $\eta = 0.18$
corresponding to physical flavor $N=2$; (e): MFL; (f): non-FL
realized at an AFM QCP. The blue, red, green, black, and magenta
curves correspond to $\xi_{\mathbf{k}}=0.001, 0.002, 0.003, 0.004,
0.005$, respectively. \label{Fig:EDC}}
\end{figure*}

We have solved the above flow equations and plot the
$\ell$-dependence of $Z_f$ in Fig.~\ref{Fig:VRGZfUSM}. As $\ell$
grows, $Z_f$ approaches to some finite value as $\ell \rightarrow
+\infty$. An immediate conclusion is that the unusual Weyl fermions
are ordinary Landau-type quasiparticles, and the FL theory is well
applicable. It is clear that usual WSM exhibits distinct behaviors
than double- and triple-WSMs. The distinction originates from the
difference in the fermion dispersion.

\section{Observable quantities \label{Sec:ObserQuant}}

As demonstrated in the last section, the non-FL state of double- and
triple-WSMs is qualitatively different from normal FLs and
conventional non-FLs. In this section, we discuss the possible
experimental signatures of this state. The residue $Z_f$ is not
directly measurable. Here we will compute a number of observable
quantities, including the spectral function, DOS, specific heat, and
conductivities, and analyze their low-energy properties. The
calculational details can be found in
Appendices~\ref{App:ObservableQuantitiesFree} and
\ref{App:ObservableQuantitiesInteraction}.

\subsection{Spectral function}

We first consider the fermion spectral function
$A(\omega,\mathbf{k})$ and the damping rate $\Gamma(\omega)$. In the
non-interacting limit, the spectral function is just a
$\delta$-function, namely
$$A_0(\omega,\mathbf{k}) = \delta(\omega-\xi_{\mathbf{k}}),$$ where
$\xi_{\mathbf{k}}$ is the kinetic energy, which displays an
infinitely sharp peak. After including quantum many-body effects, it
becomes
\begin{eqnarray}
A(\omega,\mathbf{k}) = \frac{1}{\pi} \frac{|\mathrm{Im}
\Sigma^{R}(\omega)|}{\left[\omega+\mathrm{Re}
\Sigma^{R}(\omega)-\xi_{\mathbf{k}}\right]^{2} +
\left[\mathrm{Im}\Sigma^{R}(\omega)\right]^{2}}.
\end{eqnarray}
In Fig.~\ref{Fig:EDC}, we list four different cases: 3D metal as a
normal FL; unconventional non-FL state revealed in this paper; MFL;
conventional non-FL. As shown in Fig.~\ref{Fig:EDC}(a), the energy
distribution curve (EDC) for 3D metal has a very sharp peak. Around
an AFM QCP, depicted in Fig.~\ref{Fig:EDC}(f), the EDC exhibits an
obviously asymmetric shape with a long tail. We observe from
Fig.~\ref{Fig:EDC}(e) that the EDC of MFL still displays a weak
asymmetry. For the unusual non-FL states demonstrated in
Figs.~\ref{Fig:EDC}(b)-(d), the asymmetry of EDC is even weaker than
that of MFL, and the asymmetry is gradually weakened as $\eta$
decreases. Nevertheless, the peak width is larger than that of
normal FLs, which is consistent with the fact that $Z_f = 0$.
Comparing the EDCs shown in Figs.~\ref{Fig:EDC}(a)-(f), we can infer
that, the non-FL state realized in double- and triple-WSMs stays in
between MFL and all the normal FLs. ARPES experiments
\cite{Damascelli03, Valla99, Kaminski01, Richard09, Miao16,
Siegel11, Pan12, Xu14, Kondo15} could be performed to detect the
peculiar properties of EDCs of the unconventional non-FL state.

The fermion damping rate $\Gamma(\omega)$ can be extracted from the
ARPES data of EDC or momentum distribution curve (MDC)
\cite{Damascelli03, Valla99}. It is especially efficient to fit MDC
with the Lorentzian peak \cite{Damascelli03, Valla99, Kaminski01,
Richard09, Siegel11, Pan12, Xu14, Miao16}. High-resolution ARPES
measurements have been extensively applied to determine the fermion
damping rate in many correlated electron systems, including cuprates
\cite{Damascelli03, Valla99, Kaminski01}, iron pnictides
\cite{Richard09, Miao16}, graphene \cite{Siegel11}, and topological
insulators \cite{Pan12, Xu14}. We expect that the unconventional
damping rate, i.e., $\Gamma(\omega) \sim \omega
/\left[\ln\left(\omega_{0}/\omega \right)\right]^{1-\eta}$, obtained
in our work could be investigated by using this experimental tool in
the future.

\subsection{DOS}

\begin{figure}[htbp]
\center
\includegraphics[width=3.3in]{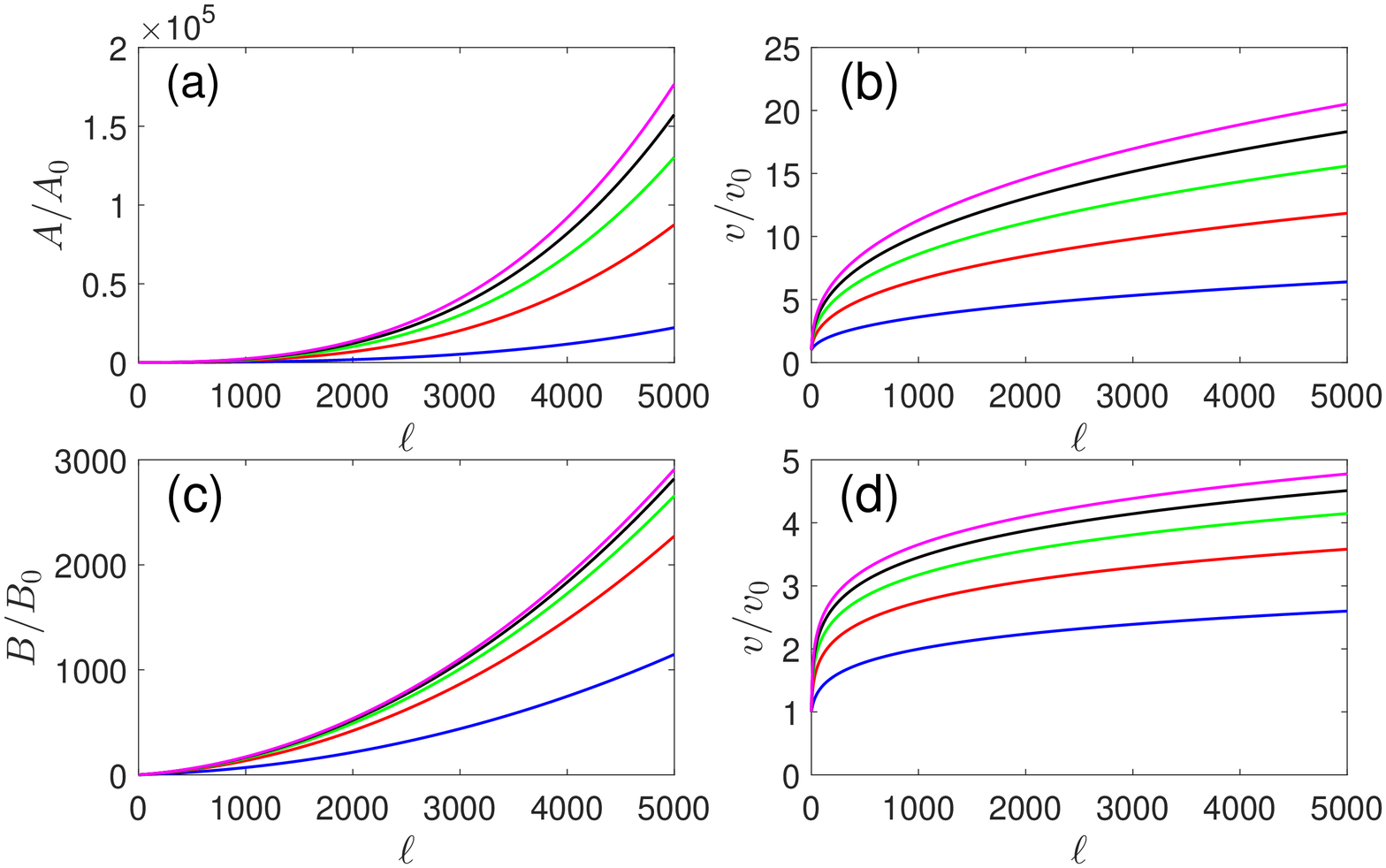}
\caption{(a) and (b): Flows of $A$ and $v$ in double-WSM with
$\beta_{d0}=1$, $\gamma_{d0}=0.2$, and $N=2$. (c) and (d): Flows of
$B$ and $v$ in triple-WSM with $\beta_{t0}=1$, $\gamma_{t0}=0.2$,
and $N=2$. In (a)-(d), blue, red, green, black, and magenta lines
correspond to $\alpha_{0}=0.1$, $0.5$, $1$, $1.5$, and $2$,
respectively. \label{Fig:VRGDispersion}}
\end{figure}

\begin{figure}[htbp]
\center
\includegraphics[width=3.3in]{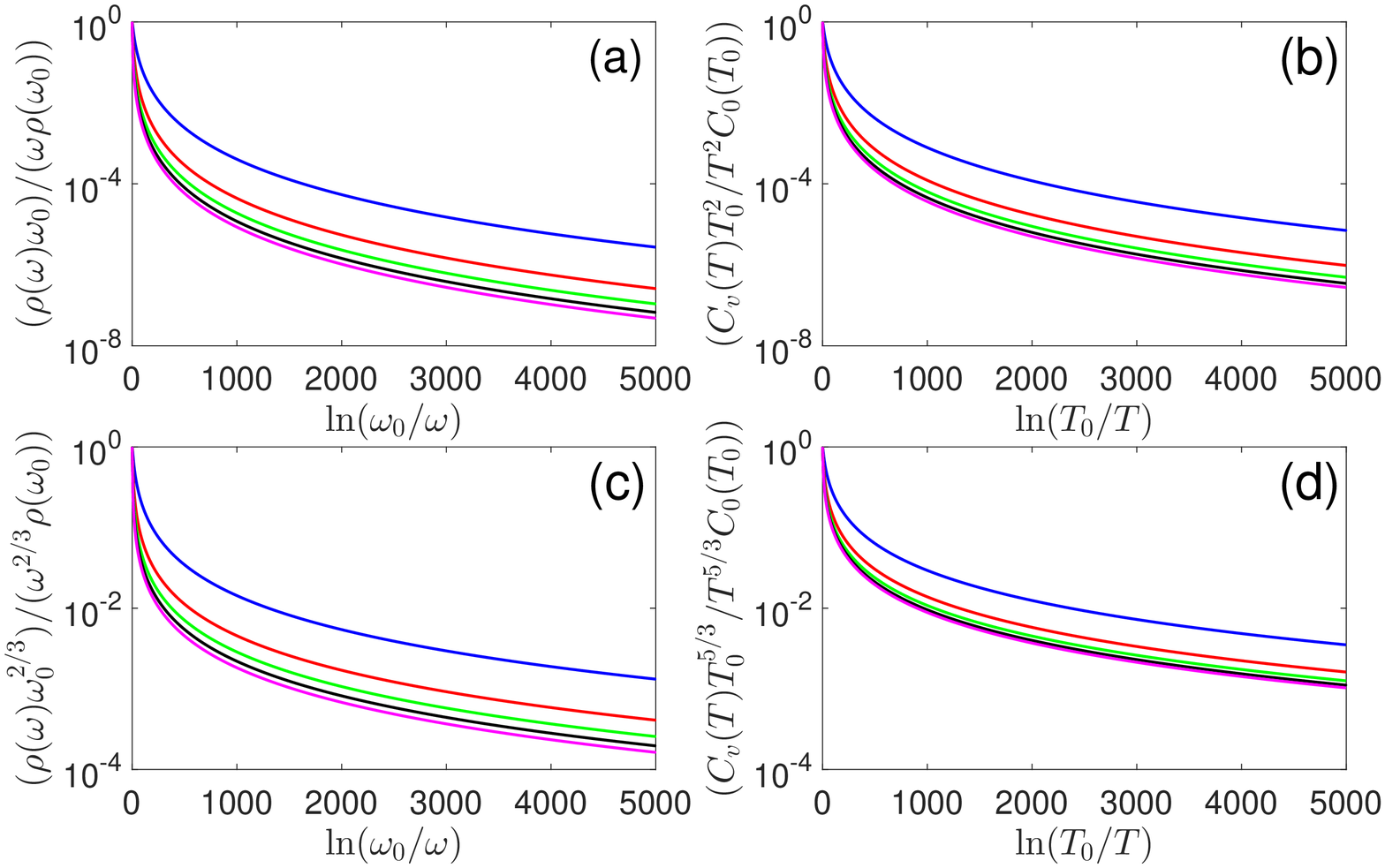}
\caption{DOS and specific heat of double- and triple-WSMs after
including the Coulomb interaction corrections. (a) and (b): Results
for double-WSM with $\beta_{d0}=1$, $\gamma_{d0}=0.2$, and $N=2$.
(c) and (d): Results for triple-WSM with $\beta_{t0}=1$,
$\gamma_{t0}=0.2$, and $N=2$. In (a)-(d), blue, red, green, black,
and magenta lines correspond to $\alpha_{0}=0.1$, $0.5$, $1$, $1.5$,
and $2$, respectively. \label{Fig:VRGDOSCv}}
\end{figure}

The DOS of free double-Weyl fermions is linear in energy, i.e.,
\begin{eqnarray}
\rho_{d}(\omega)  = \frac{N}{8\pi vA}|\omega|.
\end{eqnarray}
The Coulomb interaction leads to singular renormalization of the
parameters $A$ and $v$. The flows of $A$ and $v$ can be obtained
from the RG equations, with results shown in
Figs.~\ref{Fig:VRGDispersion}(a) and (b). Both $A$ and $v$ increase
as $\ell$ is growing, and it is obvious that $A$ increases more
rapidly than $v$. After incorporating the anomalous dimension of
fermion field and the renormalization of $A$ and $v$, we get the
following RG equation for $\rho_{d}$:
\begin{eqnarray}
\frac{d\ln(\rho_{d}(\omega))}{d\ln(\omega)}
&\sim& 1-C_{d1}+C_{d2}+C_{d3}.
\end{eqnarray}
As depicted in Fig.~\ref{Fig:VRGDOSCv}(a), the double-Weyl fermion
DOS is suppressed by Coulomb interaction.

For free triple-Weyl fermions, the DOS is given by
\begin{eqnarray}
\rho_{t}(\omega) = \frac{N\Gamma\left(1/3\right) }{12\pi^{3/2}
\Gamma\left(5/6\right)vB^{2/3}}|\omega|^{2/3}.
\end{eqnarray}
The $\ell$-dependence of $B$ and $v$ is plotted in
Figs.~\ref{Fig:VRGDispersion}(c) and (d). Both $B$ and $v$ increase
with growing $\ell$, but at different speeds. Including the
interaction corrections modifies $\rho_{t}(\omega)$, and yields
\begin{eqnarray}
\frac{d\ln(\rho_{t}(\omega))}{d\ln(\omega)} &\sim&
\frac{2}{3}-\frac{2}{3}C_{t1}+\frac{2}{3}C_{t2}+C_{t3}.
\end{eqnarray}
The energy dependence of $\rho_{t}(\omega)$ is depicted in
Fig.~\ref{Fig:VRGDOSCv}(c). The DOS of triple-Weyl fermions is also
suppressed by the Coulomb interaction.

\subsection{Specific heat}

The specific heat for free double- and triple-Weyl fermions are
\begin{eqnarray}
C_{v}^{d}(T) &=& \frac{9\zeta(3)N}{4\pi vA}T^{2},\\
C_{v}^{t}(T) &=& \frac{40a_{t}N}{9\pi^{3/2}vB^{2/3}}T^{5/3},
\end{eqnarray}
respectively. Here $a_{t}$ is a constant. Taking into account the
renormalization of parameter $A$, $B$, and $v$, we derive the
following equations:
\begin{eqnarray}
\frac{d\ln(C_{v}^{d}(T))}{d\ln(T)} &\sim& 2-2C_{d1}+C_{d2}+C_{d3},
\\
\frac{d\ln(C_{v}^{t}(T))}{d\ln(T)} &\sim&
\frac{5}{3}-\frac{5}{3}C_{t1}+\frac{2}{3}C_{t2}+C_{t3}.
\end{eqnarray}
The corresponding numerical results are presented in
Figs.~\ref{Fig:VRGDOSCv}(b) and (d). It is clear that the specific
heat is suppressed by the Coulomb interaction in both systems.

\subsection{Conductivities}

Transport properties might provide important information on the
experimental signatures of non-FL state. Here we are particularly
interested in the conductivities. We will first estimate the DC
conductivities, and then calculate the dynamical conductivities.

\subsubsection{Estimation of DC conductivities}

To estimate the DC conductivity, a heuristic method is to invoke the
Einstein relation \cite{GiulianiBook, ColemanBook, Hosur12,
ChenXieGroup16}
\begin{eqnarray}
\sigma \sim e^{2}v^{2}\rho(E)\tau(E),
\end{eqnarray}
where $v$ is the mean fermion velocity, $\rho$ the DOS, and $\tau$
the mean fermion lifetime. $E$ stands for either energy or
temperature. Since $\tau$ is inversely proportional to the damping
rate $\Gamma$, i.e.,
\begin{eqnarray}
\tau \sim \frac{1}{\Gamma},
\end{eqnarray}
the Einstein relation can be re-written as
\begin{eqnarray}
\sigma\sim e^{2}v^{2}\rho(E)/\Gamma(E).
\end{eqnarray}
In conventional metals, the fermion damping rate is $\Gamma(E)\sim
E^{2}$ \cite{GiulianiBook, ColemanBook}, whereas $v$ and $\rho$ are
both constants in the low-energy regime. Taking $E=T$, one can
easily obtain
\begin{eqnarray}
\sigma\sim T^{-2}.
\end{eqnarray}
This indicates that the resistivity
\begin{eqnarray}
R \sim \frac{1}{\sigma} \sim T^{2},
\end{eqnarray}
which is a well-known characteristic of conventional metals
\cite{GiulianiBook, ColemanBook}. We will utilize this procedure to
estimate the DC conductivities of double- and triple-WSMs.

For double-WSM, the DOS is given by $\rho(E)\sim E/(vA)$. The
damping rate is $\Gamma(E)\sim E/[\ln(E_{0}/E)]^{1-\eta}$, which can
be further simplified to $\Gamma(E)\sim E$ if the weak
logarithmic-like factor is neglected. The fermion velocity within
the $x$-$y$ plane takes the form
\begin{eqnarray}
v_{\bot}\sim Ak_{\bot} \sim \sqrt{A E}.
\end{eqnarray}
By using the Einstein relation, we find that the conductivity within
the $x$-$y$ plane can be approximately expressed as
\begin{eqnarray}
\sigma_{\bot\bot}^{d}(T)\sim T.
\end{eqnarray}
The fermion velocity along the $z$-axis is $v$, thus the
corresponding conductivity is independent of $T$, namely
\begin{eqnarray}
\sigma_{zz}^{d}(T)\sim T^{0}\sim \mathrm{Const}.
\end{eqnarray}

For triple-WSM, the DOS depends on energy in the form $\rho(E)\sim
E^{2/3}/(vB^{2/3})$. The damping rate behaves as $\Gamma(E)\sim
E/(\ln(E_{0}/E))^{1-\eta}$, which is approximated by $\Gamma(E)\sim
E$. The fermion velocity within the $x$-$y$ plane reads
\begin{eqnarray}
v_{\bot}\sim Bk_{\bot}^{2}\sim B^{1/3}E^{2/3}.
\end{eqnarray}
Employing the Einstein relation, we obtain the conductivity within
the $x$-$y$ plane
\begin{eqnarray}
\sigma_{\bot\bot}^{t}(T)\sim T,
\end{eqnarray}
which is qualitatively the same as the one for double-WSM. However,
different from double-WSM, the conductivity along $z$-axis is
$T$-dependent, given by
\begin{eqnarray}
\sigma_{zz}^{t}(T)\sim T^{-1/3}.
\end{eqnarray}

In the above estimation, we neglect the weak logarithmic-like factor
that exists in the damping rate and the renormalized parameters
($A$, $B$, and $v$). If these corrections are incorporated, the
conductivities could receive weak logarithmic-like corrections in
their temperature dependence.

\subsubsection{Dynamical conductivities}

\begin{figure}[htbp]
\center
\includegraphics[width=3.3in]{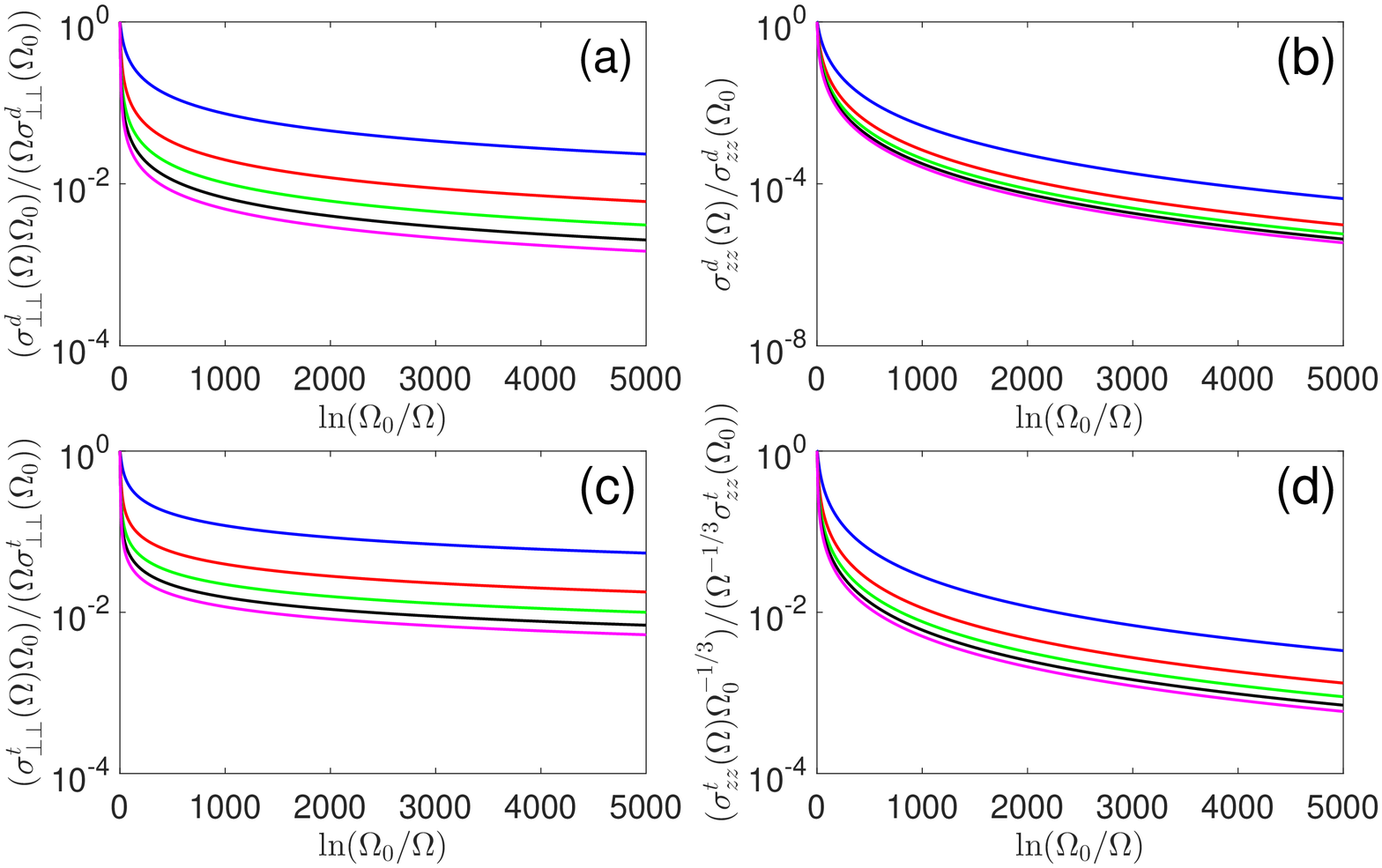}
\caption{Dynamical conductivities of double- and triple-WSMs
considering the influence of Coulomb interaction. (a) and (b):
$\sigma_{\bot\bot}^{d}$ and $\sigma_{zz}^{d}$ for double-WSM with
$\beta_{d0}=1$, $\gamma_{d0}=0.2$, and $N=2$. (c) and (d):
$\sigma_{\bot\bot}^{t}$ and $\sigma_{zz}^{t}$ for triple-WSM with
$\beta_{t0}=1$, $\gamma_{t0}=0.2$, and $N=2$. In (a)-(d), blue, red,
green, black, and magenta lines correspond to $\alpha_{0}=0.1$,
$0.5$, $1$, $1.5$, and $2$, respectively. \label{Fig:VRGConduct}}
\end{figure}

We next analyze the properties of dynamical conductivities by
including the energy dependence. By using the Kubo formula, we get
the dynamical conductivities for free double-Weyl fermions within
the $x$-$y$ plane and along the $z$ axis:
\begin{eqnarray}
\sigma_{\bot\bot}^{d}(\Omega,T) &=& c_{1}^{d}\frac{e^{2}}{v}
\delta\left(\Omega\right)T^{2} + c_{2}^{d} \frac{e^{2}}{v}|\Omega|
\nonumber \\
&&\times\tanh\left(\frac{|\Omega|}{4T}\right),
\label{Eq:SigmaxxDWSMMainText} \\
\sigma_{zz}^{d}(\Omega,T) &=& c_{3}^{d}\frac{ve^{2}}{A}
\delta\left(\Omega\right)T + c_{4}^{d}\frac{v e^{2}}{A}\nonumber
\\
&&\times\tanh\left(\frac{|\Omega|}{4T}\right).
\label{Eq:SigmazzDWSMMainText}
\end{eqnarray}
The constants $c_{1}^{d}$, $c_{2}^{d}$, $c_{3}^{d}$, and $c_{4}^{d}$
are given in Appendix~\ref{App:DWSMDynamicalConduct}. The first
terms on the right-hand sides of Eqs.~(\ref{Eq:SigmaxxDWSMMainText})
and (\ref{Eq:SigmazzDWSMMainText}) indicate the presence of Drude
peak. At zero temperature, the dynamical conductivities become
\begin{eqnarray}
\sigma_{\bot\bot}^{d}(\Omega) &=& c_{2}^{d}\frac{e^{2}}{v}|\Omega|,
\\
\sigma_{zz}^{d}(\Omega) &=& c_{4}^{d}\frac{ve^{2}}{A}.
\end{eqnarray}
Incorporating the interaction corrections, we derive the RG
equations for $\sigma_{\bot\bot}^{d}$ and $\sigma_{zz}^{d}$
\begin{eqnarray}
\frac{d\ln(\sigma_{\bot\bot}^{d}(\Omega))}{d\ln(\Omega)}
&\sim&1+C_{d1}+C_{d3}, \\
\frac{d\ln(\sigma_{zz}^{d}(\Omega))}{d\ln(\Omega)}
&\sim&2C_{d1}+ C_{d2}-C_{d3}.
\end{eqnarray}
The solutions of these two equations are shown in
Figs.~\ref{Fig:VRGConduct}(a) and (b). Both $\sigma_{\bot\bot}^{d}$
and $\sigma_{zz}^{d}$ are suppressed by the Coulomb interaction.

For free triple-Weyl fermions, the dynamical conductivities within
the $x$-$y$ plane and along the $z$-axis are
\begin{eqnarray}
\sigma_{\bot\bot}^{t}(\Omega,T) &=& c_{1}^{t}\frac{e^{2}}{v}
\delta\left(\Omega\right)T^{2} + c_{2}^{t}\frac{e^{2}}{v}|\Omega|
\nonumber \\
&&\times\tanh\left(\frac{|\Omega|}{4T}\right),
\label{Eq:SigmaxxTWSMMainText} \\
\sigma_{zz}^{t}(\Omega,T)&=&c_{3}^{t}\frac{ve^{2}}{B^{2/3}}
\delta\left(\Omega\right)T^{2/3}+c_{4}^{t} \frac{ve^{2}}{B^{2/3}}
\frac{1}{\left|\Omega\right|^{1/3}} \nonumber \\
&&\times\tanh\left(\frac{|\Omega|}{4T}\right).
\label{Eq:SigmazzTWSMMinText}
\end{eqnarray}
The constants $c_{1}^{t}$, $c_{2}^{t}$, $c_{3}^{t}$, and $c_{4}^{t}$
are shown in Appendix~\ref{App:TWSMDynamicalConduct}. At zero
temperature, they behave as
\begin{eqnarray}
\sigma_{\bot\bot}^{t}(\Omega)&=&c_{2}^{t}\frac{e^{2}}{v}|\Omega|,
\\
\sigma_{zz}^{t}(\Omega) &=& c_{4}^{t} \frac{ve^{2}}{B^{2/3}}
\frac{1}{\left|\Omega\right|^{1/3}}.
\end{eqnarray}
The Coulomb interaction alters the above behaviors and gives rise to
the following equations
\begin{eqnarray}
\frac{d\ln\left(\sigma_{\bot\bot}^{t}(\Omega)\right)}{d\ln(\Omega)}
&\sim & 1+C_{t1}+C_{t3},
\\
\frac{d\ln\left(\sigma_{zz}^{t}(\Omega)\right)}{d\ln(\Omega)} &\sim&
-\frac{1}{3}+\frac{7}{3}C_{t1}+\frac{2}{3}C_{t2}-C_{t3}.
\end{eqnarray}
The detailed energy-dependence of $\sigma_{\bot\bot}^{t}$ and
$\sigma_{zz}^{t}$ can be found from Figs.~\ref{Fig:VRGConduct}(c)
and (d).

\section{Experimental detection \label{Sec:ExpDetection}}

\begin{figure}[htbp]
\center
\includegraphics[width=3.05in]{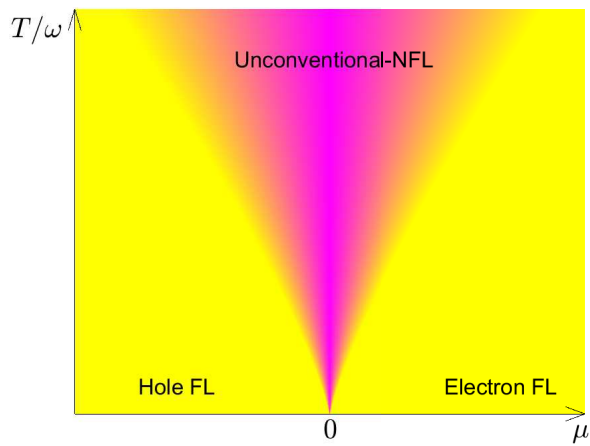}
\caption{Phase diagram for multi-WSM in the parameter space spanned
by $\mu$ and $T/\omega$. The unconventional non-FL state can be
probed in the whole quantum critical regime.
\label{Fig:PhaseDiagramMu}}
\end{figure}

It was recently suggested that HgCr$_{2}$Se$_{4}$ \cite{Fang12} and
SrSi$_{2}$ \cite{Huang16} are two candidate materials of double-WSM.
Theoretical studies \cite{LiuZunger16} proposed that cubic-Dirac SM
(cubic-DSM), in which the fermion dispersion is linear in one
momentum component and cubic in the rest two, might be realized in
some materials, including Rb(MoTe)$_{3}$ and Tl(MoTe)$_{3}$ that
belong to the A(MoX)$_{3}$ family. Similar to triple-WSM, such
cubic-DSMs should also display the unconventional non-FL behavior.
The physical fermion flavor is $N=2$ in HgCr$_{2}$Se$_{4}$
\cite{Fang12}, thus $\eta\approx 0.18$. In SrSi$_{2}$, the flavor of
double-Weyl fermions is $N = 8$ \cite{Huang16}, for which we get
$\eta \approx 0.05$. In Rb(MoTe)$_{3}$ and Tl(MoTe)$_{3}$, there is
only one type of cubic Dirac fermion, formed by the degeneracy of
two species of triple-Weyl fermions \cite{LiuZunger16}. Accordingly,
we find that $\eta \approx 0.18$ in these two materials.

The unconventional non-FL state is induced by the long-range Coulomb
interaction, thus its signature is sharpest at zero chemical
potential, i.e., $\mu=0$, which can be realized by tuning the Fermi
level to the band-touching point. In real materials, $\mu$ is
usually not strictly zero, but takes a finite value. For finite
$\mu$, the Coulomb interaction becomes short-ranged due to the
static screening caused by finite zero-energy DOS. It is necessary
to emphasize that the unconventional non-FL state still has
observable effects at finite $\mu$. At energies below $\mu$, namely
$\omega<\mu$, the system exhibits ordinary FL behavior. At energies
above $\mu$, namely namely $\omega > \mu$, the static screening
becomes relatively unimportant, and non-FL behavior re-emerges. As
$\omega$ further grows, the Coulomb interaction might be surpassed
by other interactions, such as electron-phonon coupling. Thus,
non-FL behavior actually occurs in a finite range of intermediate
energies $\mu < \omega < \omega_c$, where $\omega_c$ is a
material-dependent upper energy scale. This energy range narrows
when $\mu$ increases, and finally disappears once $\mu$ becomes
large enough. For sufficiently small $\mu$, there are always
observable signatures of the unconventional non-FL state. To
intuitively understand what happens at finite $\mu$, one could
regard $\mu$ as a tuning parameter: $\mu=0$ defines a special QCP
that exhibits an unconventional non-FL ground state at lowest energy
limit. This QCP is broadened into a finite quantum critical regime
at finite $T/\omega$, as schematically shown in
Fig.~\ref{Fig:PhaseDiagramMu}. Unconventional non-FL behavior
emerges in this regime, and can be explored at finite $T$ with
$T>\mu$ or at finite $\omega$ with $\omega>\mu$.

It is technically possible reduce $\mu$ gradually, by either doping
manipulation or gate voltage control \cite{Siegel11, Elias11,
Chae12, Yu13, Crossno16, LiuZK14, LiCaiZhen15}. Some unusual quantum
many-body effects have already been observed in SM samples
prepared at small $\mu$ \cite{Elias11, Siegel11, Chae12, Yu13,
Crossno16, Faugeras15}. For instance, RG studies predicted that the
fermion velocity is singularly renormalized by long-range Coulomb
interaction in 2D DSM \cite{Gonzalez99, Hofmann14}. This prediction
was recently confirmed in graphene with small $\mu$ by a number of
different experimental tools, including ARPES \cite{Siegel11},
Shubnikov-de Haas oscillations \cite{Elias11}, scanning tunneling
spectroscopy \cite{Chae12}, quantum capacitance measurements
\cite{Yu13}, and Landau level spectroscopy \cite{Faugeras15}.

Remarkably, one could fix the Fermi level of SM at exactly
the band-touching point via the mechanism of symmetry protection
\cite{Armitage18, Ruan16A, Ruan16B, Nie17, Du18}, which naturally
realizes ideal SM with $\mu=0$. Such symmetry-protected ideal
WSM state is proposed to exist in several materials, including
strained HgTe and Heusler compounds \cite{Ruan16A}, some
chalcopyrite compounds \cite{Ruan16B}, GdSI \cite{Nie17}, and CuF
\cite{Du18}. If symmetry-protected ideal double- and triple-WSMs
were discovered in the future, it would be more feasible to probe
the unconventional non-FL behavior.

\section{Summary \label{Sec:Summary}}

In summary, we have studied two important quantities, the
quasiparticle residue and the Landau damping rate, in double- and
triple-WSMs. The result is that, the residue $Z_{f}$ flows to zero
in the lowest energy limit, but at a lower speed than that of the
MFL. Based on the energy dependence of $Z_f$, we have obtained the
real and imaginary parts of the retarded fermion self-energy, and
then computed the spectral function. Interestingly, the Landau
damping rate is weaker than the one of MFL. These results imply that
the Coulomb interaction leads to a weaker-than-marginal breakdown of
the FL theory in both double- and triple-WSMs. The unconventional
non-FL state can be experimentally probed by measuring the fermion
self-energy and the spectral function. We have also calculated
several other observable quantities, including DOS, specific heat,
and conductivities, and discussed the impact of the anomalous
interaction corrections. We would emphasize that this non-FL state
has observable effects even at finite chemical potential.

Notice that the Coulomb interaction is long-ranged in all the DSMs
and WSMs that host isolated band-touching points. However, this new
type of non-FL state emerges only in double- and triple-WSMs, but
has never been found neither in the 2D/3D DSMs nor in the usual
WSMs. In order to make a comparison between different types of WSMs,
we have computed the residue $Z_f$ of usual Weyl fermions and found
that it flows to a finite value as the energy is lowered down to
zero. The difference indicates that the influence of long-range
Coulomb interaction in SMs is closely linked to the energy
dispersion of the fermion excitations.

The impact of Coulomb interaction has recently been studied in
double-WSM \cite{Lai15, Jian15} and triple-WSM
\cite{WangLiuZhang17B, ZhangShiXin17} by means of RG method. These
RG studies are based on the instantaneous approximation (neglecting
the energy dependence of Coulomb interaction), and thus do not
provide any information about the residue $Z_f$ and fermion damping
rate $\Gamma(\omega)$. Different from Refs.~\cite{Lai15, Jian15,
WangLiuZhang17B, ZhangShiXin17}, we have incorporated the dynamical
screening of Coulomb interaction in our RG analysis and obtained
both the residue and damping rate.

\section*{ACKNOWLEDGEMENTS}

We acknowledge the support by the National Key R\&D Program of China
under Grants 2016YFA0300404 and 2017YFA0403600, and the support by
the National Natural Science Foundation of China under Grants
11574285, 11504379, 11674327, U1532267, and U1832209. J.R.W. is also
supported by the Natural Science Foundation of Anhui Province under
Grant 1608085MA19.

\appendix

\section{Calculation of polarization functions \label{App:Polarization}}

Although the Coulomb interaction is long-ranged at zero chemical
potential, it is dynamically screened by the collective
particle-hole excitations. The dynamical screening effect is
embodied in the energy/momentum dependence of the polarization
functions. The aim of this Appendix is to calculate the
polarizations $\Pi_{d,t}(\Omega,\mathbf{q})$ for double- and
triple-Weyl fermions.

\subsection{Polarization $\Pi_{d}(\Omega,\mathbf{q})$ of double-Weyl
fermions}

For double-Weyl fermions, the polarization is given by
\begin{eqnarray}
\Pi_{d}(i\Omega,\mathbf{q}) &=& -N\int\frac{d\omega}{2\pi}
\int\frac{d^3\mathbf{k}}{(2\pi)^{3}}
\mathrm{Tr}\left[G_{d0}(i\omega,\mathbf{k})\right.\nonumber
\\
&&\left.\times G_{d0}\left(i\omega+i\Omega,\mathbf{k} +
\mathbf{q}\right)\right]. \label{Eq:PolaDoubleDefAppen}
\end{eqnarray}
Substituting the fermion propagator
Eq.~(\ref{Eq:FreeFermonPropagatorDoubleWeylDef}) into
Eq.~(\ref{Eq:PolaDoubleDefAppen}), we get
\begin{eqnarray}
\Pi_{d}(i\Omega,\frac{q_{x}}{\sqrt{A}},\frac{q_{y}}{\sqrt{A}},
\frac{q_{z}}{v}) = \frac{2N}{Av}\int\frac{d\omega}{2\pi}
\int\frac{d^3\mathbf{k}}{(2\pi)^{3}}\frac{F_{d1}^{A}}{F_{d1}^{B}},
\end{eqnarray}
where
\begin{eqnarray}
F_{d1}^{A} &=& \omega\left(\omega+\Omega\right) -
\sum_{i=1}^{3}d_{i}(\mathbf{k}) d_{i}(\mathbf{k}+\mathbf{q}),
\\
F_{d1}^{B} &=& \left(\omega^{2}+E_{\mathbf{k}}^{2}\right)
\left[\left(\omega+\Omega\right)^{2} +
E_{\mathbf{k}+\mathbf{q}}^{2}\right],
\end{eqnarray}
with $d_{1}(\mathbf{k})=k_{x}^{2}-k_{y}^{2}$, $d_{2}(\mathbf{k}) =
2k_{x}k_{y}$, $d_{3}(\mathbf{k})=k_{z}$ and $E_{\mathbf{k}} =
\sqrt{k_{\bot}^{4}+k_{z}^{2}}$. This expression is obtained by
making the following re-scaling transformations:
\begin{eqnarray}
&&k_{x}\rightarrow\frac{k_{x}}{\sqrt{A}},\quad
q_{x}\rightarrow\frac{q_{x}}{\sqrt{A}},\quad
k_{y}\rightarrow\frac{k_{y}}{\sqrt{A}},\nonumber
\\
&&q_{y}\rightarrow\frac{q_{y}}{\sqrt{A}},\quad
k_{z}\rightarrow\frac{k_{z}}{v},\quad
q_{z}\rightarrow\frac{q_{z}}{v}.
\end{eqnarray}
Employing the Feynman integral
\begin{eqnarray}
\frac{1}{XY} = \int_{0}^{1} dx\frac{1}{\left[xX +
(1-x)Y\right]^{2}},\label{Eq:FeynmanParameterization}
\end{eqnarray}
the polarization can be further written as
\begin{eqnarray}
\Pi_{d}(i\Omega,\frac{q_{x}}{\sqrt{A}},\frac{q_{y}}{\sqrt{A}},\frac{q_{z}}{v})
&=& \frac{2N}{Av}\int_{0}^{1}dx\int\frac{dk_{x}}{2\pi}
\frac{dk_{y}}{2\pi}F_{d2}^{A}\nonumber
\\
&&\times\int\frac{d^2\mathbf{K}}{(2\pi)^{2}}
\frac{1}{\left(F_{d2}^{B}\right)^{2}},
\end{eqnarray}
where
\begin{eqnarray}
F_{d2}^{A} &=& -x(1-x)\left(\Omega^2 - q_{z}^{2}\right)
-d_{1}\left(\mathbf{k}-\frac{\mathbf{q}}{2}\right)
d_{1}\left(\mathbf{k}+\frac{\mathbf{q}}{2}\right)\nonumber
\\
&&-d_{2}\left(\mathbf{k}-\frac{\mathbf{q}}{2}\right)d_{2}
\left(\mathbf{k}+\frac{\mathbf{q}}{2}\right),
\\
F_{d2}^{B}&=&K^{2}+x(1-x)\left(\Omega^2+q_{z}^{2}\right) + (1-x)
\left(\mathbf{k}-\frac{\mathbf{q}}{2}\right)_{\bot}^{4} \nonumber \\
&& +x\left(\mathbf{k}+\frac{\mathbf{q}}{2}\right)_{\bot}^{4},
\end{eqnarray}
and $\mathbf{K}=(\omega,k_z{})$. The transformations
$k_{x}\rightarrow k_{x} - \frac{q_{x}}{2}$ and $k_{y}\rightarrow
k_{y} - \frac{q_{y}}{2}$ have been made in the above derivation. We
then integrate over $\mathbf{K}$ by using the formula
\begin{eqnarray}
\int\frac{d^dQ}{(2\pi)^{d}}\frac{1}{\left(Q^2+\Delta\right)^{n}} =
\frac{\Gamma\left(n-d/2\right)}{(4\pi)^{d/2}\Gamma(n)}
\frac{1}{\Delta^{n-d/2}},\label{Eq:AppenIntegralFormula}
\end{eqnarray}
and obtain
\begin{eqnarray}
\Pi_{d}(i\Omega,\frac{q_{x}}{\sqrt{A}},\frac{q_{y}}{\sqrt{A}},\frac{q_{z}}{v})
&=& \frac{N}{2\pi Av}\int_{0}^{1}dx \int \frac{dk_{x}}{2\pi}
\frac{dk_{y}}{2\pi} \nonumber \\
&&\times\left(\frac{F_{d3}^{A}}{F_{d3}^{B}} + 1\right),
\end{eqnarray}
where
\begin{eqnarray}
F_{d3}^{A} &=& -x(1-x)\left(\Omega^2 - q_{z}^{2}\right) -
\left(k_{\bot}^{2} - \frac{q_{\bot}^{2}}{4}\right)^{2} \nonumber
\\
&&+ \left(k_{y}q_{x} - k_{x}q_{y}\right)^{2}, \\
F_{d3}^{B} &=& x(1-x)\left(\Omega^2+q_{z}^{2}\right) +
\left(k_{\bot}^{2} + \frac{q_{\bot}^{2}}{4}\right)^{2} \nonumber \\
&& +\left(k_{x}q_{x}+k_{y}q_{y}\right)^{2}-2(1-2x)
\left(k_{\bot}^{2} + \frac{q_{\bot}^{2}}{4}\right) \nonumber \\
&&\times\left(k_{x}q_{x} + k_{y}q_{y}\right).
\end{eqnarray}
This polarization has already been regularized by replacing
$\Pi_{d}(i\Omega,\frac{q_{x}}{\sqrt{A}},
\frac{q_{y}}{\sqrt{A}},\frac{q_{z}}{v})$ with
$\Pi_{d}(i\Omega,\frac{q_{x}}{\sqrt{A}},
\frac{q_{y}}{\sqrt{A}},\frac{q_{z}}{v}) - \Pi_{d}(0,0,0,0)$. We now
introduce the polar coordinates, and re-write the polarization as
\begin{eqnarray}
\Pi_{d}(i\Omega,\frac{q_{\bot}}{\sqrt{A}},q_{z}) &=&
\frac{N}{8\pi^{3}Av} \int_{0}^{1}dx
\int_{0}^{\Lambda_{UV}}dk_{\bot}k_{\bot}\int_{0}^{2\pi} d\theta
\nonumber \\
&&\times\left(\frac{F_{d4}^{A}}{F_{d4}^{B}} + 1\right),
\label{Eq:PolaDWGeneralForm}
\end{eqnarray}
where
\begin{eqnarray}
F_{d4}^{A}&=&-x(1-x)\left(\Omega^2-q_{z}^{2}\right) -
\left(k_{\bot}^{2}-\frac{q_{\bot}^{2}}{4}\right)^{2}\nonumber
\\
&&+k_{\bot}^{2}q_{\bot}^{2}\sin^{2}(\theta),
\\
F_{d4}^{B}&=&x(1-x)\left(\Omega^2 +
q_{z}^{2}\right) +
\left(k_{\bot}^{2}+\frac{q_{\bot}^{2}}{4}\right)^{2}\nonumber
\\
&& +
k_{\bot}^{2}q_{\bot}^{2}  \cos^{2}(\theta)- 2(1-2x)\left(k_{\bot}^{2} +
\frac{q_{\bot}^{2}}{4}\right)\nonumber
\\
&&\times k_{\bot}q_{\bot}\cos(\theta).
\end{eqnarray}
$\theta$ is the angle between $\mathbf{k}_{\bot}$ and
$\mathbf{q}_{\bot}$, and $\Lambda_{UV}$ is an UV cutoff. This
polarization is formally very complicated, and cannot be directly
used to calculate fermion self-energy. To proceed, we will analyze
its asymptotic behaviors in several different limits, and then
obtain an appropriate approximate expression. This strategy is
widely employed in the study of interacting fermion systems
\cite{GiulianiBook, ColemanBook, Varma02, Isobe16}.

\subsubsection{$q_{\bot}=0$}

In the limit $q_{\bot}=0$, the polarization takes the form
\begin{eqnarray}
\Pi_{d}(i\Omega,0,\frac{q_{z}}{v})
&=&\frac{N}{2\pi^{2} Av}q_{z}^{2}\int_{0}^{1}dx
x(1-x)\int_{0}^{\Lambda_{UV}}dk_{\bot}k_{\bot}\nonumber
\\
&&\times\frac{1}{x(1-x)\left(\Omega^2+q_{z}^{2}\right) + k_{\bot}^{4}}.
\end{eqnarray}
The upper limit of the integration over $k_{\bot}$ can be taken to
be infinity, which leads to
\begin{eqnarray}
\Pi_{d}(i\Omega,0,\frac{q_{z}}{v}) &=& \frac{N}{64
Av}\frac{q_{z}^{2}}{\sqrt{\Omega^2 + q_{z}^{2}}}.
\label{Eq:PolaDWLimit1}
\end{eqnarray}

\subsubsection{$\Omega=0$ and $q_{z}=0$}

Setting $\Omega = 0$ and $q_{z} = 0$, integrating over $x$, and then
making the transformation $k_{\bot} = q_{\bot}y$, we obtain
\begin{eqnarray}
\Pi_{d}(0,\frac{q_{\bot}}{\sqrt{A}},0) = \frac{N
q_{\bot}^{2}}{8\pi^{3} Av}\int_{0}^{\frac{\pi}{2}} d\theta
\int_{0}^{\frac{\Lambda_{UV}}{q_{\bot}}}dy F_{d5},
\end{eqnarray}
where
\begin{eqnarray}
F_{d5} &=& 4y + 2\frac{y^{2}\sin^{2}\theta -\left(y^2 -
\frac{1}{4}\right)^{2}}{\left(y^2+\frac{1}{4}\right)\cos\theta}
\nonumber \\
&&\times\ln\left(\frac{y^2+\frac{1}{4} + y\cos\theta}{y^2 +
\frac{1}{4}-y\cos\theta}\right).
\end{eqnarray}
In the limit $y \rightarrow \infty$, it is easy to verify that
\begin{eqnarray}
F_{d5}\rightarrow \frac{16-8\cos(2\theta)}{3y}.
\end{eqnarray}
In the low-energy regime, $\Pi(0,\frac{q_{\bot}}{\sqrt{A}},0)$ is
given by
\begin{eqnarray}
\Pi_{d}(0,\frac{q_{\bot}}{\sqrt{A}},0) &\approx&
\frac{Nq_{\bot}^{2}}{8\pi^{3} Av} \left[a_{1}+a_{2}+\frac{8\pi}{3}
\ln\left(\frac{\Lambda_{UV}}{q_{\bot}}\right)\right],\nonumber\\
\end{eqnarray}
where
\begin{eqnarray}
a_{1} &=& \int_{0}^{\frac{\pi}{2}}d\theta\int_{0}^{1}dyF_{d5},
\\
a_{2} &=& \int_{0}^{\frac{\pi}{2}}d\theta\int_{1}^{+\infty}dy
\left[F_{d5} - \frac{16-8\cos(2\theta)}{3y}\right].
\end{eqnarray}
Through numerical calculation, we find that
\begin{eqnarray}
a_{1}\approx3.83504, \qquad
a_{2}\approx-1.4367.
\end{eqnarray}
Retaining the leading contribution yields
\begin{eqnarray}
\Pi_{d}(0,\frac{q_{\bot}}{\sqrt{A}},0) =
\frac{Nq_{\bot}^{2}}{3\pi^{2} Av}
\ln\left(\frac{\Lambda_{UV}}{q_{\bot}}\right).
\label{Eq:PolaDWLimit2}
\end{eqnarray}

\subsubsection{$q_{z}=0$ and $q_{\bot}^{2} \ll \Omega$}

In the limit $q_{z} = 0$ and $q_{\bot}^{2} \ll \Omega$ , the
polarization can be approximated by
\begin{eqnarray}
\Pi_{d}(i\Omega,\frac{q_{\bot}}{\sqrt{A}},0) &=&
\frac{Nq_{\bot}^{2}}{2\pi^{2}
Av}\int_{0}^{\Lambda_{UV}}dk_{\bot}k_{\bot}^{3}
\left[\left(1+2\frac{k_{\bot}^{4}}{\Omega^2}\right)\right.\nonumber
\\
&&\left.\times\int_{0}^{1}dx\frac{1} {x(1-x)\Omega^2+k_{\bot}^{4}} -
\frac{2}{\Omega^2}\right].\nonumber\\
\end{eqnarray}
Carrying out the integration over $x$ and using the transformation
$k_{\bot}=\sqrt{|\Omega|}y$, we get
\begin{eqnarray}
\Pi_{d}(i\Omega,\frac{q_{\bot}}{\sqrt{A}},0) =
\frac{Nq_{\bot}^{2}}{\pi^{2} Av}
\int_{0}^{\frac{\Lambda_{UV}}{\sqrt{|\Omega|}}}dy F_{d6},
\end{eqnarray}
where
\begin{eqnarray}
F_{d6} = y^3\left[\frac{1+2y^4}{\sqrt{1 + 4y^4}}
\ln\left(\frac{\sqrt{1+4y^4}+1}{\sqrt{1 + 4y^4} - 1}\right) -
1\right].
\end{eqnarray}
As $y \rightarrow \infty$, the integrand becomes
\begin{eqnarray}
F_{d6}\rightarrow\frac{1}{3y},
\end{eqnarray}
which allows us to express
$\Pi_{d}(i\Omega,\frac{q_{\bot}}{\sqrt{A}},0)$ in the form
\begin{eqnarray}
\Pi_{d}(i\Omega,\frac{q_{\bot}}{\sqrt{A}},0) \approx
\frac{Nq_{\bot}^{2}}{\pi^{2} Av}\left[a_{3} + a_{4} + \frac{1}{3}
\ln\left(\frac{\Lambda_{UV}}{\sqrt{|\Omega|}}\right)\right],
\nonumber \\
\end{eqnarray}
where
\begin{eqnarray}
a_{3} &=& \int_{0}^{1}dyF_{d6}, \\
a_{4} &=& \int_{1}^{+\infty}dy\left(F_{d6}-\frac{1}{3y}\right).
\end{eqnarray}
Numerical calculations show that
\begin{eqnarray}
a_{3}\approx 0.192007,\quad a_{4}\approx -0.0114519.
\end{eqnarray}
Retaining the leading term, we approximately obtain
\begin{eqnarray}
\Pi_{d}(i\Omega,\frac{q_{\bot}}{\sqrt{A}},0) \approx
\frac{Nq_{\bot}^{2}}{3\pi^{2} Av}
\ln\left(\frac{\Lambda_{UV}}{\sqrt{|\Omega|}}\right).
\label{Eq:PolaDWLimit3}
\end{eqnarray}

\subsubsection{Ansatz for $\Pi_{d}(i\Omega,\mathbf{q})$}

Based on the polarization calculated in different limits, as shown
in Eqs.~(\ref{Eq:PolaDWLimit1}), (\ref{Eq:PolaDWLimit2}), and
(\ref{Eq:PolaDWLimit3}), we find that the polarization can be
approximated by the following \emph{Anstaz}:
\begin{eqnarray}
\Pi_{d}(i\Omega,\frac{q_{\bot}}{\sqrt{A}},\frac{q_{z}}{v}) &\approx&
N\left[\frac{q_{\bot}^{2}}{3\pi^{2} Av} \ln\left(
\frac{\Lambda_{UV}}{\left(\Omega^2 + q_{\bot}^{4}
\right)^{1/4}} + 1\right)\right. \nonumber \\
&&\left.+\frac{1}{64 Av} \frac{q_{z}^{2}}{\sqrt{\Omega^2 +
q_{z}^{2}}}\right]. \label{Eq:PoladDWAnsatz1}
\end{eqnarray}
Making the re-scaling transformations
\begin{eqnarray}
\frac{q_{x}}{\sqrt{A}}\rightarrow q_{x},\,\,
\frac{q_{y}}{\sqrt{A}}\rightarrow q_{y},\,\,
\frac{\Lambda_{UV}}{\sqrt{A}}\rightarrow \Lambda_{UV},\,\,
\frac{q_{z}}{v}\rightarrow q_{z},
\end{eqnarray}
we further have
\begin{eqnarray}
\Pi_{d}(i\Omega,q_{\bot},q_{z}) &\approx&
N\left[\frac{q_{\bot}^{2}}{3\pi^{2} v} \ln\left(\frac{\sqrt{A}
\Lambda_{UV}}{\left(\Omega^2 + A^{2}q_{\bot}^{4} \right)^{1/4}} + 1
\right)\right.\nonumber \\
&&\left. + \frac{1}{64 A} \frac{v q_{z}^{2}}{\sqrt{\Omega^2 +
v^{2}q_{z}^{2}}}\right]. \label{Eq:PolaDWAnsatz2}
\end{eqnarray}
This function will be used to compute the fermion self-energy.

\begin{figure*}[htbp]
\center
\includegraphics[width=2in]{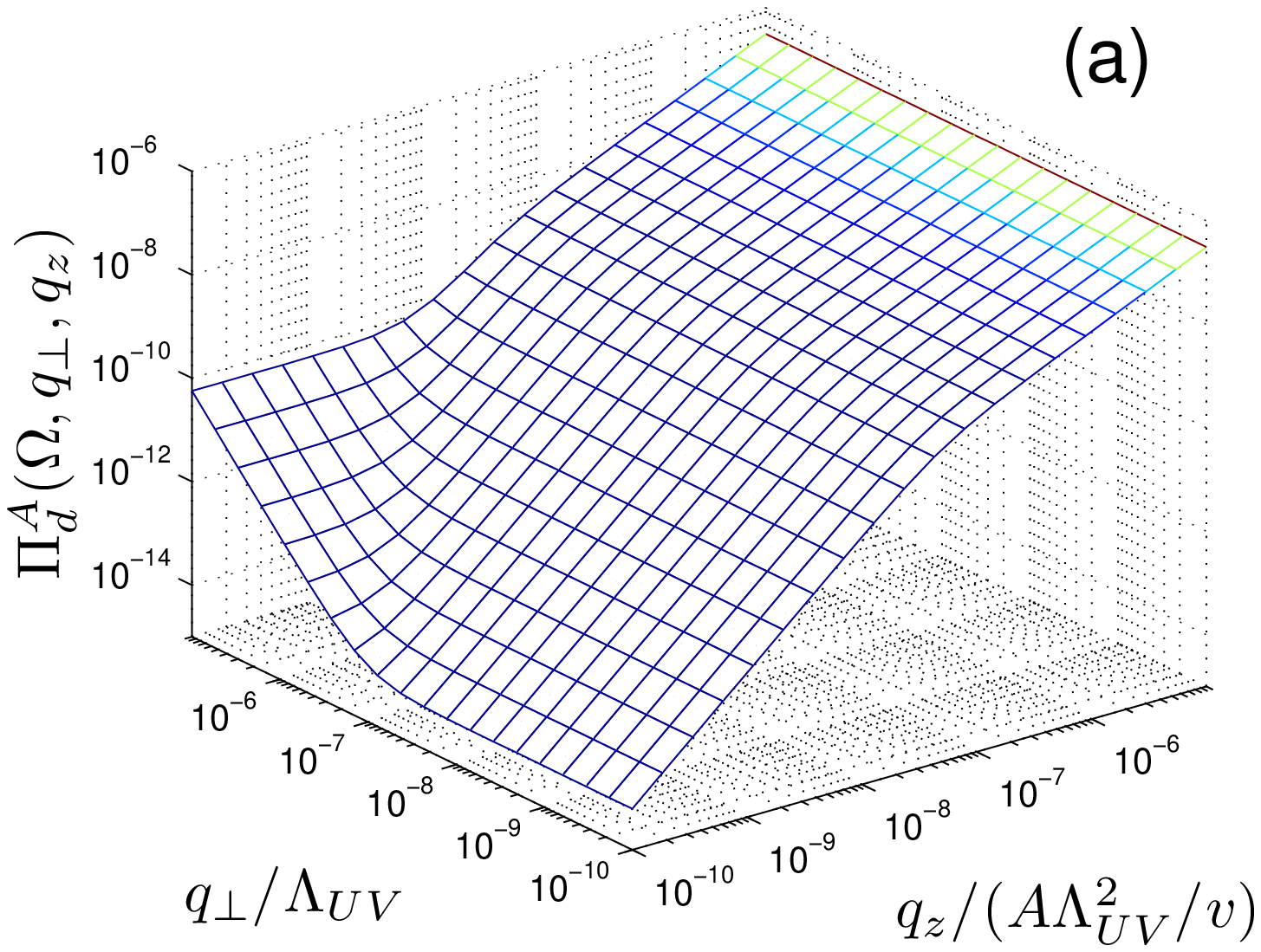}
\includegraphics[width=2in]{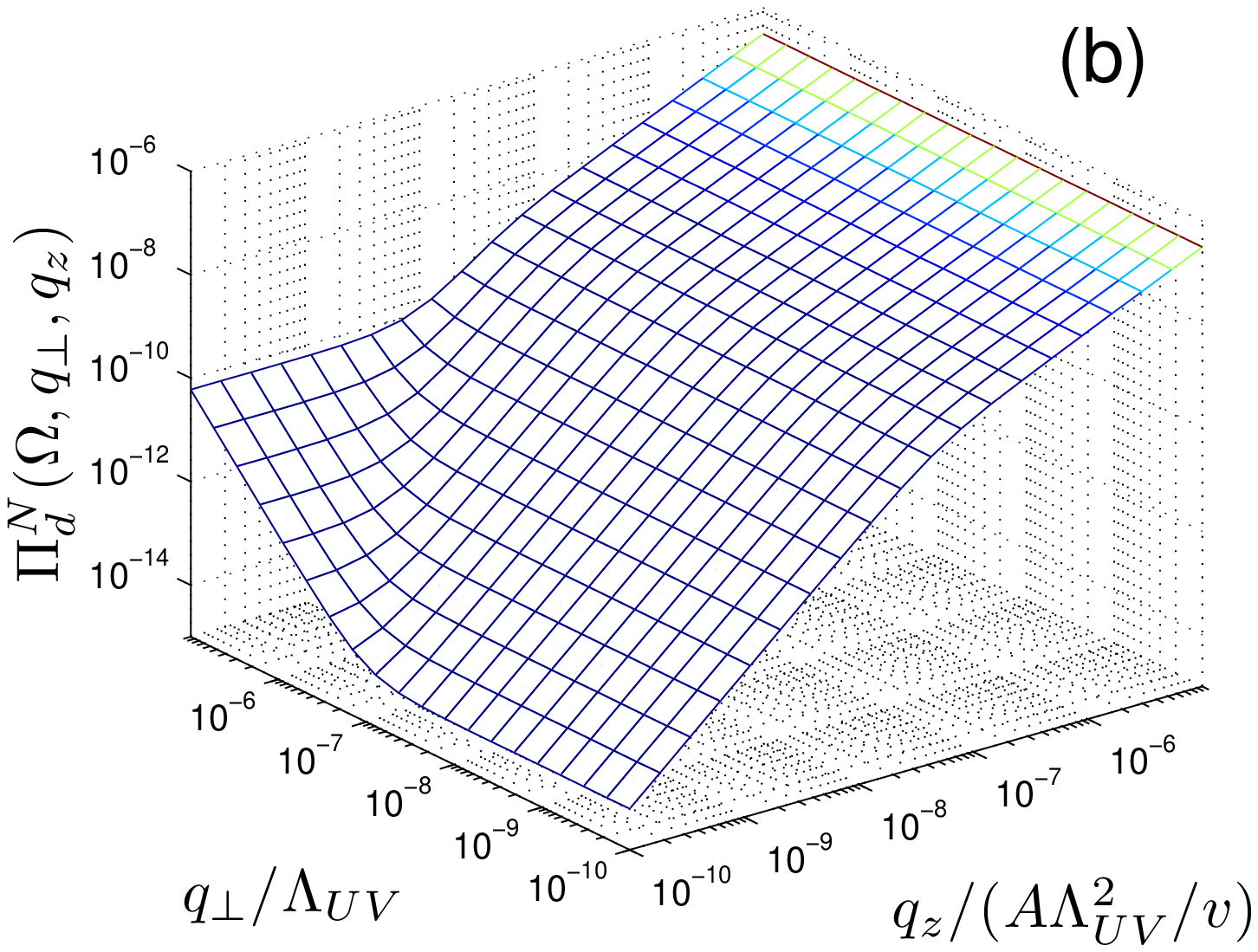}
\includegraphics[width=2in]{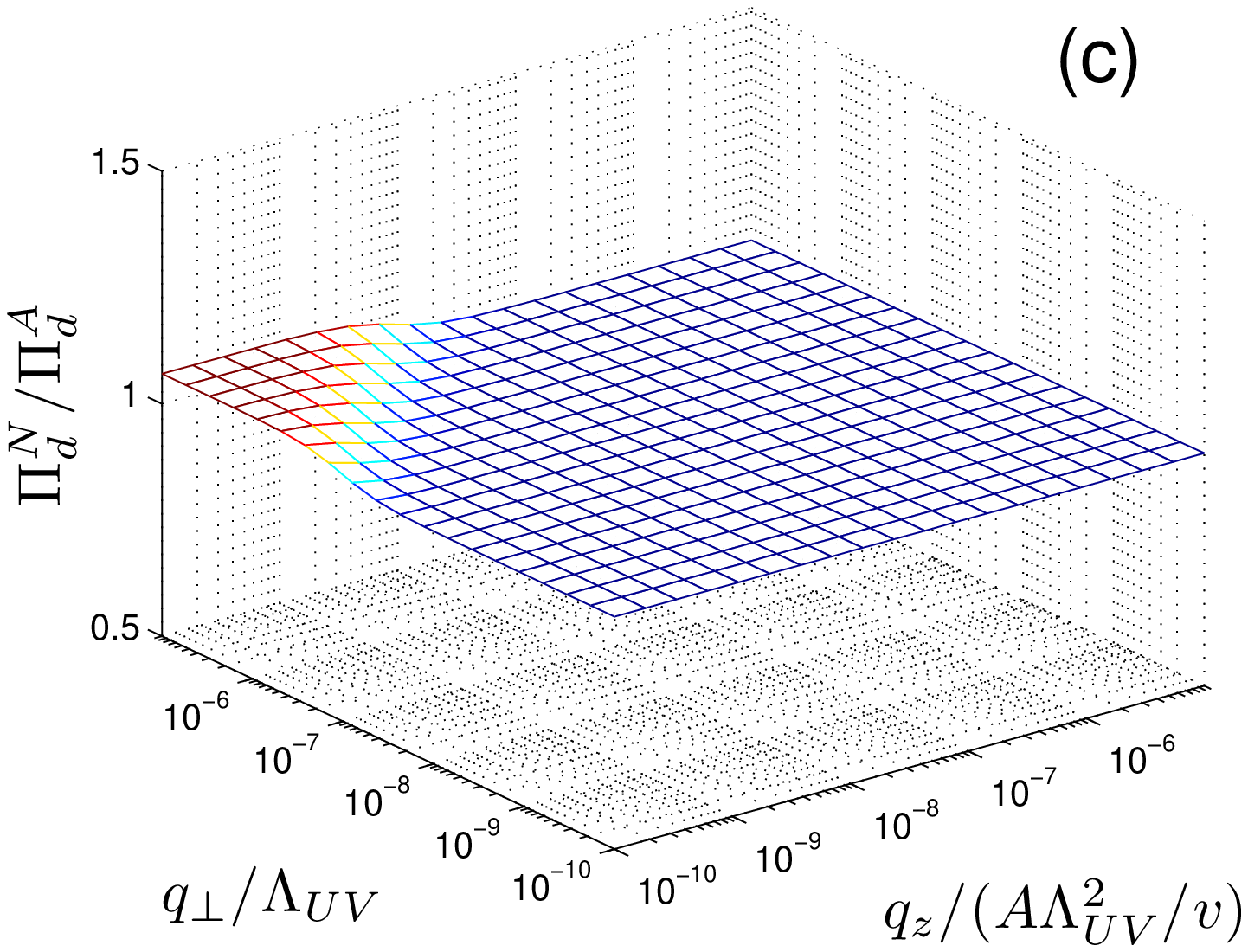}
\includegraphics[width=2in]{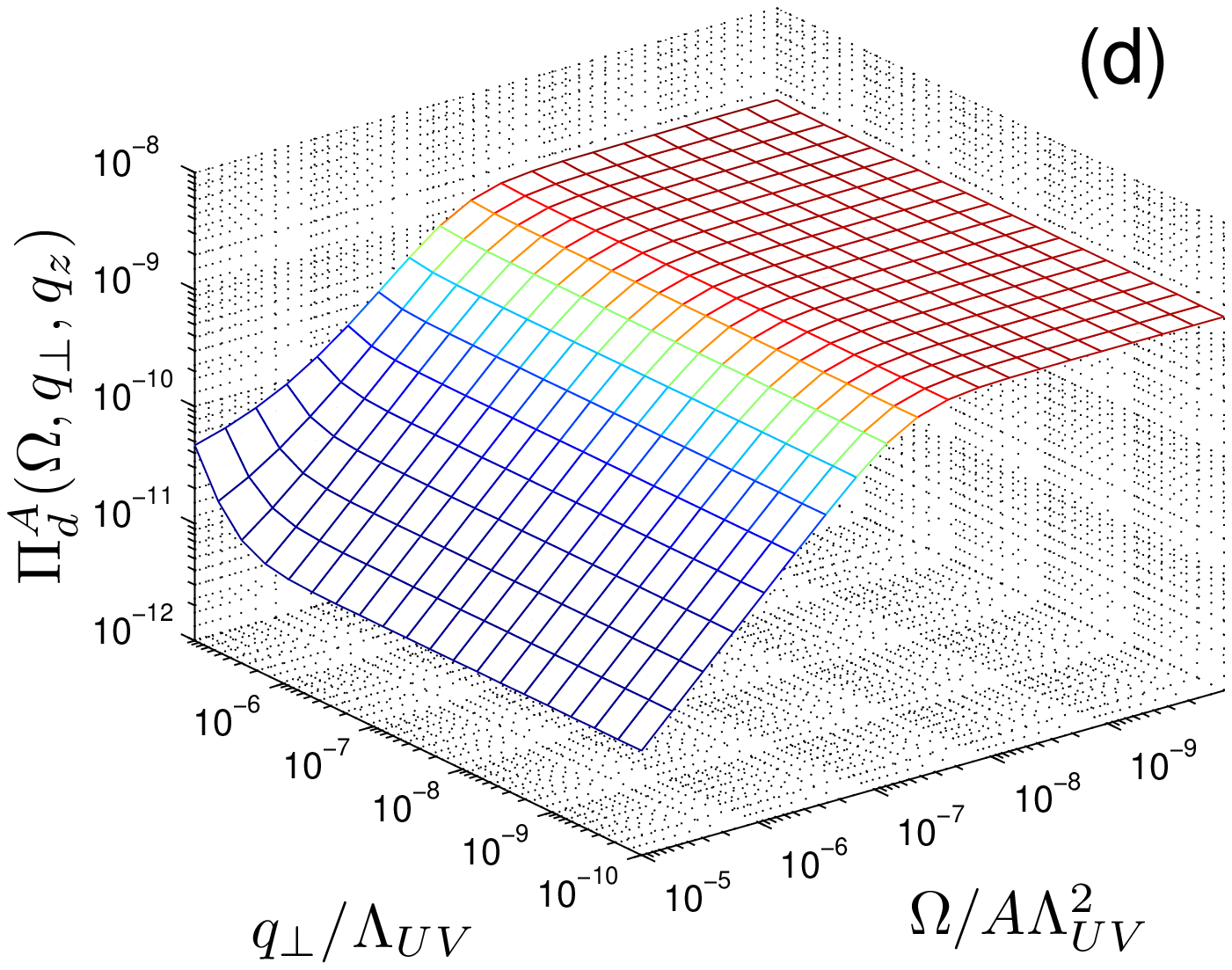}
\includegraphics[width=2in]{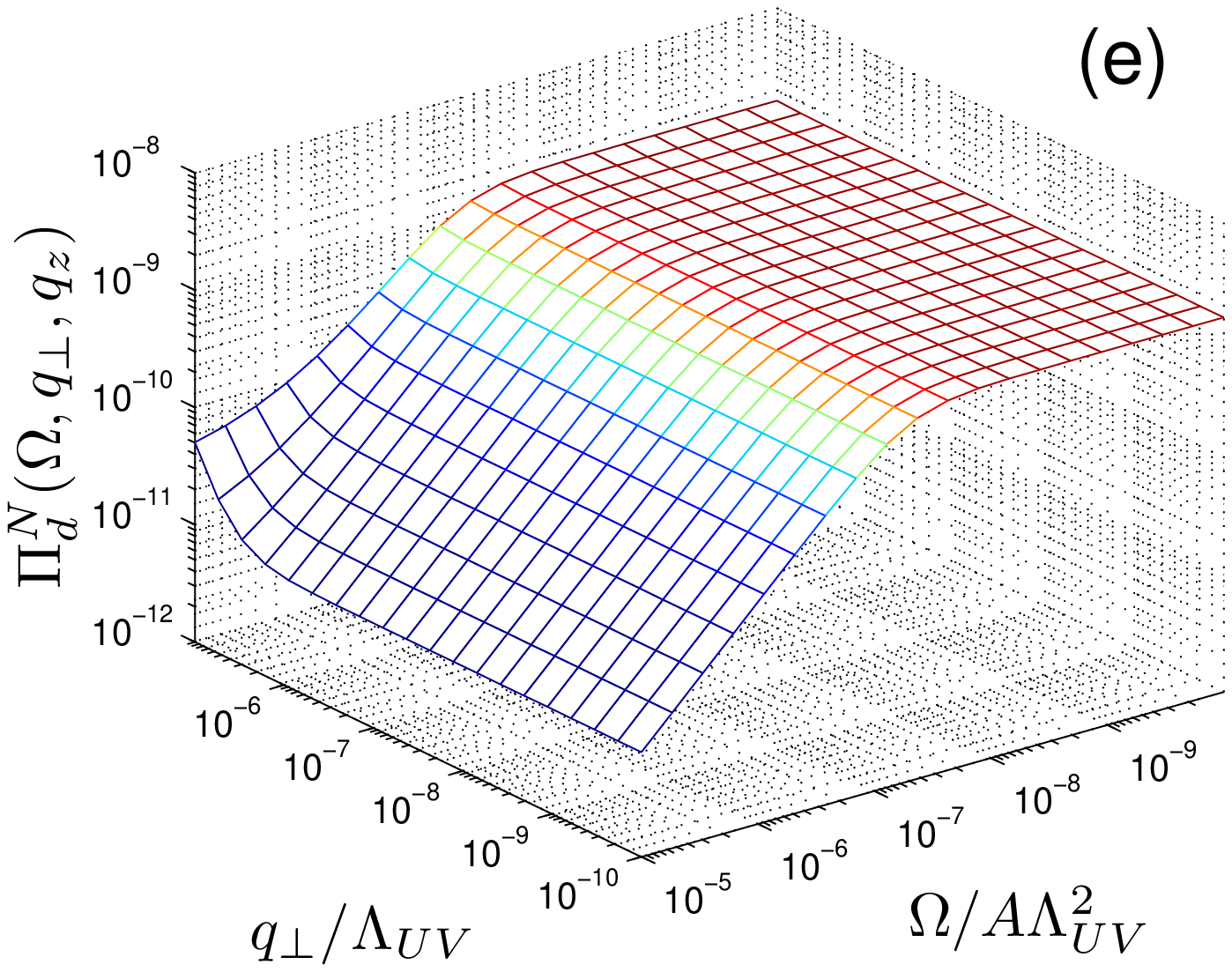}
\includegraphics[width=2in]{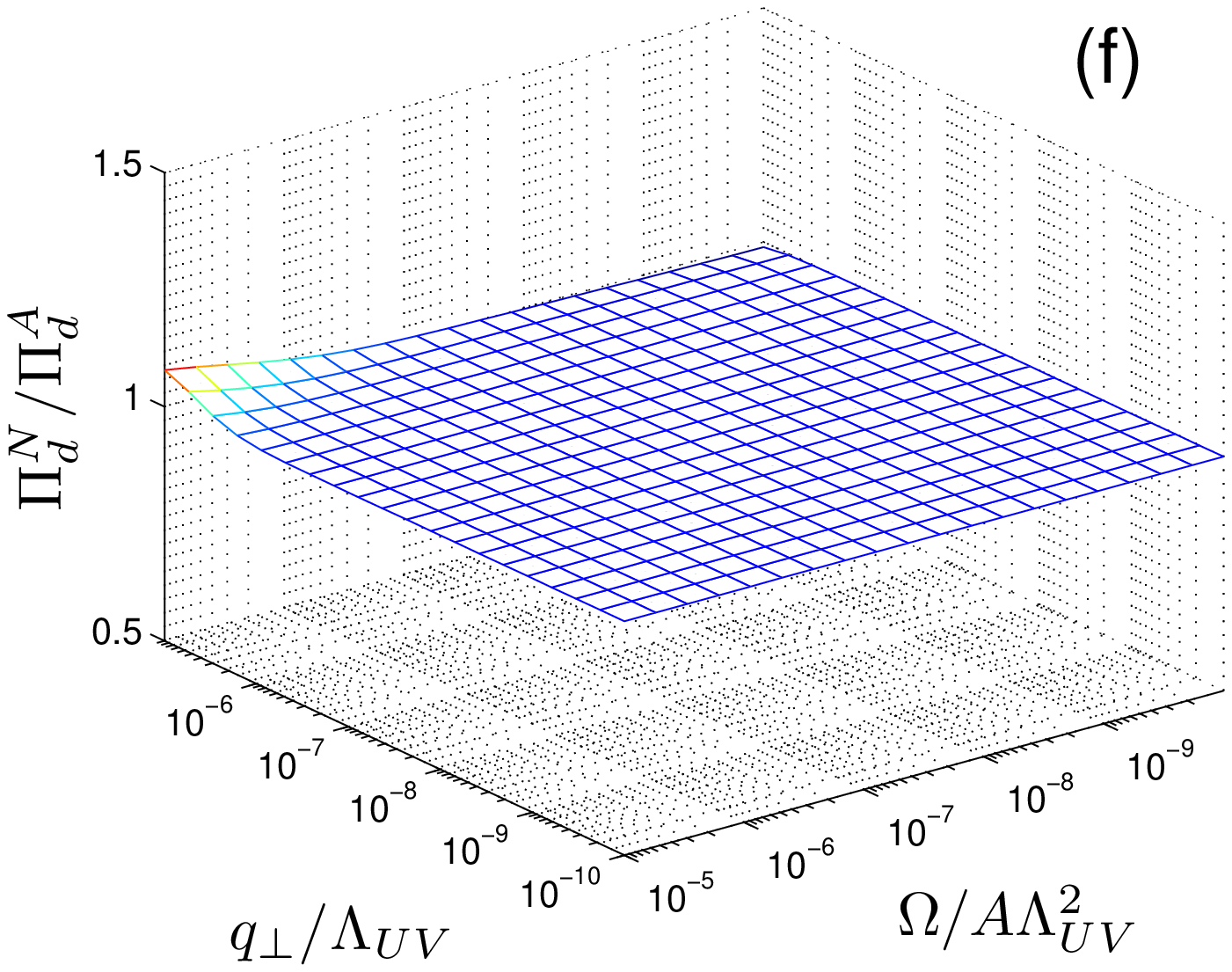}
\includegraphics[width=2in]{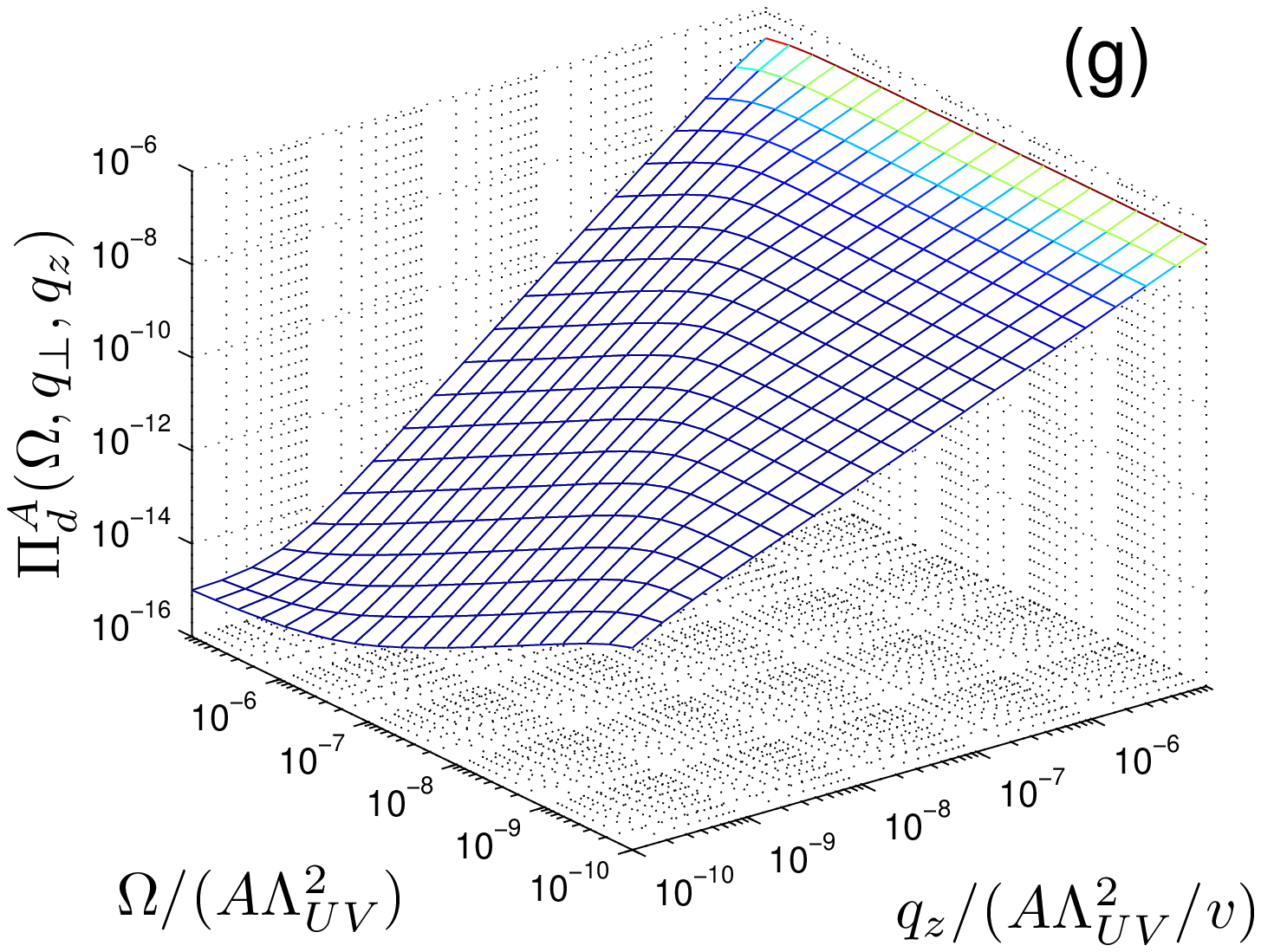}
\includegraphics[width=2in]{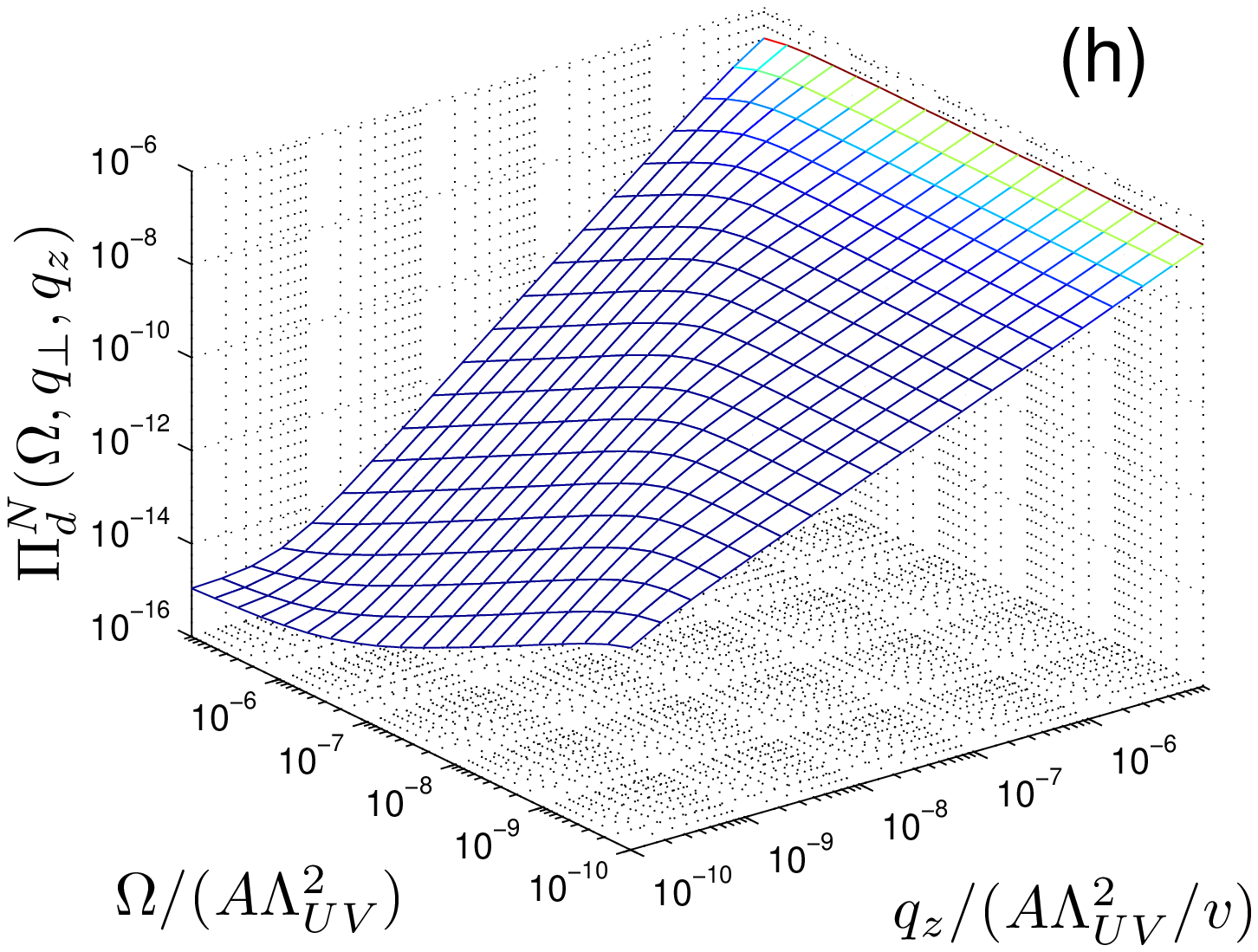}
\includegraphics[width=2in]{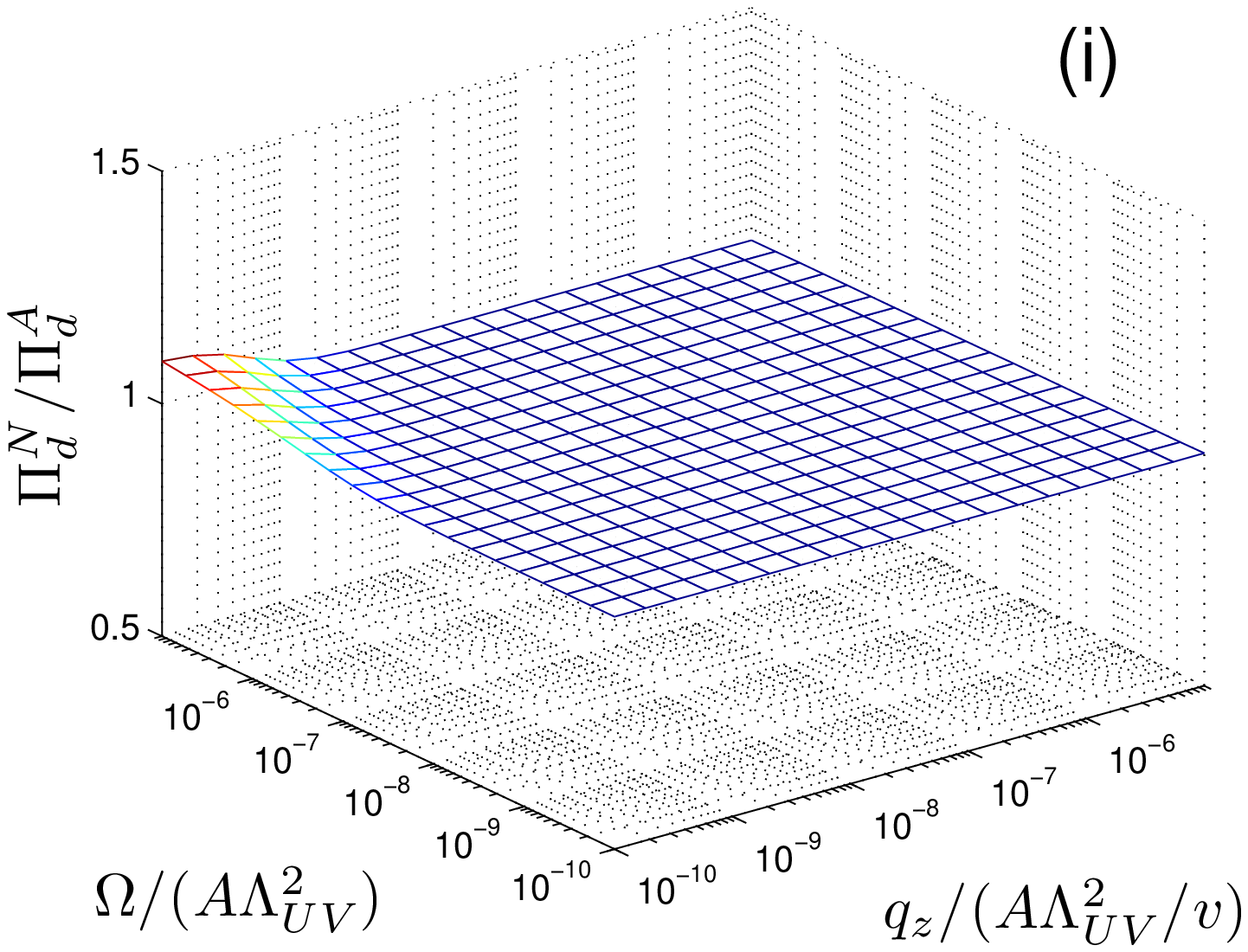}
\caption{Comparing the approximate expression of polarization with
the exact polarization for double-Weyl fermions. $\Pi_{d}^{A}$
represents the approximate expression of polarization given by
Eq.~(\ref{Eq:PolaDWAnsatz2}). $\Pi_{d}^{N}$ is the numerical result
of the polarization as shown in Eq.~(\ref{Eq:PolaDWGeneralForm}).
Dependence of $\Pi_{d}^{A}$, $\Pi_{d}^{N}$ and
$\Pi_{d}^{N}/\Pi_{d}^{A}$ on $q_{\bot}$ and $q_{z}$ with
$\Omega/A\Lambda_{UV}^{2}=10^{-5}$ are shown in (a), (b), and (c);
Dependence of $\Pi_{d}^{A}$, $\Pi_{d}^{N}$ and
$\Pi_{d}^{N}/\Pi_{d}^{A}$ on $\Omega$ and $q_{\bot}$ with
$q_{z}/(A\Lambda_{UV}^{2}/v)=10^{-5}$ are displayed in (d), (e), and
(f); Dependence of $\Pi_{d}^{A}$, $\Pi_{d}^{N}$ and
$\Pi_{d}^{N}/\Pi_{d}^{A}$ on $\Omega$ and $q_{z}$ with
$q_{\bot}/\Lambda_{UV}=10^{-5}$ are presented in (g), (h), and (i).
\label{Fig:PolaCompareDW} }
\end{figure*}

In Fig.~\ref{Fig:PolaCompareDW}, we present a direct comparison
between the approximate analytical expression
Eq.~(\ref{Eq:PolaDWAnsatz2}) and the exact one-loop polarization
Eq.~(\ref{Eq:PolaDWGeneralForm}). One can see that
Eq.~(\ref{Eq:PolaDWAnsatz2}) is very close to the exact one-loop
polarization of double-Weyl fermions in the low-energy region.

\subsection{Polarization $\Pi_{t}(i\Omega,\mathbf{q})$ of triple-Weyl fermions}

For triple-WSM, the polarization is given by
\begin{eqnarray}
\Pi_{t}(i\Omega,\mathbf{q}) &=& -N\int\frac{d\omega}{2\pi}\int
\frac{d^3\mathbf{k}}{(2\pi)^{3}}
\mathrm{Tr}\left[G_{t0}(i\omega,\mathbf{k})\right.\nonumber
\\
&&\left.\times G_{t0}\left(i\omega+i\Omega,\mathbf{k} +
\mathbf{q}\right)\right]. \label{Eq:PolaTripleDefAppen}
\end{eqnarray}
Substituting the fermion propagator
Eq.~(\ref{Eq:FreeFermonPropagatorTripleWeylDef}) into
Eq.~(\ref{Eq:PolaTripleDefAppen}), we find that
\begin{eqnarray}
\Pi_{t}(i\Omega,\frac{q_{x}}{B^{1/3}},\frac{q_{y}}{B^{1/3}},
\frac{q_{z}}{v}) = \frac{2N}{B^{2/3}v}\int\frac{d\omega}{2\pi}
\int\frac{d^3\mathbf{k}}{(2\pi)^{3}}\frac{F_{t1}^{A}}{F_{t1}^{B}},
\nonumber \\
\end{eqnarray}
where
\begin{eqnarray}
F_{t1}^{A} &=& \omega(\omega+\Omega) - \sum_{i=1}^{3}
g_{i}(\mathbf{k}) g_{i}(\mathbf{k} + \mathbf{q}), \\
F_{t1}^{B} &=& \left(\omega^{2}+E_{\mathbf{k}}^{2}\right)
\left[(\omega+\Omega)^{2}+E_{\mathbf{k}+\mathbf{q}}^{2}\right].
\end{eqnarray}
Here, $g_{1}(\mathbf{k}) = \left(k_{x}^{3}-3k_{x}k_{y}^{2}\right)$,
$g_{2}(\mathbf{k}) = \left(k_{y}^{3}-3k_{y}k_{x}^{2}\right)$,
$g_{3}(\mathbf{k}) = k_{z}$, and $E_{\mathbf{k}} =
\sqrt{k_{\bot}^{6}+k_{z}^{2}}$. The above calculations are completed
by using the following re-scaling manipulations
\begin{eqnarray}
&& k_{x}\rightarrow\frac{k_{x}}{B^{1/3}},\quad
k_{y}\rightarrow\frac{k_{y}}{B^{1/3}},\quad
q_{x}\rightarrow\frac{q_{x}}{B^{1/3}},\nonumber
\\
&& q_{y}\rightarrow\frac{q_{y}}{B^{1/3}},\quad
k_{z}\rightarrow\frac{k_{z}}{v},\quad\quad
q_{z}\rightarrow\frac{q_{z}}{v}.
\end{eqnarray}
Introducing Feynman integral, we further re-write the polarization
as
\begin{eqnarray}
\Pi_{t}(i\Omega,\frac{q_{x}}{B^{1/3}},\frac{q_{y}}{B^{1/3}},
\frac{q_{z}}{v}) &=&
\frac{2N}{B^{2/3}v}\int_{0}^{1}dx\int\frac{dk_{x}}{2\pi}
\frac{dk_{y}}{2\pi}F_{t2}^{A}\nonumber
\\
&&\times\int\frac{d^2\mathbf{K}}{(2\pi)^{2}}
\frac{1}{\left(F_{t2}^{B}\right)^{2}},
\end{eqnarray}
where $\mathbf{K}=(\omega,k_z)$ and
\begin{eqnarray}
F_{t2}^{A} &=& -x(1-x)\left(\Omega^2-q_{z}^{2}\right) -
g_{1}\left(\mathbf{k}-\frac{\mathbf{q}}{2}\right)
g_{1}\left(\mathbf{k}+\frac{\mathbf{q}}{2}\right) \nonumber \\
&& -g_{2}\left(\mathbf{k}-\frac{\mathbf{k}}{2}\right)
g_{2}\left(\mathbf{k} + \frac{\mathbf{q}}{2}\right), \\
F_{t2}^{B} &=& K^{2}+x(1-x)\left(\Omega^2+q_{z}^{2}\right) +
(1-x)\left(\mathbf{k}-\frac{\mathbf{q}}{2}\right)_{\bot}^{6}
\nonumber \\
&& + x\left(\mathbf{k}+\frac{\mathbf{q}}{2}\right)_{\bot}^{6}.
\end{eqnarray}
Transformations $k_{x}\rightarrow k_{x}-\frac{q_{x}}{2}$ and
$k_{y}\rightarrow k_{y}-\frac{q_{y}}{2}$ have been adopted in the
above calculation. Making use of
Eq.~(\ref{Eq:AppenIntegralFormula}), we integrate over $\mathbf{K}$
and obtain
\begin{eqnarray}
\Pi_{t}(i\Omega,\frac{q_{x}}{B^{1/3}},
\frac{q_{y}}{B^{1/3}},\frac{q_{z}}{v}) &=& \frac{N}{2\pi
B^{2/3}v}\int_{0}^{1}dx\int\frac{dk_{x}}{2\pi}
\frac{dk_{y}}{2\pi}\nonumber
\\
&&\times \left(\frac{F_{t3}^{A}}{F_{t3}^{B}} + 1\right).
\end{eqnarray}
where
\begin{eqnarray}
F_{t3}^{A}&=&-x(1-x)\left(\Omega^2 -
q_{z}^{2}\right) -
\left(k_{\bot}^{2}-\frac{q_{\bot}^{2}}{4}\right)^{3}\nonumber
\\
&& +3\left(k_{\bot}^{2} - \frac{q_{\bot}^{2}}{4}\right)
\left(k_{x}q_{y}-k_{y}q_{x}\right)^{2},
\\
F_{t3}^{B}&=&x(1-x)\left(\Omega^2+q_{z}^{2}\right) +
\left(k_{\bot}^{2} + \frac{q_{\bot}^{2}}{4}\right)^3 \nonumber
\\
&&+ 3\left(k_{\bot}^{2} + \frac{q_{\bot}^{2}}{4}\right)
\left(k_{x}q_{x} + k_{y}q_{y}\right)^{2}\nonumber
\\
&&-(1-2x)\Bigg[3\left(k_{\bot}^{2} + \frac{q_{\bot}^{2}}{4}
\right)^{2}\left(k_{x}q_{x} + k_{y}q_{y}\right)\nonumber
\\
&&+ \left(k_{x}q_{x} + k_{y}q_{y}\right)^{3}\Bigg].
\end{eqnarray}
In order to regularize the polarization, we have used
$\Pi_{t}(i\Omega,\frac{q_{x}}{B^{1/3}},
\frac{q_{y}}{B^{1/3}},\frac{q_{z}}{v}) - \Pi_{t}(0,0,0,0)$
to replace $\Pi_{t}(i\Omega,\frac{q_{x}}{B^{1/3}},
\frac{q_{y}}{B^{1/3}},\frac{q_{z}}{v})$. In the polar
coordinates, the polarization can be further converted to
\begin{eqnarray}
\Pi_{t}(i\Omega,\frac{q_{\bot}}{B^{1/3}},\frac{q_{z}}{v}) &=&
\frac{N}{8\pi^3
B^{2/3}v}\int_{0}^{1}dx\int_{0}^{\Lambda_{UV}}
dk_{\bot}k_{\bot}\nonumber
\\
&&\times\int_{0}^{2\pi}d \theta\left(\frac{F_{4t}^{A}}{F_{4t}^{B}} +
1\right), \label{Eq:PolaTWGeneralForm}
\end{eqnarray}
where
\begin{eqnarray}
F_{4t}^{A}&=&-x(1-x)
\left(\Omega^2-q_{z}^{2}\right)
-\left(k_{\bot}^{2}-\frac{q_{\bot}^{2}}{4}\right)^{3}\nonumber
\\
&&+3\left(k_{\bot}^{2}-\frac{q_{\bot}^{2}}{4}\right) k_{\bot}^{2}
q_{\bot}^{2}\sin^2\theta,
\\
F_{4t}^{B}&=&x(1-x)\left(\Omega^2+q_{z}^{2}\right) +
\left(k_{\bot}^{2} + \frac{q_{\bot}^{2}}{4}\right)^3\nonumber
\\
&& +3\left(k_{\bot}^{2}+\frac{q_{\bot}^{2}}{4}\right)
k_{\bot}^{2}q_{\bot}^{2}\cos^2\theta\nonumber
\\
&&-(1-2x)\Bigg[3\left(k_{\bot}^{2} +
\frac{q_{\bot}^{2}}{4}\right)^{2}k_{\bot}q_{\bot}
\cos\theta\nonumber
\\
&&+k_{\bot}^{3}q_{\bot}^{3}\cos^3\theta\Bigg].
\end{eqnarray}
Similar to the case of double-WSM, we need to analyze the asymptotic
behavior of the above polarization in several limits.

\subsubsection{$q_{\bot}=0$}

In the limit $q_{\bot}=0$, the polarization takes the form
\begin{eqnarray}
\Pi_{t}(i\Omega,0,\frac{q_{z}}{v})
&=&\frac{N}{2\pi^2 B^{2/3}v}q_{z}^{2}\int_{0}^{1}dxx(1-x)
\int_{0}^{\Lambda_{UV}} dk_{\bot}\nonumber
\\
&&\times k_{\bot}\frac{1}{x(1-x)
\left(\Omega^2+q_{z}^{2}\right)+k_{\bot}^6}.
\end{eqnarray}
Extending the upper limit of the integration over $k_{\bot}$ to
infinity, we find that
\begin{eqnarray}
\Pi_{t}(i\Omega,0,\frac{q_{z}}{v}) = c_{t}\frac{N}{B^{2/3}v}
\frac{q_{z}^{2}}{\left(\Omega^2 + q_{z}^{2}\right)^{2/3}},
\label{Eq:PolaTWLimit1}
\end{eqnarray}
where
\begin{eqnarray}
c_{t} = \frac{2^{1/3}\pi^{1/2}}{90\Gamma \left(5/6\right)
\Gamma(2/3)}.
\end{eqnarray}

\subsubsection{$\Omega=0$ and $q_{z}=0$}

When $\Omega=0$ and $q_{z}=0$, after integrating over $x$ and making
the transformation $k_{\bot} = q_{\bot}y$, the polarization becomes
\begin{eqnarray}
\Pi_{t}(0,\frac{q_{\bot}}{B^{1/3}},0) = \frac{Nq_{\bot}^{2}}{4\pi^3
B^{2/3}v} \int_{0}^{\frac{\pi}{2}}d\theta
\int_{0}^{\frac{\Lambda_{UV}}{q_{\bot}}} dy F_{t5},
\end{eqnarray}
where
\begin{eqnarray}
F_{t5} &=& 2y + 3\frac{-\left(y^2-\frac{1}{4}\right)^{3} +
3\left(y^2-\frac{1}{4}\right) y^2\sin^2 \theta}{3\left(y^2 +
\frac{1}{4}\right)^{2}\cos \theta + y^2\cos^3\theta}\nonumber
\\
&&\times\ln\left(\frac{y^2+\frac{1}{4}+y\cos \theta} {y^2 +
\frac{1}{4}-y\cos \theta}\right).
\end{eqnarray}
As $y\rightarrow \infty$, the integrand can be simplified as
\begin{eqnarray}
F_{t5}\rightarrow \frac{6-3\cos\left(2\theta\right)}{y}.
\end{eqnarray}
In the low-energy region this polarization is simplified to
\begin{eqnarray}
\Pi_{t}(0,\frac{q_{\bot}}{B^{1/3}},0) &\approx& \frac{N
q_{\bot}^{2}}{4\pi^3 B^{2/3}v}\Big[a_{5} + a_{6}\nonumber
\\
&& + 3\pi\ln\left(\frac{\Lambda_{UV}}{q_{\bot}}\right)\Big],
\end{eqnarray}
where
\begin{eqnarray}
a_{5} &=& \int_{0}^{\frac{\pi}{2}}d\theta\int_{0}^{1} dy F_{t5},
\\
a_{6} &=& \int_{0}^{\frac{\pi}{2}}d\theta\int_{1}^{+\infty}dy
\left[F_{t5} - \frac{6-3\cos\left(2\theta\right)}{y}\right].
\end{eqnarray}
Numerical calculations lead to
\begin{eqnarray}
a_{5}\approx 1.9333, \quad a_{6}\approx -3.28925.
\end{eqnarray}
Retaining the leading term, we get
\begin{eqnarray}
\Pi_{t}(0,\frac{q_{\bot}}{B^{1/3}},0) = \frac{3N
q_{\bot}^{2}}{4\pi^2 B^{2/3}v}\ln
\left(\frac{\Lambda_{UV}}{q_{\bot}}\right).
\label{Eq:PolaTWLimit2}
\end{eqnarray}

\subsubsection{$q_{z}=0$ and $q_{\bot}^{3} \ll \Omega$}

For $q_{z} = 0$ and $q_{\bot}^{3} \ll \Omega$, the polarization can
be approximated by
\begin{eqnarray}
\Pi_{t}(i\Omega,\frac{q_{\bot}}{B^{1/3}},0) &=& \frac{9N
q_{\bot}^{2}}{8\pi^2 B^{2/3}v}\int dk_{\bot}k_{\bot}^{5}
\left[\left(1 + 2\frac{k_{\bot}^{6}}{\Omega^2}\right)\right.
\nonumber \\
&& \left.\times\int_{0}^{1}dx \frac{1}{x(1-x)\Omega^2 + k_{\bot}^6}
- \frac{2}{\Omega^2}\right]. \nonumber \\
\end{eqnarray}
Performing the integration over $x$ and employing the transformation
$k_{\bot}=|\Omega|^{1/3}y$, we further obtain
\begin{eqnarray}
\Pi_{t}(i\Omega,\frac{q_{\bot}}{B^{1/3}},0) =
\frac{9Nq_{\bot}^{2}}{4\pi^2 B^{2/3}v}
\int_{0}^{\frac{\Lambda_{UV}}{|\Omega|^{1/3}}}dy F_{t6},
\end{eqnarray}
where
\begin{eqnarray}
F_{t6} = y^5 \left[\frac{1+2y^{6}}{\sqrt{1+4y^6}} \ln
\left(\frac{\sqrt{1+4y^6}+1}{\sqrt{1+4y^6}-1}\right)-1\right].
\end{eqnarray}
As $y \rightarrow \infty$, the integrand becomes
\begin{eqnarray}
F_{t6}\rightarrow \frac{1}{3y},
\end{eqnarray}
thus we simplify the polarization to
\begin{eqnarray}
\Pi_{t}(i\Omega,\frac{q_{\bot}}{B^{1/3}},0) &\approx& \frac{9N
q_{\bot}^{2}}{4\pi^2 B^{2/3}v}\Big[a_{7}+a_{8}\nonumber
\\
&& +
\frac{1}{3} \ln\left(\frac{\Lambda_{UV}}{|\Omega|^{1/3}}
\right)\Big],
\end{eqnarray}
where
\begin{eqnarray}
a_{7} &=& \int_{0}^{1} dyF_{t6}, \\
a_{8} &=& \int_{1}^{+\infty} dy\left(F_{t6}-\frac{1}{3y}\right).
\end{eqnarray}
The values of $a_7$ and $a_8$ are
\begin{eqnarray}
a_{7}\approx 0.128005,\quad a_{8} \approx -0.0076346.
\end{eqnarray}
Keeping the leading contribution, the polarization is finally given
by
\begin{eqnarray}
\Pi_{t}(\Omega,\frac{q_{\bot}}{B^{1/3}},0) \approx \frac{3N
q_{\bot}^{2}}{4\pi^2 B^{2/3}v}
\ln\left(\frac{\Lambda_{UV}}{|\Omega|^{1/3}}\right).
\label{Eq:PolaTWLimit3}
\end{eqnarray}


\subsubsection{Ansatz for $\Pi_{t}(i\Omega,\mathbf{q})$}

Based on the polarization calculated in different limits, as shown
in Eqs.~(\ref{Eq:PolaTWLimit1}), (\ref{Eq:PolaTWLimit2}), and
(\ref{Eq:PolaTWLimit3}), the polarization can be approximated by the
\emph{anstaz}
\begin{eqnarray}
\Pi_{t}(i\Omega,q_{\bot},q_{z}) &=& N\left[\frac{3
q_{\bot}^{2}}{4\pi^2v} \ln\left(\frac{B^{1/3}
\Lambda_{UV}}{\left(\Omega^2 + B^{2}q_{\bot}^{6}\right)^{1/6}} +
1\right)\right.\nonumber \\
&&\left.+c_{t}\frac{1}{B^{2/3}}\frac{vq_{z}^{2}}{\left(\Omega^2 +
v^2q_{z}^{2}\right)^{2/3}}\right].\label{Eq:PolaTWAnsatz3}
\end{eqnarray}
According to the numerical results (presented only in the published
version, but not here due to space restriction), we can see that the
above approximate analytical expression Eq.~(\ref{Eq:PolaTWAnsatz3})
is very close to the exact one-loop polarization of triple-Weyl
fermions in the low-energy region.

\section{Fermion self-energy \label{App:SelfEnergy}}

We now compute the fermion self-energy functions caused by the
Coulomb interaction, first for double-Weyl fermions and then for
triple-Weyl fermions.

\begin{widetext}
\subsection{Self-energy of double-Weyl fermions \label{App:SelfEnergyDW}}

To the leading order of $1/N$ expansion, the self-energy of
double-Weyl fermions due to Coulomb interaction is defined as
\begin{eqnarray}
\Sigma_{d}(i\omega,\mathbf{k})&=&\int'\frac{d\Omega}{2\pi}
\frac{d^{3}\mathbf{q}}{(2\pi)^{3}}
G_{d0}(i\omega+i\Omega,\mathbf{k} + \mathbf{q})
V_{d}(i\Omega,\mathbf{q}),\label{Eq:DWFermionSelfEnergyDefAppen}
\end{eqnarray}
where the dressed Coulomb interaction function is
\begin{eqnarray}
V_{d}(i\Omega,\mathbf{q}) &=& \frac{1}{V_{0}^{-1}(\mathbf{q}) +
\Pi_{d}(i\Omega,\mathbf{q})}
= \frac{1}{\frac{q_{\bot}^{2}+\zeta q_{z}^{2}}{4\pi\alpha v} +
\frac{N q_{\bot}^{2}}{3\pi^{2} v}
\ln\left(\frac{\sqrt{A}\Lambda_{UV}}{\left(\Omega^2 +
A^{2}q_{\bot}^{4}\right)^{1/4}}+1\right) + \frac{N}{64
A}\frac{vq_{z}^{2}}{\sqrt{\Omega^2+v^{2}q_{z}^{2}}}}.
\end{eqnarray}
We then substitute Eq.~(\ref{Eq:FreeFermonPropagatorDoubleWeylDef})
into Eq.~(\ref{Eq:DWFermionSelfEnergyDefAppen}), and expand
$\Sigma_{d}(i\omega,\mathbf{k})$ in powers of small values of
$i\omega$, $k_{x}$, $k_{y}$, and $k_{z}$. To the leading order, the
self-energy can be expressed as
\begin{eqnarray}
\Sigma_{d}(i\omega,\mathbf{k}) &\approx& i\omega\Sigma_{d1} -
A\left[d_{1}(\mathbf{k})\sigma_{x} +
d_{2}(\mathbf{k})\sigma_{y}\right]\Sigma_{d2} -
vk_{z}\sigma_{z}\Sigma_{d3}, \label{Eq:DWSelfEnergyResultAppen}
\end{eqnarray}
where
\begin{eqnarray}
\Sigma_{d1} &=& \frac{1}{8\pi^3}\int' d\Omega
dq_{\bot}q_{\bot}dq_{z} \frac{\Omega^2-A^2q_{\bot}^{4}-v^2
q_{z}^{2}}{\left(\Omega^2 + A^2q_{\bot}^{4} + v^2
q_{z}^{2}\right)^{2}}V_{d}(i\Omega,\mathbf{q}),
\label{Eq:DWE1Appen} \\
\Sigma_{d2} &=& \frac{1}{8\pi^3}\int' d\Omega dq_{\bot}
q_{\bot}dq_{z}\left[\frac{\Omega^2 - 4A^2q_{\bot}^{4} + v^2
q_{z}^{2}}{\left(\Omega^2 + A^2q_{\bot}^{4} + v^2 q_{z}^{2}
\right)^{2}} + \frac{4 A^4 q_{\bot}^{8}}{\left(\Omega^2 +
A^2q_{\bot}^{4} + v^2q_{z}^{2}\right)^{3}}\right]
V_{d}(i\Omega,\mathbf{q}), \label{Eq:DWE2Appen}
\\
\Sigma_{d3}&=&\frac{1}{8\pi^3}\int' d\Omega dq_{\bot}q_{\bot}dq_{z}
\frac{\Omega^2+A^2q_{\bot}^{4}-v^2q_{z}^{2}}{\left(\Omega^2 +
A^2q_{\bot}^{4} + v^2q_{z}^{2}\right)^{2}}
V_{d}(i\Omega,\mathbf{q}). \label{Eq:DWE3Appen}
\end{eqnarray}
Here, we adopt the following integration ranges:
\begin{eqnarray}
-\infty < \Omega < \infty, \qquad b\Lambda<E_{d}<\Lambda,
\label{Eq:DWRGSchemeAAppen}
\end{eqnarray}
with $E_{d} = \sqrt{A^2q_{\bot}^{4}+v^2q_{z}^{2}}$
where $b=e^{-\ell}$.  If we define
\begin{eqnarray}
E = \sqrt{A^2q_{\bot}^{4}+v^{2}q_{z}^{2}},\quad \kappa =
\frac{Aq_{\bot}^{2}}{v\left|q_{z}\right|},
\label{Eq:DWIntegalTransformA}
\end{eqnarray}
then we can write $q_{\bot}$ and $q_{z}$ as
\begin{eqnarray}
q_{\bot} = \frac{\sqrt{\kappa}\sqrt{E}}{\sqrt{A}\left(1 +
\kappa^2\right)^{1/4}},\quad \left|q_{z}\right| =
\frac{E}{v\sqrt{1+\kappa^2}}. \label{Eq:DWIntegalTransformB}
\end{eqnarray}
Therefore, the integration over $q_{\bot}$ and $q_{z}$ can be
converted to the integration over $E$ and $\kappa$, by invoking the
relation
\begin{eqnarray}
dq_{\bot}d|q_{z}|&=&
\left|\left|
\begin{array}{cc}
\frac{\partial q_{\bot}}{\partial E} & \frac{\partial
q_{\bot}}{\partial \kappa}
\\
\frac{\partial |q_{z}|}{\partial E} & \frac{\partial
|q_{z}|}{\partial \kappa}
\end{array}
\right|\right|dEd\kappa
= \frac{\sqrt{E}}{2v\sqrt{A}\sqrt{\kappa}
\left(1+\kappa^2\right)^{3/4}}dEd\kappa.
\label{Eq:DWIntegralTranformMeasure}
\end{eqnarray}
Using the transformations given by
Eqs.~(\ref{Eq:DWIntegalTransformB}) and
(\ref{Eq:DWIntegralTranformMeasure}), we calculate
Eqs.~(\ref{Eq:DWE1Appen})-(\ref{Eq:DWE3Appen}) along with the RG
scheme (\ref{Eq:DWRGSchemeAAppen}), and eventually obtain
\begin{eqnarray}
\Sigma_{d1} = C_{d1}\ell,\quad \Sigma_{d2} = C_{d2}\ell,
\quad \Sigma_{d3} = C_{d3}\ell,
\end{eqnarray}
where
\begin{eqnarray}
C_{d1} &=& \frac{1}{8\pi^{3}}\int_{-\infty}^{+\infty}dx
\int_{0}^{+\infty}d\kappa \frac{1}{\left(1 +
\kappa^2\right)^{1/2}}\frac{x^2-1}{\left(x^2+1\right)^{2}}
\mathcal{G}_{d}(x,\kappa), \label{Eq:ExpressionCd1}
\\
C_{d2}&=&\frac{1}{8\pi^3}\int_{-\infty}^{+\infty}dx
\int_{0}^{+\infty}d\kappa \frac{1}{\left(1+\kappa^2\right)^{3/2}}
\left[\frac{x^2\left(1+\kappa^2\right) -
4\kappa^2+1}{\left(x^2+1\right)^{2}} +
\frac{4\kappa^4}{\left(1+\kappa^2\right)\left(x^{2} +
1\right)^{3}}\right]\mathcal{G}_{d}(x,\kappa),
\label{Eq:ExpressionCd2}
\\
C_{d3}&=&\frac{1}{8\pi^3}\int_{-\infty}^{+\infty}dx
\int_{0}^{+\infty}d\kappa \frac{1}{\left(1+\kappa^2\right)^{3/2}}
\frac{x^2\left(1+\kappa^2\right)+\kappa^2-1}{\left(x^2 +
1\right)^{2}}\mathcal{G}_{d}(x,\kappa), \label{Eq:ExpressionCd3}
\end{eqnarray}
with
\begin{eqnarray}
\mathcal{G}_{d}^{-1}(x,\kappa) = \frac{1}{4\pi\alpha}\left(\kappa +
\frac{\beta_{d}}{\left(1+\kappa^2\right)^{1/2}}\right) +
N\left[\frac{\kappa}{3\pi^{2}}\ln\left(\frac{\gamma_{d}e^{\frac{\ell}{2}}
\left(1+\kappa^2\right)^{1/4}}{\left(x^2 \left(1+\kappa^2\right) +
\kappa^{2}\right)^{1/4}}+1\right) + \frac{1}{64} \frac{1}{\sqrt{x^2
\left(1+\kappa^2\right) + 1}}\right]. \label{Eq:ExpressionGd}
\end{eqnarray}
Here, $\beta_{d}$ and $\gamma_{d}$ are defined by $\beta_{d} =
\frac{\zeta A\Lambda}{v^{2}}$ and
$\gamma_{d}=\frac{\sqrt{A}\Lambda_{UV}}{\sqrt{\Lambda}}$,
respectively.

\subsection{Self-energy of triple-Weyl fermions \label{App:SelfEnergyTW}}

To the leading order of $1/N$ expansion, the self-energy of
triple-Weyl fermions induced by the long-range Coulomb interaction
is given by
\begin{eqnarray}
\Sigma_{t}(i\omega,\mathbf{k}) = \int'\frac{d\Omega}{2\pi}
\frac{d^{3}\mathbf{q}}{(2\pi)^{3}}
G_{t0}(i\omega+i\Omega,\mathbf{k} + \mathbf{q})
V_{t}(i\Omega,\mathbf{q}), \label{Eq:TWFermionSelfEnergyDefAppen}
\end{eqnarray}
where
\begin{eqnarray}
V_{t}(i\Omega,\mathbf{q}) &=& \frac{1}{V_{0}^{-1}(\mathbf{q}) +
\Pi_{t}(i\Omega,\mathbf{q})}
= \frac{1}{\frac{q_{\bot}^{2}+\zeta q_{z}^{2}}{4\pi\alpha v} +
\frac{3N q_{\bot}^{2}}{4\pi^2v}
\ln\left(\frac{B^{1/3}\Lambda_{UV}}{\left(\Omega^2 +
B^{2}q_{\bot}^{6}\right)^{1/6}} + 1\right) +
c_{t}\frac{N}{B^{2/3}}\frac{vq_{z}^{2}}{\left(\Omega^2 +
v^2q_{z}^{2}\right)^{2/3}}}.
\end{eqnarray}
After substituting Eq.~(\ref{Eq:FreeFermonPropagatorTripleWeylDef})
into Eq.~(\ref{Eq:TWFermionSelfEnergyDefAppen}), and expanding
$\Sigma_{t}(\omega,\mathbf{k})$ in powers of small $i\omega$,
$k_{x}$, $k_{y}$, and $k_{z}$ up to the leading order, we get
\begin{eqnarray}
\Sigma_{t}(i\omega,\mathbf{k}) \approx i\omega\Sigma_{t1} -
B\left[g_{1}(\mathbf{k})\sigma_{x} + g_{2}(\mathbf{k})
\sigma_{y}\right]\Sigma_{t2} - vk_{z}\sigma_{z}\Sigma_{t3},
\end{eqnarray}
where
\begin{eqnarray}
\Sigma_{t1}&=&\frac{1}{8\pi^{3}}\int'd\Omega dq_{\bot}q_{\bot}dq_{z}
\frac{\Omega^{2}-B^{2}q_{\bot}^{6} - v^2q_{z}^{2}}{\left(\Omega^{2}
+ B^{2}q_{\bot}^{6} + v^2q_{z}^{2}\right)^{2}}
V_{t}(i\Omega,\mathbf{q}),
\label{Eq:TWE1Appen} \\
\Sigma_{t2}&=&\frac{1}{8\pi^{3}}\int'd\Omega dq_{\bot}q_{\bot}dq_{z}
\left[\frac{\Omega^{2}-18B^{2}q_{\bot}^{6}+v^2q_{z}^{2}}{\left(\Omega^{2}
+ B^{2}q_{\bot}^{6}+v^2q_{z}^{2}\right)^{2}} + \frac{45B^{4}
q_{\bot}^{12}}{\left(\Omega^{2} + B^{2}q_{\bot}^{6} + v^2q_{z}^{2}
\right)^{3}}-\frac{27B^{6}q_{\bot}^{18}}{\left(\Omega^{2} +
B^{2}q_{\bot}^{6}+v^2q_{z}^{2}\right)^{4}}\right]\nonumber
\\
&&\times V_{t}(i\Omega,\mathbf{q}),\label{Eq:TWE2Appen} \\
\Sigma_{t3}&=&\frac{1}{8\pi^{3}}\int'd\Omega dq_{\bot}q_{\bot}dq_{z}
\frac{\Omega^{2}+B^{2}q_{\bot}^{6}-v^2q_{z}^{2}}
{\left(\Omega^{2}+B^{2}q_{\bot}^{6}+v^2q_{z}^{2}\right)^{2}}
V_{t}(i\Omega,\mathbf{q}). \label{Eq:TWE3Appen}
\end{eqnarray}
To make RG analysis, we consider the following range in energy:
\begin{eqnarray}
-\infty < \Omega < \infty, \qquad b\Lambda < E_{t} < \Lambda,
\label{Eq:TWRGSchemeAAppen}
\end{eqnarray}
with $E_{t} = \sqrt{B^2q_{\bot}^{6} + v^2q_{z}^{2}}$. Making use of
the definition
\begin{eqnarray}
E = \sqrt{B^2q_{\bot}^{6}+v^{2}q_{z}^{2}},\quad \kappa =
\frac{Bq_{\bot}^{3}}{v\left|q_{z}\right|},
\label{Eq:TWIntegalTransformA}
\end{eqnarray}
we re-express $q_{\bot}$ and $q_{z}$ as
\begin{eqnarray}
q_{\bot} = \frac{\kappa^{1/3}
E^{1/3}}{B^{1/3} \left(1 +
\kappa^{2}\right)^{1/6}},\quad |q_{z}| =
\frac{E}{v\sqrt{1+\kappa^{2}}}.\label{Eq:TWIntegalTransformB}
\end{eqnarray}
Now the integration over $q_{\bot}$ and $q_{z}$ can be transformed
to the integration over $E$ and $\kappa$ through the relation
\begin{eqnarray}
dq_{\bot}d|q_{z}| = \left|\left|
\begin{array}{cc}
\frac{\partial q_{\bot}}{\partial E} & \frac{\partial
q_{\bot}}{\partial \kappa}
\\
\frac{\partial |q_{z}|}{\partial E} & \frac{\partial
|q_{z}|}{\partial \kappa}
\end{array}
\right|\right|dEd\kappa = \frac{E^{1/3}}{3vB^{1/3}
\kappa^{2/3}\left(1+\kappa^{2}\right)^{2/3}}dE d\kappa.
\label{Eq:TWIntegralTranformMeasure}
\end{eqnarray}
By virtue of Eqs.~(\ref{Eq:TWIntegalTransformB}) and
(\ref{Eq:TWIntegralTranformMeasure}), we calculate
Eqs.~(\ref{Eq:TWE1Appen})-(\ref{Eq:TWE3Appen}) and obtain
\begin{eqnarray}
\Sigma_{t1} = C_{t1}\ell,\quad \Sigma_{t2} =
C_{t2}\ell, \quad \Sigma_{t3} = C_{t3}\ell,
\end{eqnarray}
where
\begin{eqnarray}
C_{t1}&=&\frac{1}{12\pi^{3}}\int_{-\infty}^{+\infty}dx
\int_{0}^{+\infty}d\kappa \frac{1}{\kappa^{1/3}
\left(1+\kappa^{2}\right)^{1/2}} \frac{x^{2}-1}{\left(x^{2} +
1\right)^{2}} \mathcal{G}_{t}(x,\kappa), \label{Eq:ExpressionCt1}
\\
C_{t2} &=& \frac{1}{12\pi^{3}}\int_{-\infty}^{+\infty}dx
\int_{0}^{+\infty}d\kappa \frac{1}{\kappa^{1/3}
\left(1+\kappa^{2}\right)^{3/2}} \left[\frac{x^{2}
\left(1+\kappa^{2}\right) - 18\kappa^{2}+1}{\left(x^{2} +
1\right)^{2}} + \frac{45\kappa^{4}}{\left(x^{2} + 1
\right)^{3}\left(1+\kappa^{2}\right)}- \frac{27
\kappa^{6}}{\left(x^{2}+1\right)^{4}
\left(1+\kappa^{2}\right)^{2}}\right]\nonumber
\\
&&\times\mathcal{G}_{t}(x,\kappa),
\label{Eq:ExpressionCt2}
\\
C_{t3}&=&\frac{1}{12\pi^{3}}\int_{-\infty}^{+\infty}dx
\int_{0}^{+\infty}d\kappa \frac{1}{\kappa^{1/3}\left(1 +
\kappa^{2}\right)^{3/2}} \frac{x^{2}\left(1 +
\kappa^{2}\right)+\kappa^{2}-1}{\left(x^{2} +
1\right)^{2}}\mathcal{G}_{t}(x,\kappa), \label{Eq:ExpressionCt3}
\end{eqnarray}
with
\begin{eqnarray}
\mathcal{G}_{t}^{-1}(x,\kappa) = \frac{1}{4\pi\alpha}
\left(\kappa^{2/3} + \frac{\beta_{t}}{\left(1 +
\kappa^{2}\right)^{2/3}}\right) +
N\left[\frac{3\kappa^{2/3}}{4\pi^2}
\ln\left(\frac{\gamma_{t}e^{\frac{\ell}{3}}
\left(1+\kappa^{2}\right)^{1/6}}{\left(x^{2}
\left(1+\kappa^{2}\right)+\kappa^{2}\right)^{1/6}}+1\right) +
c_{t}\frac{1}{\left(x^{2}\left(1 + \kappa^{2}\right) +
1\right)^{2/3}}\right]. \label{Eq:ExpressionGt}
\end{eqnarray}
Here, $\beta_{t}$ and $\gamma_{t}$ are defined by $\beta_{t} =
\frac{\zeta B^{2/3}\Lambda^{4/3}}{v^{2}}$ and $\gamma_{t}
= \frac{B^{1/3} \Lambda_{UV}}{\Lambda^{1/3}}$,
respectively.
\end{widetext}

\section{Deriving RG equations \label{App:DerivationRGEquation}}

We derive the coupled RG equations for double- and triple-WSMs in
order.

\subsection{Double-WSM \label{App:DerivationRGEquationDW}}

We rewrite the free action of double-Weyl fermions as
\begin{eqnarray}
S_{\psi_{d}}^{0} &=& \int\frac{d\omega}{2\pi}
\frac{d^{3}\mathbf{k}}{(2\pi)^{3}}\psi_{d}^{\dag}(\omega,\mathbf{k})
\left[i\omega-A\left(d_{1}(\mathbf{k})\sigma_{x} \right.\right.\nonumber
\\
&&\left.\left.+ d_{2}(\mathbf{k})
\sigma_{y}\right)-vk_{z}\sigma_{z}\right]\psi_{d}(\omega,\mathbf{k}).
\end{eqnarray}
Including the fermion self-energy induced by the Coulomb
interaction, this action becomes
\begin{eqnarray}
S_{\psi_{d}} &=& \int\frac{d\omega}{2\pi} \frac{d^{3}
\mathbf{k}}{(2\pi)^{3}}\psi_{d}^{\dag}(\omega,\mathbf{k})
\left[i\omega - A\left(d_{1}(\mathbf{k})\sigma_{x}\right)
\right.\nonumber
\\
&&\left.\left.+d_{2}(\mathbf{k})\sigma_{y}\right) - vk_{z}
\sigma_{z} + \Sigma_{d}(i\omega,\mathbf{k})\right]
\psi_{d}(\omega,\mathbf{k}) \nonumber
\\
&\approx& \int\frac{d\omega}{2\pi}\frac{d^{3}
\mathbf{k}}{(2\pi)^{3}} \psi_{d}^{\dag}(\omega,\mathbf{k})
\left[i\omega e^{C_{d1}\ell} - A\left(d_{1}(\mathbf{k})\sigma_{x}
\right.\right.\nonumber \\
&&\left.\left.+d_{2}(\mathbf{k})\sigma_{y}\right)
e^{C_{d2}\ell} - vk_{z}\sigma_{z}e^{C_{d3}\ell}\right]
\psi_{d}(\omega,\mathbf{k}).
\end{eqnarray}
We then make the following scaling transformations
\begin{eqnarray}
k_{x} &=& k_{x}'e^{-\frac{\ell}{2}},\label{Eq:DWRGTransformation1}
\\
k_{y} &=& k_{y}'e^{-\frac{\ell}{2}},\label{Eq:DWRGTransformation2}
\\
k_{z} &=& k_{z}'e^{-\ell},\label{Eq:DWRGTransformation3}
\\
\omega &=& \omega'e^{-\ell},\label{Eq:DWRGTransformation4}
\\
\psi_{d} &=& \psi_{d}'e^{\left(2-\frac{C_{d1}}{2}\right)\ell},
\label{Eq:DWRGTransformation5}
\\
A &=& A'e^{(C_{d1}-C_{d2})\ell},\label{Eq:DWRGTransformation6}
\\
v &=& v'e^{(C_{d1}-C_{d3})\ell},\label{Eq:DWRGTransformation7}
\end{eqnarray}
which leads to
\begin{eqnarray}
S_{\psi_{d}'} &=& \int\frac{d\omega'}{2\pi} \frac{d^{3}
\mathbf{k}'}{(2\pi)^3}\psi_{d}'^{\dag}(\omega',\mathbf{k'})
\left[i\omega'-A'\left(d_{1}(\mathbf{k}')\sigma_{x}\right.\right.\nonumber
\\
&&\left.\left.+d_{2}(\mathbf{k}')\sigma_{y}\right) -
v'k_{z}'\sigma_{z}\right]\psi_{d}'(\omega',\mathbf{k}').
\end{eqnarray}
This action has the same form as the free action, which will be used
to derive the flow equations.

From Eq.~(\ref{Eq:DWRGTransformation5}), the flow equation for the
residue $Z_f$ is
\begin{eqnarray}
\frac{dZ_{f}}{d\ell}=-C_{d1}Z_{f}. \label{Eq:RGDWZfApp}
\end{eqnarray}
According to Eqs.~(\ref{Eq:DWRGTransformation6}) and
(\ref{Eq:DWRGTransformation7}), we find that the flow equations of
$A$ and $v$ are
\begin{eqnarray}
\frac{dA}{d\ell} &=& \left(C_{d2}-C_{d1}\right)A, \label{Eq:RGDWAApp}
\\
\frac{dv}{d\ell} &=& \left(C_{d3}-C_{d1}\right)v. \label{Eq:RGDWvApp}
\end{eqnarray}
The flow equations of other parameters $\alpha$, $\beta_{d}$, and
$\gamma_{d}$ are
\begin{eqnarray}
\frac{d\alpha}{d\ell} &=& \left(C_{d1} - C_{d3}\right)\alpha,
\label{Eq:RGDWAlphaApp} \\
\frac{d\beta_{d}}{d\ell} &=& \left(C_{d1}+C_{d2}-2C_{d3} -
1\right)\beta_{d}, \label{Eq:RGDWBetaApp} \\
\frac{d\gamma_{d}}{d\ell} &=& \frac{1}{2}(C_{d2}-C_{d1})\gamma_{d}.
\label{Eq:RGDWGammaApp}
\end{eqnarray}

\subsection{Triple-WSM \label{App:DerivationRGEquationTW}}

By repeating the same computational procedure employed in the case
of double-WSM, we add the self-energy of triple-Weyl fermions to the
free action and then obtain
\begin{eqnarray}
S_{\psi_{t}} &=& \int\frac{d\omega}{2\pi} \frac{d^3
\mathbf{k}}{(2\pi)^{3}}\psi_{t}^{\dag}(\omega,\mathbf{k})
\left[i\omega e^{C_{t1}\ell} - B\left(g_{1}(\mathbf{k})
\sigma_{x}\right.\right.\nonumber
\\
&& +\left.\left.g_{2}(\mathbf{k})\sigma_{y}\right) e^{C_{t2}\ell} -
v k_{z}\sigma_{z}e^{C_{t3}\ell}\right]
\psi_{t}(\omega,\mathbf{k}).\nonumber \\
\end{eqnarray}
Making use of the scaling transformations
\begin{eqnarray}
k_{x}&=&k_{x}'e^{-\frac{\ell}{3}},\label{Eq:TWRGTransformation1}
\\
k_{y}&=&k_{y}'e^{-\frac{\ell}{3}},\label{Eq:TWRGTransformation2}
\\
k_{z}&=&k_{z}'e^{-\ell},\label{Eq:TWRGTransformation3}
\\
\omega&=&\omega'e^{-\ell},\label{Eq:TWRGTransformation4}
\\
\psi_{t}&=&\psi_{t}'e^{\left(\frac{11}{6}-\frac{C_{t1}}{2}\right)\ell},
\label{Eq:TWRGTransformation4} \\
B&=&B'e^{(C_{t1}-C_{t2})\ell},\label{Eq:TWRGTransformation5} \\
v&=&v'e^{(C_{t1}-C_{t3})\ell},\label{Eq:TWRGTransformation6}
\end{eqnarray}
the above action is converted to
\begin{eqnarray}
S_{\psi_{t}'} &=& \int\frac{d\omega'}{2\pi} \frac{d^{3}
\mathbf{k}'}{(2\pi)^{3}}
\psi_{t}'^{\dag}(\omega',\mathbf{k'})\left[i\omega' -
B'\left(g_{1}(\mathbf{k}')\sigma_{x}\right.\right.\nonumber
\\
&&\left.\left.+g_{2}(\mathbf{k}')\sigma_{y}\right) -
v'k_{z}'\sigma_{z}\right]\psi'_{t}(\omega',\mathbf{k}'),
\end{eqnarray}
which recovers the same form as the free action. From the
transformations
(\ref{Eq:TWRGTransformation4})-(\ref{Eq:TWRGTransformation6}), we
get the flow equations for $Z_{f}$, $B$, and $v$
\begin{eqnarray}
\frac{dZ_{f}}{d\ell}&=&-C_{t1}Z_{f},\label{Eq:RGTWZfApp}
\\
\frac{dB}{d\ell}&=&(C_{t2}-C_{t1})B,\label{Eq:RGTWBApp}
\\
\frac{dv}{d\ell}&=&(C_{t3}-C_{t1})v.\label{Eq:RGTWVfApp}
\end{eqnarray}
The flow equations of $\alpha$, $\beta_{t}$, and $\gamma_{t}$ are
given by
\begin{eqnarray}
\frac{d\alpha}{d\ell}&=&\left(C_{t1}-C_{t3}\right)\alpha,
\label{Eq:RGTWAlphaApp}
\\
\frac{d\beta_{t}}{d\ell} &=& \left(\frac{4}{3}C_{t1} +
\frac{2}{3}C_{t2}-2C_{t3}-\frac{4}{3}\right)\beta_{t},
\label{Eq:RGTWBetaApp}
\\
\frac{d\gamma_{t}}{d\ell} &=& \frac{1}{3}(C_{t2} -
C_{t1})\gamma_{t}.\label{Eq:RGTWGammaApp}
\end{eqnarray}

\section{Impact of Finite chemical potential \label{App:FiniteMu}}

As discussed in the main text, the unconventional non-FL state can
by experimentally explored by measuring the fermion damping rate and
spectral function in several candidate materials for double- and
triple-WSMs. To observe the predicted non-FL behavior, the sample
needs to be carefully prepared. In particular, the chemical
potential $\mu$ should be made sufficiently small, because the
signature is sharpest at $\mu = 0$. At finite $\mu$, the fermion DOS
takes a finite value, and as such leads to static screening of
long-range Coulomb interaction. We now make a brief remark on the
impact of finite $\mu$ on the unconventional non-FL behavior.

For a double-WSM prepared at finite $\mu$, the Matsubara
fermion propagator becomes
\begin{eqnarray}
G_{d0}(i\omega_{n},\mathbf{k}) = \frac{1}{i\omega_{n} + \mu -
\mathcal{H}_{d}(\mathbf{k})}.
\end{eqnarray}
where $\mathcal{H}_{d}(\mathbf{k}) = Ad_{1}(\mathbf{k})\sigma_{x} +
Ad_{2}(\mathbf{k})\sigma_{y}+ vk_{z}\sigma_{z}$. The retarded
fermion propagator has the form
\begin{eqnarray}
G_{d0}^{\mathrm{ret}}(\omega,\mathbf{k}) = \frac{1}{\omega +
\mu-\mathcal{H}_{d}(\mathbf{k}) + i\delta},
\end{eqnarray}
which gives rise to the following spectral function
\begin{eqnarray}
\mathcal{A}_{d}(\omega,\mathbf{k}) &=& -\frac{1}{\pi} \mathrm{Tr}
\left[\mathrm{Im}\left[G_{d0}^{\mathrm{ret}}(\omega,
\mathbf{k})\right]\right]\nonumber \\
&=& 2|\omega+\mu|\delta\left(\left(\omega + \mu\right)^{2} -
E_{d}^{2}(\mathbf{k})\right),
\end{eqnarray}
where $E_{d}(\mathbf{k})=\sqrt{A^{2}k_{\bot}^{4}+v^{2}k_{z}^{2}}$.
The fermion DOS is given by
\begin{eqnarray}
\rho_{d}(\omega) = N\int\frac{d^3\mathbf{k}}{(2\pi)^{3}}
\mathcal{A}_{d}(\omega,\mathbf{k}) = \frac{N}{8\pi vA}|\omega+\mu|.
\end{eqnarray}
Here, we have carried out the transformations shown in
Eqs.~(\ref{Eq:DWIntegalTransformA})-(\ref{Eq:DWIntegralTranformMeasure}).
In the limit $\omega\rightarrow 0$, we have
\begin{eqnarray}
\rho_{d}(0) = \frac{N|\mu|}{8\pi vA}.
\end{eqnarray}
In the limits of $\Omega = 0$ and $\mathbf{q} = 0$, the polarization
function behaves as
\begin{eqnarray}
\Pi_{d}(0,0) = N\int\frac{d^3\mathbf{k}}{(2\pi)^{3}}
\frac{1}{E_{d}(\mathbf{k})}\theta\left(|\mu|-E_{d}(\mathbf{k})\right).
\end{eqnarray}
Substituting $E_{d}(\mathbf{k})$ into this formula and employing the
transformations
(\ref{Eq:DWIntegalTransformA})-(\ref{Eq:DWIntegralTranformMeasure}),
we obtain
\begin{eqnarray}
\Pi_{d}(0,0) = \frac{N|\mu|}{8\pi vA} = \rho_{d}(0).
\end{eqnarray}

In the case of triple-WSM, one can similarly get the
retarded fermion propagator
\begin{eqnarray}
G_{t0}^{\mathrm{ret}}(\omega,\mathbf{k}) =
\frac{1}{\omega+\mu-\mathcal{H}_{t}(\mathbf{k})+i\delta}.
\end{eqnarray}
where $\mathcal{H}_{t}(\mathbf{k}) = Bg_{1}(\mathbf{k})\sigma_{x} +
Bg_{2}(\mathbf{k})\sigma_{y}+vk_{z}\sigma_{z}$. The spectral
function is
\begin{eqnarray}
\mathcal{A}_{t}(\omega,\mathbf{k}) &=& -\frac{1}{\pi}\mathrm{Tr}
\left[\mathrm{Im} \left[G_{t0}^{\mathrm{ret}}(\omega,\mathbf{k})
\right]\right]\nonumber
\\
&=& 2|\omega+\mu|\delta\left(\left(\omega+\mu\right)^{2} -
E_{t}^{2}(\mathbf{k})\right),
\end{eqnarray}
where $E_{t}(\mathbf{k})=\sqrt{B^{2}k_{\bot}^{6} + v^{2}k_{z}^{2}}$.
The fermion DOS is
\begin{eqnarray}
\rho_{t}(\omega)&=& N\int\frac{d^3\mathbf{k}}{(2\pi)^{3}}
\mathcal{A}_{t}(\omega,\mathbf{k})\nonumber
\\
&=& \frac{N\Gamma\left(1/3\right) |\omega + \mu|^{2/3}}{12\pi^{3/2}
\Gamma\left(5/6\right)vB^{2/3}},
\end{eqnarray}
which reduces to
\begin{eqnarray}
\rho_{t}(0) = \frac{N\Gamma\left(1/3\right)
|\mu|^{2/3}}{12\pi^{3/2}
\Gamma\left(5/6\right)vB^{2/3}}
\end{eqnarray}
in the lowest energy limit. Therefore, we have
\begin{eqnarray}
\Pi_{t}(0,0) &=& N\int\frac{d^3\mathbf{k}}{(2\pi)^{3}}
\frac{1}{E_{t}(\mathbf{k})}\theta\left(|\mu| -
E_{t}(\mathbf{k})\right)\nonumber \\
&=& \frac{3}{2}\rho_{t}(0).
\end{eqnarray}

From the above results, we know that the polarizations $\Pi_{d,t}$
always take certain finite value in the zero-energy
(long-wavelength) limit. Consequently, the Coulomb interaction,
described by the dressed function
\begin{eqnarray}
V_{d,t}(i\Omega,\mathbf{q}) = \frac{1}{V_{0}^{-1}(\mathbf{q}) +
\Pi_{d,t}(i\Omega,\mathbf{q})},
\end{eqnarray}
is statically screened and becomes short-ranged. Here, we provide a
qualitative analysis for the behavior of fermion self-energy
$\Sigma_{d,t}(i\omega,\mathbf{k})$ at $\mu \neq 0$.

At energies below the scale set by $\mu$, the screened Coulomb
interaction is relatively unimportant and only produces ordinary FL
behavior. In contrast, at energies beyond the scale of $\mu$ the
static screening effect is unimportant, and the Coulomb interaction
still induces unconventional non-FL behavior. Thus, increasing the
energy scale drives a crossover from the FL regime to the non-FL
regime. If one varies the temperature $T$, there is an analogous
crossover between the FL and non-FL regimes: the system exhibits FL
behavior at $kT<\mu$ and non-FL behavior $kT>\mu$. We notice that
the crossover from a usual FL state to a singular FL state, in which
$Z_{f}$ approaches to a finite value in the lowest energy limit but
the fermion velocity receives singular renormalization, has been
studied in DSMs at finite chemical potential \cite{Sheehy07,
Setiawan15}. In the double- and triple-WSMs considered in this
paper, the unconventional non-FL state always has observable effects
as long as the chemical potential is not large enough, as explained
in Sec.~\ref{Sec:ExpDetection} in more detail.

\section{Observable quantities for free fermions  \label{App:ObservableQuantitiesFree}}

We now calculate the specific heat and dynamical conductivities for
free double- and triple-Weyl fermions. The interaction induced
corrections will be included later.

\subsection{Specific heat}

\subsubsection{Double-WSM}

In the Matsubara formalism, the propagator of free double-Weyl
fermions reads
\begin{eqnarray}
G_{d0}(i\omega_{n},\mathbf{k}) = \frac{-i\omega_{n} +
\mathcal{H}_{d}(\mathbf{k})}{\omega_{n}^{2}+E_{d}^{2}(\mathbf{k})},
\end{eqnarray}
where $\omega_{n}=(2n+1)\pi T$ with $n$ being integers. The
corresponding free energy is given by
\begin{eqnarray}
F_{f}^{d}(T) = -2NT\sum_{\omega_{n}}\int
\frac{d^3\mathbf{k}}{(2\pi)^3} \ln\left[\left(\omega_{n}^2 +
E_{d}^{2}(\mathbf{k})\right)^{\frac{1}{2}}\right].
\end{eqnarray}
Carrying out frequency summation, one obtains
\begin{eqnarray}
F_{f}^{d}(T) &=& -2N\int\frac{d^3\mathbf{k}}{(2\pi)^3}
\Big[E_{d}(\mathbf{k}) + 2T\nonumber
\\
&&\times\ln\left(1 +
e^{-\frac{E_{d}(\mathbf{k})}{T}}\right)\Big],
\end{eqnarray}
which is divergent due to the first term in the bracket. To
regularize this divergence, we re-define $F_{f}^{d}(T) - F_{f}^{d}(0)$ as
$F_{f}^{d}(T)$ and then get
\begin{eqnarray}
F_{f}^{d}(T) &=& -4NT\int\frac{d^3\mathbf{k}}{(2\pi)^3}
\ln\left(1+e^{-\frac{E_{d(\mathbf{k})}}{T}}\right)\nonumber
\\
&=& -\frac{2NT}{\pi^{2}}\int dk_{\bot} d|k_{z}| k_{\bot}
\ln\left(1+e^{-\frac{E_{d}(\mathbf{k})}{T}}\right).
\end{eqnarray}
Utilizing the integration transformations shown in
Eqs.~(\ref{Eq:DWIntegalTransformA})
-(\ref{Eq:DWIntegralTranformMeasure}), the free energy becomes
\begin{eqnarray}
F_{f}^{d}(T) &=& -\frac{NT}{\pi^{2}vA}\int_{0}^{+\infty} dEE
\ln\left(1+e^{-\frac{E}{T}}\right) \nonumber \\
&&\times \int_{0}^{+\infty}d\kappa \frac{1}{\left(1+\kappa^2\right)}
\nonumber \\
&=& -\frac{3\zeta(3)N}{8\pi vA}T^{3}.
\end{eqnarray}
Then it is easy to get the specific heat
\begin{eqnarray}
C_{v}^{d}(T) &=& -T\frac{\partial^2 F_{f}^{d}(T)}{\partial T^2}
\nonumber \\
&=& \frac{9\zeta(3)N}{4\pi vA}T^{2}. \label{Eq:SpecificHeatDef}
\end{eqnarray}

\subsubsection{Triple-WSM}

For triple-WSM, the free energy can be expressed as
\begin{eqnarray}
F_{f}^{t}(T) = -\frac{2NT}{\pi^{2}}\int dk_{\bot}d|k_{z}|k_{\bot}
\ln\left(1+e^{-\frac{E_{t}(\mathbf{k})}{T}}\right).
\end{eqnarray}
Carrying out the transformations of
Eqs.~(\ref{Eq:TWIntegalTransformA})-(\ref{Eq:TWIntegralTranformMeasure}),
we re-write $F_{f}^{t}(T)$ in the form
\begin{eqnarray}
F_{f}^{t}(T) &=& -\frac{2NT}{3\pi^{2}v B^{2/3}} \int_{0}^{+\infty}
dE E^{2/3} \ln\left(1 + e^{-\frac{E}{T}}\right)\nonumber \\
&&\times\int_{0}^{+\infty} d\kappa \frac{1}{\kappa^{1/3}
\left(1+\kappa^{2}\right)^{5/6}}\nonumber \\
&=&-\frac{c_{t}N}{\pi^{3/2}v B^{2/3}}T^{8/3},
\end{eqnarray}
where the constant $a_t$ is
\begin{eqnarray}
a_{t} = \frac{1}{18}\left(4-2^{1/3}\right) \zeta\left(8/3\right)
\frac{\Gamma\left(1/3\right)\Gamma
\left(2/3\right)}{\Gamma\left(5/6\right)}.
\end{eqnarray}
The specific heat is
\begin{eqnarray}
C_{v}^{t}(T) &=& -T\frac{\partial^2 F_{f}^{t}(T)}{\partial T^2}
\nonumber \\
&=& \frac{40a_{t}N}{9\pi^{3/2}vB^{2/3}}T^{5/3}.
\end{eqnarray}

\subsection{Dynamical Conductivities}

The energy-dependence dynamical conductivities will be computed by
using the Kubo formula.

\subsubsection{Double-WSM \label{App:DWSMDynamicalConduct}}

In the Matsubara formalism, the current-current correlation function
for double-Weyl fermions can be written as
\begin{eqnarray}
\Pi_{ij}^{d}(i\Omega_{m}) &=& -e^{2}T\sum_{\omega_{n}}\int
\frac{d^3\mathbf{k}}{(2\pi)^{3}}\mathrm{Tr}\left[\gamma_{i}^{d}(\mathbf{k})
G_{d0}(i\omega_{n},\mathbf{k})\right.\nonumber
\\
&&\left.\times\gamma_{j}^{d}(\mathbf{k})
G_{d0}(i\omega_{n}+i\Omega_{m},\mathbf{k})\right].
\end{eqnarray}
Here, $\gamma_{i}^{d}$ is given by
\begin{eqnarray}
\gamma_{i}^{d} = \frac{\partial \mathcal{H}_{d}}{\partial k_{i}},
\end{eqnarray}
with $\mathcal{H}_{d}$ being the Hamiltonian density
\begin{eqnarray}
\mathcal{H}_{d} = A\left(k_{x}^{2}-k_{y}^{2}\right)\sigma_{x}
+2Ak_{x}k_{y}\sigma_{y}+vk_{z}\sigma_{z}.
\end{eqnarray}
It is easy to verify that
\begin{eqnarray}
\gamma_{x}^{d} &=& \frac{\partial \mathcal{H}_{d}}{\partial k_{x}} =
2Ak_{x}\sigma_{x}+2Ak_{y}\sigma_{y}, \label{Eq:gammaxDefDWSM}
\\
\gamma_{y}^{d}&=&\frac{\partial \mathcal{H}_{d}}{\partial k_{y}} = - 2Ak_{y}
\sigma_{x} + 2Ak_{x}\sigma_{y},
\label{Eq:gammayDefDWSM}
\\
\gamma_{z}^{d} &=& \frac{\partial \mathcal{H}_{d}}{\partial k_{z}} =
v\sigma_{z}. \label{Eq:gammazDefDWSM}
\end{eqnarray}
Symmetry consideration reveals that the following identity
\begin{equation}
\Pi_{xx}^{d} = \Pi_{yy}^{d}\equiv\Pi_{\bot\bot}^{d}
\end{equation}
is satisfied. Therefore, we only need to calculate $\Pi_{xx}^{d}$
and $\Pi_{zz}^{d}$, which are defined as follows
\begin{eqnarray}
\Pi_{xx}^{d}(i\Omega_{m}) &=& -e^{2}T\sum_{\omega_{n}}\int
\frac{d^3\mathbf{k}}{(2\pi)^{3}}\mathrm{Tr}\left[\gamma_{x}^{d}
G_{d0}(i\omega_{n},\mathbf{k})\gamma_{x}^{d}\right.\nonumber
\\
&&\left.\times G_{d0}(i\omega_{n}+i\Omega_{m},\mathbf{k})\right],
\label{Eq:PolaxxDefDWSM}
\\
\Pi_{zz}^{d}(i\Omega_{m})&=&-e^{2}T\sum_{\omega_{n}}\int
\frac{d^3\mathbf{k}}{(2\pi)^{3}}\mathrm{Tr}\left[\gamma_{z}^{d}
G_{d0}(i\omega_{n},\mathbf{k})\gamma_{z}^{d}\right.\nonumber
\\
&&\left.\times G_{d0}(i\omega_{n}+i\Omega_{m},\mathbf{k})\right].
\label{Eq:PolazzDefDWSM}
\end{eqnarray}
Employing the spectral representation
\begin{equation}
G_{d0}\left(i\omega_n,\mathbf{k}\right)=-\int_{-\infty}^{+\infty}
\frac{d\omega_1}{\pi}\frac{\mathrm{Im}\left[G_{d0}^{\mathrm{ret}}
\left(\omega_1,\mathbf{k}\right)\right]}{i\omega_n-\omega_1},
\end{equation}
we re-write $\Pi_{xx}^{d}$ and $\Pi_{zz}^{d}$ as follows
\begin{widetext}
\begin{eqnarray}
\Pi_{xx}^{d}(i\Omega_{m}) &=& -4A^{2}e^{2}\int
\frac{d^3\mathbf{k}}{(2\pi)^{3}}\int_{-\infty}^{+\infty}
\frac{d\omega_1}{\pi}\int_{-\infty}^{+\infty}\frac{d\omega_2}{\pi}
\Big\{k_{x}^{2} \mathrm{Tr}
\left[\sigma_{x}\mathrm{Im}\left[G_{d0}^{\mathrm{ret}}
\left(\omega_1,\mathbf{k}\right)\right]\sigma_{x}\mathrm{Im}
\left[G_{d0}^{\mathrm{ret}}
\left(\omega_2,\mathbf{k}\right)\right]\right]\nonumber
\\
&&+k_{y}^{2}\mathrm{Tr}\left[\sigma_{y} \mathrm{Im}
\left[G_{d0}^{\mathrm{ret}}\left(\omega_1,\mathbf{k}\right)\right]
\sigma_{y} \mathrm{Im}\left[G_{d0}^{\mathrm{ret}}
\left(\omega_2,\mathbf{k}\right)\right]\right]\Big\}
\frac{n_F\left(\omega_1\right)-n_F\left(\omega_2\right)}{\omega_1 -
\omega_2+i\Omega_{m}}, \label{Eq:PolaxxMidADWSM}
\\
\Pi_{zz}^{d}(i\Omega_{m}) &=& -v^{2}e^{2}\int
\frac{d^3\mathbf{k}}{(2\pi)^{3}} \int_{-\infty}^{+\infty}
\frac{d\omega_1}{\pi}\int_{-\infty}^{+\infty} \frac{d\omega_2}{\pi}
\mathrm{Tr}\left[\sigma_{z}\mathrm{Im}\left[G_{d0}^{\mathrm{ret}}
\left(\omega_1,\mathbf{k}\right)\right]\sigma_{z}
\mathrm{Im}\left[G_{d0}^{\mathrm{ret}}
\left(\omega_2,\mathbf{k}\right)\right]\right]\nonumber
\\
&&\times \frac{n_F\left(\omega_1\right) -
n_F\left(\omega_2\right)}{\omega_1-\omega_2 + i\Omega_{m}},
\label{Eq:PolazzMidADWSM}
\end{eqnarray}
where $n_{F}(x)=\frac{1}{e^{x/T}+1}$. The expression of
$\mathrm{Im}\left[G_{d0}^{\mathrm{ret}}(\omega,\mathbf{k})\right]$
is given by
\begin{eqnarray}
\mathrm{Im}\left[G_{d0}^{\mathrm{ret}}(\omega,\mathbf{k})\right] &=&
-\pi\mathrm{sgn}(\omega)\left(\omega+\mathcal{H}_{d}\right)
\frac{1}{2E_{d}(\mathbf{k})}\left[\delta\left(\omega+E_{d}(\mathbf{k})\right)
+\delta\left(\omega-E_{d}(\mathbf{k})\right)\right].
\end{eqnarray}
We then carry out analytical continuation $i\Omega_{m}\rightarrow
\Omega + i\delta$, and get the imaginary parts:
\begin{eqnarray}
\mathrm{Im}\left[\Pi_{xx}^{d,\mathrm{ret}}(\Omega,T)\right] &=&
4A^{2}e^{2}\int\frac{d^3\mathbf{k}}{(2\pi)^{3}}\int_{-\infty}^{+\infty}
\frac{d\omega_1}{\pi}\Big\{k_{x}^{2}
\mathrm{Tr}\left[\sigma_{x}\mathrm{Im}\left[G_{d0}^{\mathrm{ret}}
\left(\omega_1,\mathbf{k}\right)\right]\sigma_{x}
\mathrm{Im}\left[G_{d0}^{\mathrm{ret}} \left(\omega_1 +
\Omega,\mathbf{k}\right)\right]\right] \nonumber \\
&&+k_{y}^{2}\mathrm{Tr}\left[\sigma_{y} \mathrm{Im}
\left[G_{d0}^{\mathrm{ret}} \left(\omega_1,\mathbf{k}\right)\right]
\sigma_{y} \mathrm{Im}\left[G_{d0}^{\mathrm{ret}} \left(\omega_1 +
\Omega,\mathbf{k}\right)\right]\right]\Big\}
\left[n_F\left(\omega_1\right)-n_F\left(\omega_1+\Omega\right)\right],
\label{Eq:PolaxxMidBDWSM}
\\
\mathrm{Im}\left[\Pi_{zz}^{t,\mathrm{ret}}(\Omega,T)\right] &=&
v^{2}e^{2}\int \frac{d^3\mathbf{k}}{(2\pi)^{3}}
\int_{-\infty}^{+\infty} \frac{d\omega_1}{\pi}
\mathrm{Tr}\left[\sigma_{z}\mathrm{Im} \left[G_{d0}^{\mathrm{ret}}
\left(\omega_1,\mathbf{k}\right)\right] \sigma_{z}
\mathrm{Im}\left[G_{d0}^{\mathrm{ret}}\left(\omega_1 +
\Omega,\mathbf{k}\right)\right]\right]\nonumber
\\
&&\times\left[n_F\left(\omega_1\right) - n_F
\left(\omega_1+\Omega\right)\right]. \label{Eq:PolazzMidBDWSM}
\end{eqnarray}
\end{widetext}
The formula $\frac{1}{x+i\delta} =
\mathcal{P}\frac{1}{x}-i\pi\delta(x)$, $\mathcal{P}$ stands for
principal value, has been used in the above computation. According
to Kubo formula, the dynamical conductivities are defined as
\begin{eqnarray}
\sigma_{xx}^{d}(\Omega,T)&=&\frac{\mathrm{Im}
\left[\Pi_{xx}^{d,\mathrm{ret}}(\Omega,T)\right]}{\Omega},
\label{Eq:SigmaxxDefDWSM}
\\
\sigma_{zz}^{d}(\Omega,T)&=&\frac{\mathrm{Im}
\left[\Pi_{zz}^{d,\mathrm{ret}}(\Omega,T)\right]}{\Omega}.
\label{Eq:SigmazzDefDWSM}
\end{eqnarray}

After carrying out analytical calculations, we arrive at the
following compact expressions of conductivities
\begin{eqnarray}
\sigma_{xx}^{d}(\Omega,T) &=& c_{1}^{d}\frac{e^{2}}{v}
\delta\left(\Omega\right)T^{2} \nonumber \\
&& + c_{2}^{d}\frac{e^{2}}{v}|\Omega|
\tanh\left(\frac{|\Omega|}{4T}\right), \label{Eq:SigmaxxResultDWSM}
\\
\sigma_{zz}^{d}(\Omega,T) &=& c_{3}^{d}\frac{ve^{2}}{A} \delta
\left(\Omega\right)T \nonumber \\
&& + c_{4}^{d}\frac{ve^{2}}{A} \tanh
\left(\frac{|\Omega|}{4T}\right), \label{Eq:SigmazzResultDWSM}
\end{eqnarray}
where
\begin{eqnarray}
c_{1}^{d}&=&\frac{1}{6\pi }\int_{0}^{+\infty} dx
x^{2}\frac{1}{\sinh^{2}\left(\frac{x}{2}\right)}, \\
c_{2}^{d}&=&\frac{1}{12\pi}, \\
c_{3}^{d}&=&\frac{1}{32}\int_{0}^{+\infty} dx
x\frac{1}{\sinh^{2}\left(\frac{x}{2}\right)},
\\
c_{4}^{d}&=&\frac{1}{64}.
\end{eqnarray}
The first term in the right-hand side of
Eqs.~(\ref{Eq:SigmaxxResultDWSM}) and (\ref{Eq:SigmazzResultDWSM})
represents the Drude peak.

\subsubsection{Triple-WSM \label{App:TWSMDynamicalConduct}}

In the Matsubara formalism, the current-current correlation function
for triple-Weyl fermions is
\begin{eqnarray}
\Pi_{ij}^{t}(i\Omega_{m}) &=& -e^{2}T\sum_{\omega_{n}}\int \frac{d^3
\mathbf{k}}{(2\pi)^{3}}\mathrm{Tr}\left[\gamma_{i}^{t}(\mathbf{k})
G_{t0}(i\omega_{n},\mathbf{k})\right.\nonumber
\\
&&\left.\times\gamma_{j}^{t}(\mathbf{k})
G_{t0}(i(\omega_{n}+\Omega_{m}),\mathbf{k})\right],
\end{eqnarray}
where $\gamma_{i}^{t} = \frac{\partial\mathcal{H}_{t}}{\partial
k_{i}}$ with the Hamiltonian density being $\mathcal{H}_{t} =
B\left(k_{x}^{3}-3k_{x}k_{y}^{2}\right) \sigma_{x} +
B\left(k_{y}^{3}-3k_{y}k_{x}^{2}\right) \sigma_{y} +
vk_{z}\sigma_{z}$. It is easy to get
\begin{eqnarray}
\gamma_{x}^{t} &=& \frac{\partial \mathcal{H}_{t}}{\partial k_{x}} =
3B\left(k_{x}^{2} - k_{y}^{2}\right)\sigma_{x} -
6Bk_{x}k_{y}\sigma_{y}, \label{Eq:gammaxDefTWSM} \\
\gamma_{y}^{t}&=&\frac{\partial \mathcal{H}_{t}}{\partial k_{y}} = -
6Bk_{x}k_{y} \sigma_{x} + 3B\left(k_{y}^{2}-k_{x}^{2}\right)
\sigma_{y}, \label{Eq:gammayDefTWSM}
\\
\gamma_{z}^{t} &=& \frac{\partial \mathcal{H}_{t}}{\partial k_{z}} =
v\sigma_{z}. \label{Eq:gammazDefTWSM}
\end{eqnarray}
Due to the relation $\Pi_{xx}^{t} = \Pi_{yy}^{t}\equiv
\Pi_{\bot\bot}^{t}$, we only calculate $\Pi_{xx}^{t}$ and
$\Pi_{zz}^{t}$, which take the form
\begin{eqnarray}
\Pi_{xx}^{t}(i\Omega_{m}) &=& -e^{2}T\sum_{\omega_{n}}\int
\frac{d^3\mathbf{k}}{(2\pi)^{3}}\mathrm{Tr}\left[\gamma_{x}^{t}
G_{t0}(i\omega_{n},\mathbf{k})\gamma_{x}^{t}\right.\nonumber
\\
&&\left.\times G_{t0}(i\omega_{n}+i\Omega_{m},\mathbf{k})\right],
\label{Eq:PolaxxDefTWSM}
\\
\Pi_{zz}^{t}(i\Omega_{m})&=&-e^{2}T\sum_{\omega_{n}}\int
\frac{d^3\mathbf{k}}{(2\pi)^{3}}\mathrm{Tr}\left[\gamma_{z}^{t}
G_{t0}(i\omega_{n},\mathbf{k})\gamma_{z}^{t}\right.\nonumber
\\
&&\left.\times G_{t0}(i\omega_{n}+i\Omega_{m},\mathbf{k})\right].
\label{Eq:PolazzDefTWSM}
\end{eqnarray}
After performing a series of calculations, we find that the
imaginary parts of retarded correlation functions $\Pi_{xx}^{t}$ and
$\Pi_{zz}^{t}$ have the forms
\begin{widetext}
\begin{eqnarray}
\mathrm{Im}\left[\Pi_{xx}^{t,\mathrm{ret}}(\Omega, T)\right] &=&
9B^{2}e^{2}\int\frac{d^3\mathbf{k}}{(2\pi)^{3}}\int_{-\infty}^{+\infty}
\frac{d\omega_1}{\pi}\Big\{\left(k_{x}^{2}-k_{y}^{2}\right)^{2}
\mathrm{Tr}\left[\sigma_{x}\mathrm{Im}\left[G_{t0}^{\mathrm{ret}}
\left(\omega_1,\mathbf{k}\right)\right]\sigma_{x}
\mathrm{Im}\left[G_{t0}^{\mathrm{ret}} \left(\omega_1 +
\Omega,\mathbf{k}\right)\right]\right]\nonumber
\\
&&+4k_{x}^{2}k_{y}^{2}\mathrm{Tr}\left[\sigma_{y}
\mathrm{Im}\left[G_{t0}^{\mathrm{ret}}
\left(\omega_1,\mathbf{k}\right)\right]\sigma_{y}
\mathrm{Im}\left[G_{t0}^{\mathrm{ret}} \left(\omega_1 +
\Omega,\mathbf{k}\right)\right]\right]\Big\}
\left[n_F\left(\omega_1\right)-n_F\left(\omega_1+\Omega\right)\right],
\label{Eq:PolaxxMidBTWSM}
\\
\mathrm{Im}\left[\Pi_{zz}^{t,\mathrm{ret}}(\Omega,T)\right] &=&
v^{2}e^{2}\int\frac{d^3\mathbf{k}}{(2\pi)^{3}}
\int_{-\infty}^{+\infty} \frac{d\omega_1}{\pi}
\mathrm{Tr}\left[\sigma_{z}\mathrm{Im}\left[G_{t0}^{\mathrm{ret}}
\left( \omega_1,\mathbf{k}\right)\right]\sigma_{z}
\mathrm{Im}\left[G_{t0}^{\mathrm{ret}}\left(\omega_1 +
\Omega,\mathbf{k}\right)\right]\right]\nonumber
\\
&&\times\left[n_F\left(\omega_1\right) -
n_F\left(\omega_1+\Omega\right)\right]. \label{Eq:PolazzMidBTWSM}
\end{eqnarray}
The expression of
$\mathrm{Im}\left[G_{t0}^{\mathrm{ret}}(\omega,\mathbf{k})\right]$
reads as
\begin{eqnarray}
\mathrm{Im}\left[G_{t0}^{\mathrm{ret}}(\omega,\mathbf{k})\right] &=&
-\pi\mathrm{sgn}(\omega)\left(\omega+\mathcal{H}_{t}\right)
\frac{1}{2 E_{t}(\mathbf{k})}\left[\delta
\left(\omega+E_{t}(\mathbf{k})\right) + \delta
\left(\omega-E_{t}(\mathbf{k})\right)\right].
\label{Eq:ImPartPropagatorTWSM}
\end{eqnarray}
\end{widetext}
The conductivities are given by
\begin{eqnarray}
\sigma_{xx}^{t}(\Omega,T)&=&\frac{\mathrm{Im}
\left[\Pi_{xx}^{t,\mathrm{ret}}(\Omega,T)\right]}{\Omega},
\label{Eq:SigmaxxDefTWSM}
\\
\sigma_{zz}^{t}(\Omega,T)&=&\frac{\mathrm{Im}
\left[\Pi_{zz}^{t,\mathrm{ret}}(\Omega,T)\right]}{\Omega}.
\label{Eq:SigmazzDefTWSM}
\end{eqnarray}
Substituting Eq.~(\ref{Eq:ImPartPropagatorTWSM}) into
Eqs.~(\ref{Eq:PolaxxMidBTWSM}), (\ref{Eq:PolazzMidBTWSM}),
(\ref{Eq:SigmaxxDefTWSM}), and (\ref{Eq:SigmazzDefTWSM}) leads to
the following expressions
\begin{eqnarray}
\sigma_{xx}^{t}(\Omega,T) &=& c_{1}^{t}\frac{e^{2}}{
v}\delta\left(\Omega\right)T^{2} + c_{2}^{t}
\frac{e^{2}}{v}|\Omega|\nonumber \\
&&\times\tanh\left(\frac{|\Omega|}{4T}\right),
\label{Eq:SigmaxxResultTWSM} \\
\sigma_{zz}^{t}(\Omega,T) &=& c_{3}^{t}\frac{v e^{2}}{B^{2/3}}
\delta\left(\Omega\right)T^{\frac{2}{3}}+c_{4}^{t} \frac{v
e^{2}}{B^{2/3}}\frac{1}{\left|\Omega\right|^{1/3}} \nonumber
\\
&&\times\tanh\left(\frac{|\Omega|}{4T}\right),
\label{Eq:SigmazzResultTWSM}
\end{eqnarray}
where
\begin{eqnarray}
c_{1}^{t} &=& \frac{1}{4\pi}\int_{0}^{+\infty} dx
x^{2}\frac{1}{\sinh^{2}\left(\frac{x}{2}\right)}, \\
c_{2}^{t} &=& \frac{1}{8\pi}, \\
c_{3}^{t} &=& \frac{\Gamma\left(1/3\right)}{40\Gamma
\left(5/6\right)\sqrt{\pi}}\int_{0}^{+\infty} dx x^{2/3}
\frac{1}{\sinh^{2}\left(\frac{x}{2}\right)}, \\
c_{4}^{t} &=& \frac{2^{1/3}\Gamma\left(1/3\right)}{120\Gamma
\left(5/6\right)\sqrt{\pi}}.
\end{eqnarray}

\section{Interaction corrections to observable quantities \label{App:ObservableQuantitiesInteraction}}

Now we compute the interaction corrections to observable quantities
by using the RG solutions of model parameters.

\subsection{DOS}

\subsubsection{Double-WSM}

For free double-Weyl fermions, the DOS is
\begin{eqnarray}
\rho_{d}(\omega) \sim \frac{\omega}{v A}.
\end{eqnarray}
Upon including interaction corrections, the constants $A$ and $v$
become $\omega$-dependent. We derive the following RG equation for
$\rho_{d}(\omega)$
\begin{eqnarray}
\frac{d\ln(\rho_{d}(\omega))}{d\ln(\omega)}\sim 1+C_{d1} +
\frac{d\ln\left(\frac{1}{v A}\right)}{d\ln(\omega)}.
\label{Eq:RhoScalingOmegaDWSMA}
\end{eqnarray}
On the right-hand side, the second term comes from the anomalous
dimension of fermion field, and the third term is induced by the
fermion dispersion renormalization. Recall that $A$ and $v$ satisfy
the flow equations:
\begin{eqnarray}
\frac{dA}{d\ell} &=& \left(C_{d2}-C_{d1}\right)A,
\label{Eq:AVRGDWSMApp} \\
\frac{dv}{d\ell} &=& \left(C_{d3}-C_{d1}\right)v.
\label{Eq:VVRGDWSMApp}
\end{eqnarray}
Using the transformation $\omega = \omega_{0}e^{-\ell}$, where
$\omega_0$ is certain high-energy scale, the above equations are
converted to
\begin{eqnarray}
\frac{d\ln(A)}{d\ln(\omega)} &=& -\left(C_{d2}-C_{d1}\right),
\label{Eq:AFlowOmegaDWSM}
\\
\frac{d\ln(v)}{d\ln(\omega)} &=& -\left(C_{d3}-C_{d1}\right).
\label{Eq:VFlowOmegaDWSM}
\end{eqnarray}
We then substitute Eqs.~(\ref{Eq:AFlowOmegaDWSM}) and
(\ref{Eq:VFlowOmegaDWSM}) into Eq.~(\ref{Eq:RhoScalingOmegaDWSMA}),
and obtain the RG equation for $\rho_{d}(\omega)$
\begin{eqnarray}
\frac{d\ln(\rho_{d}(\omega))}{d\ln(\omega)} \sim
1-C_{d1}+C_{d2}+C_{d3}.
\end{eqnarray}

\subsubsection{Triple-WSM}

For free triple-Weyl fermions, the DOS satisfies
\begin{eqnarray}
\rho_{t}(\omega) \sim \frac{\omega^{2/3}}{v B^{2/3}}.
\label{Eq:RhoScalingOmegaTWSMA}
\end{eqnarray}
The Coulomb interaction results in the following equation
\begin{eqnarray}
\frac{d\ln(\rho_{t}(\omega))}{d\ln(\omega)} \sim \frac{2}{3} +
C_{t1} + \frac{d\ln\left(\frac{1}{v B^{2/3}}\right)}{d\ln(\omega)}.
\end{eqnarray}
We employ the transformation $\omega = \omega_{0}e^{-\ell}$ again,
and get the RG equations for $B$ and $A$
\begin{eqnarray}
\frac{d\ln(B)}{d\ln(\omega)} &=& -\left(C_{t2}-C_{t1}\right),
\label{Eq:BFlowOmegaTWSM} \\
\frac{d\ln(v)}{d\ln(\omega)} &=& -\left(C_{t3}-C_{t1}\right).
\label{Eq:VFlowOmegaTWSM}
\end{eqnarray}
Substituting Eqs.~(\ref{Eq:BFlowOmegaTWSM}) and
(\ref{Eq:VFlowOmegaTWSM}) into Eq.~(\ref{Eq:RhoScalingOmegaTWSMA})
yields
\begin{eqnarray}
\frac{d\ln(\rho_{t}(\omega))}{d\ln(\omega)} \sim 1 -
\frac{2}{3}C_{t1} + \frac{2}{3}C_{t2}+C_{t3}.
\end{eqnarray}

\subsection{Specific heat}

\subsubsection{Double-WSM}

The specific heat of free double-Weyl fermions exhibits the
following $T$-dependence
\begin{eqnarray}
C_{v}^{d}(T) \sim \frac{T^{2}}{v A}.
\end{eqnarray}
The Coulomb interaction renormalized $A$ and $v$, and as such leads
to
\begin{eqnarray}
\frac{d\ln(C_{v}^{d}(T))}{d\ln(T)} \sim 2 +
\frac{d\ln\left(\frac{1}{v A}\right)}{d\ln(T)}.
\label{Eq:CvScalingTemperatureDWSMA}
\end{eqnarray}
It is worth mentioning that the anomalous dimension of fermion field
does not qualitatively modify the specific heat. The qualitative
interaction correction to specific heat originates merely from the
fermion dispersion renormalization. To get the $T$-dependence of $A$
and $v$, we need to use the transformation $T=T_{0}e^{-\ell}$, where
$T_{0}$ is certain high temperature scale. It is easy to convert
Eqs.~(\ref{Eq:AVRGDWSMApp}) and (\ref{Eq:VVRGDWSMApp}) into
\begin{eqnarray}
\frac{d\ln(A)}{d\ln(T)} &=& -\left(C_{d2}-C_{d1}\right),
\label{Eq:AFlowTemperatureDWSM}
\\
\frac{d\ln(v)}{d\ln(T)} &=& -\left(C_{d3}-C_{d1}\right).
\label{Eq:VFlowTemperatureDWSM}
\end{eqnarray}
Substituting Eqs.~(\ref{Eq:AFlowTemperatureDWSM}) and
(\ref{Eq:VFlowTemperatureDWSM}) into
Eq.~(\ref{Eq:CvScalingTemperatureDWSMA}), we obtain
\begin{eqnarray}
\frac{d\ln(C_{v}^{d}(T))}{d\ln(T)} \sim 2 - 2C_{d1}+C_{d2}+C_{d3}.
\end{eqnarray}

\subsubsection{Triple-WSM}

The specific heat for free triple-Weyl fermions is
\begin{eqnarray}
C_{v}^{t}(T) \sim \frac{T^{5/3}}{v B^{2/3}}.
\end{eqnarray}
One can show that the renormalized $B$ and $v$ satisfy the following
equations
\begin{eqnarray}
\frac{d\ln(B)}{d\ln(T)} &=& -\left(C_{t2}-C_{t1}\right),
\label{Eq:BFlowTemperatureTWSM}
\\
\frac{d\ln(v)}{d\ln(T)} &=& -\left(C_{t3}-C_{t1}\right).
\label{Eq:VFlowTemperatureTWSM}
\end{eqnarray}
The renormalized specific heat is found to have the form
\begin{eqnarray}
\frac{d\ln(C_{v}^{t}(T))}{d\ln(T)} &\sim& \frac{5}{3} +
\frac{d\ln\left(\frac{1}{v B^{2/3}}\right)}{d\ln(T)} \nonumber
\label{Eq:CvScalingTemperatureTWSMA} \\
&\sim& \frac{5}{3} -\frac{5}{3}C_{t1} + \frac{2}{3}C_{t2}+C_{t3}.
\end{eqnarray}

\subsection{Dynamical Conductivities}

The dynamical conductivities can be calculated by following the
steps adopted to compute DOS.

\subsubsection{Double-WSM}

The dynamical conductivity for free double-Weyl fermions within the
$x$-$y$ plane is
\begin{eqnarray}
\sigma_{\bot\bot}^{d}(\Omega) \sim \frac{e^{2}}{v}|\Omega|.
\end{eqnarray}
After incorporating the corrections due to the Coulomb interaction,
we find that $\sigma_{\bot\bot}^{d}(\Omega)$ satisfies the following
RG equation
\begin{eqnarray}
\frac{d\ln(\sigma_{\bot\bot}^{d}(\Omega))}{d\ln(\Omega)} &\sim&
1+2C_{d1}+\frac{d\ln\left(\frac{e^{2}}{v}\right)}{\ln(\Omega)}
\nonumber \\
&\sim& 1+C_{d1}+C_{d3}.
\end{eqnarray}

The dynamical conductivity along $z$-axis is
\begin{eqnarray}
\sigma_{zz}^{d}(\Omega) \sim \frac{ve^{2}}{A},
\end{eqnarray}
which is altered by the Coulomb interaction to become
\begin{eqnarray}
\frac{d\ln(\sigma_{zz}^{d}(\Omega))}{d\ln(\Omega)} &\sim& 2C_{d1} +
\frac{d\ln\left(\frac{ve^{2}}{A}\right)}{\ln(\Omega)},
\nonumber \\
 &\sim& 2C_{d1} +
C_{d2}-C_{d3}.
\end{eqnarray}

\subsubsection{Triple-WSM}

The dynamical conductivities for free triple-Weyl fermions within
$x$-$y$ plane and along $z$-axis are given by
\begin{eqnarray}
\sigma_{\bot\bot}^{t}(\Omega) &\sim& \frac{e^{2}}{v}|\Omega|,
\\
\sigma_{zz}^{t}(\Omega) &\sim& \frac{ve^{2}}{B^{2/3}}
\frac{1}{\left|\Omega\right|^{1/3}}.
\end{eqnarray}
After including the interaction corrections, we find that
\begin{eqnarray}
\frac{d\ln\left(\sigma_{\bot\bot}^{t}(\Omega)\right)}{d\ln(\Omega)}
&\sim& 1+2C_{t1} +
\frac{d\ln\left(\frac{e^{2}}{v}\right)}{d\ln(\Omega)}, \nonumber
\\
&\sim & 1+C_{t1}+C_{t3},
\end{eqnarray}
and that
\begin{eqnarray}
\frac{d\ln\left(\sigma_{zz}^{t}(\Omega)\right)}{d\ln(\Omega)}&\sim&
-\frac{1}{3}+2C_{t1}+\frac{d\ln\left(\frac{ve^{2}}{B^{2/3}}\right)}{d\ln(\Omega)},
\nonumber \\
&\sim& -\frac{1}{3}+\frac{7}{3}C_{t1}+\frac{2}{3}C_{t2}-C_{t3}.
\end{eqnarray}

\end{document}